\documentclass[a4paper,twocolumn,nofootinbib,superscriptaddress]{revtex4-2}

\usepackage{xcolor}
\usepackage{hyperref}
\hypersetup{
	colorlinks,
	linkcolor={red!90!black},
	citecolor={black!10!blue},
	urlcolor={blue!80!black}
}

\usepackage{amssymb}
\usepackage{bm,amsmath}
\usepackage{graphicx}
\usepackage{array}
\usepackage[normalem]{ulem}
\newcommand{\dr}{\text{d}\mathbf{r}}
\newcommand{\bk}{\mathbf{k}}
\newcommand{\br}{\mathbf{r}}

\renewcommand{\d}{\mathrm{d}}

\begin{document}

\title{Different glassy characteristics are related to either caging or dynamical heterogeneity}

\author{Puneet Pareek}
\email{ppareek@tifrh.res.in}
\affiliation{Tata Institute of Fundamental Research, Hyderabad - 500046, India}
\author{Monoj Adhikari}
\affiliation{Tata Institute of Fundamental Research, Hyderabad - 500046, India}
\author{Chandan Dasgupta}
\affiliation{Department of Physics, Indian Institute of Science, Bangalore 560012, India}
\affiliation{International Centre for Theoretical Sciences, TIFR, Bangalore 560089, India}
\author{Saroj Kumar Nandi}
\email{saroj@tifrh.res.in}
\affiliation{Tata Institute of Fundamental Research, Hyderabad - 500046, India}

\begin{abstract}
Despite the enormous theoretical and application interests, a fundamental understanding of the glassy dynamics remains elusive. The static properties of glassy and ordinary liquids are similar, but their dynamics are dramatically different. What leads to this difference is the central puzzle of the field. Even the primary defining glassy characteristics, their implications, and if they are related to a single mechanism remain unclear. This lack of clarity is a severe hindrance to theoretical progress. Here, we combine analytical arguments and simulations of various systems in different dimensions and address these questions. Our results suggest that the myriad of glassy features are manifestations of two distinct mechanisms. Particle caging controls the mean, and coexisting slow- and fast-moving regions govern the distribution of particle displacements. All the other glassy characteristics are manifestations of these two mechanisms; thus, the Fickian yet non-Gaussian nature of glassy liquids is not surprising. We discover a crossover, from stretched exponential to a power law, in the behavior of the overlap function. This crossover is prominent in simulation data and forms the basis of our analyses. Our results have crucial implications on how the glassy dynamics data are analyzed, challenge some recent suggestions on the mechanisms governing glassy dynamics, and impose strict constraints that a correct theory of glasses must have.
\end{abstract}

\maketitle

What does a theory of glass need to explain? Is there a single defining mechanism of glasses? Are the different glassy characteristics distinct properties?
Glassy dynamics poses one of the most fascinating and challenging problems of statistical physics \cite{giulioreview}. If you supercool a liquid below its melting point without allowing it to crystallize, it becomes glass. Its dynamics becomes spatially heterogeneous \cite{cicerone1996,sillescu1999}, relaxation becomes slower than exponential \cite{williams1970}, particle displacements become non-Gaussian \cite{szamel2006,chaudhuri2007}, etc. The snapshots of a glassy and an ordinary liquid are nearly identical, yet, their dynamics are drastically different \cite{giulioreview}. 
The experimental glass transition is not a thermodynamic transition; ``{\it where and why does liquid end and glass begin}" \cite{science2005} remains a puzzle. Even the primary characteristics of the problem remain confusing \cite{hecksher2008,giulioreview,dhbook,miotto2021,rusciano2022}. On the other hand, glassiness has widespread applications for many crucial processes: the dynamics of a cellular monolayer \cite{angelini2011,garcia2015,souvik2021,sadhukhan2022}, intracellular transport \cite{parry2014,zhou2009,fabry2001}, cancer progression \cite{streitberger2019}, satisfiability problems \cite{biroli2002}, protein folding \cite{wolynes1995}, active systems in their dense regime \cite{dauchot2005,deseigne2010,dreyfus2005,palacci2010,ni2013,berthier2014}, etc. Thus, glass physics is significant from both fundamental and application perspectives.

A defining feature of glassy liquids is the rapid growth of relaxation time, $\tau$, as the temperature, $T$, decreases. Below a particular $T$, equilibration is unreachable, and the system becomes glass. We define the experimental glass transition temperature, $T_g$, when $\tau\sim 10^2-10^3$s. As $T$ approaches $T_g$, several hallmarks of glassy systems appear: First, the relaxation becomes complex. The decay of the self-intermediate scattering function, $F_s(k,t)$ at wave vector $k$ and time $t$, is slower than exponential. Data from a wide variety of systems fit with KWW (Kohlrausch-Williams-Watts) form \cite{kohlrausch1854,williams1970} that is a stretched exponential relaxation (SER): $F_s(k,t)\sim \exp[-(t/\tau_\alpha)^{\beta}]$, $\beta$ is the stretching exponent, $\tau_\alpha$ is the relaxation time, defined via $F_s(k,\tau_\alpha)=1/e$. Second, the mean-square displacement (MSD) changes from being sub-diffusive at intermediate times to diffusive at long times. Third, Stokes-Einstein (SE) relation \cite{einsteinpaper,hansenmcdonald} breaks down as we approach glass transition \cite{cicerone1996,parmar2017,shila2013}. Fourth, $G_s(r,t)$, the probability of particle displacements $r$ at $t$ is non-Gaussian at intermediate times but becomes progressively Gaussian with increasing $t$. Fifth, the overlap function, $Q(t)$, also shows SER. There are other prominent features as well.

One such feature that has emerged as a crucial aspect of glassy liquids is Dynamical heterogeneity (DH) \cite{dhbook,shell2005,ediger2000,berthier2011,yamamoto1998,franz2000,weeks2000,smarajit2015}. DH refers to the coexisting dynamic fast- and slow-relaxing regions. There are strong correlations between DH and other characteristics, such as $\beta$ \cite{cicerone1996}, non-diffusive MSD \cite{szamel2006}, SE relation violation \cite{parmar2017}, non-Gaussian $G_s(r,t)$ \cite{chaudhuri2007}, etc. But, the correlations are not conclusive. In addition, whether the distinct glassy characteristics are interrelated is unclear. This lack of clarity highlights the absence of a fundamental understanding of glassy systems.
Moreover, some recent works \cite{miotto2021,rusciano2022} have argued that glassy liquids are Fickian yet non-Gaussian (FnG); that is, MSD is diffusive, but $G_s(r,t)$ is non-Gaussian. These studies are motivated by the FnG behavior in some complex biological systems \cite{guan2014,wang2009,miotto2021}. As we will argue, and many others have shown \cite{chechkin2017,chubynsky2014,metzler2014,metzler2016,Jain2016}, FnG alone is not surprising \cite{berthiercomment}. Indeed, several works have reported this aspect for glassy systems \cite{szamel2006,das2018,rusciano2022,rusciano2023}. 
However, we emphasize that glasses and other complex systems have a crucial difference. The heterogeneity in these non-glassy systems is exogenous \cite{wang2009,skaug2013,guan2014,chechkin2017,acharya2017}. In contrast, the DH of glassy liquids is self-induced, with a finite lifetime \cite{shell2005,ediger2000}. The central question for glassy systems is how this DH affects the other features. Although, direct answers to these questions are impractical, we can still obtain indirect answers and valuable insights. It is well-known that various glassy characteristics show crossovers from one behavior to another. The crossover times should be similar if there exists a unique mechanism. Contrary to this expectation, different crossover time scales are distinct \cite{szamel2006,saltzman2008,kim2009,das2018}; this also contrasts many works that led to the expectation that DH is central to glassy dynamics \cite{cicerone1996,parmar2017,shell2005,ediger2000}.

In this work, we combine analytical arguments and large-scale simulations of diverse systems and show that two distinct mechanisms exist in glassy liquids: caging and DH. They lead to two different time scales: $\tau_F$ when MSD becomes diffusive and $\tau_G$ when $G_s(r,t)$ becomes Gaussian \cite{rusciano2022,rusciano2023}. All the other crossover times are related to either $\tau_F$ or $\tau_G$ alone. $\tau_G$ is greater than $\tau_F$; compared to $\tau_\alpha$, $\tau_G$ always remains greater, but $\tau_F$ can be smaller at lower $T$. This result calls for a deeper understanding of the time scales and, specifically, the role of caging in glassy dynamics. Note that crossover times can only be defined qualitatively \cite{berthiercomment}. Added to this difficulty is the challenge of precise measurement of $F_s(k,t)$ at very long times due to its inherent fluctuations. By contrast, the absence of fluctuations in $Q(t)$ [see the definition, Eq. (\ref{Qoft})] makes it suitable to study long-time dynamics \cite{spinglassbook,guiselin2020}. Surprisingly, the detailed dynamical properties of $Q(t)$ are not well-understood. We show that the crossover signature is prominent in $Q(t)$ and it forms the basis of our analysis of the other variables. Our results show that glassy liquids resembles an ordinary liquid beyond $\tau_G$: $F_s(k,t)$ is exponential, particle displacements are Gaussian, and SE relation gets restored. We show that the conclusions are also valid in higher dimensions and for diverse systems. The detailed analytical calculations, various definitions, simulation details, and higher dimension results are in the Appendix. We discuss the implications of our results in Sec. \ref{disc}. Our work provides a coherent deeper understanding of various glassy aspects and a clear picture that a correct theory of glass must address.

\section{Results}
\label{Results}
\subsection{Analytical arguments: two different time scales}
\label{twotimescales}
We first present our analytical arguments for the existence of two distinct time scales and their effects on various glassy observables. It will help in analyzing the simulation data presented in the next section. For a general description, we define $G_s(x,t)$ as the van-Hove function for the particle displacement $x$ in a specific direction. $G_s(x,t)$ can be easily related to $G_s(r,t)$ for various dimensions ($d$). Although $Q(t)$ is the most convenient variable to characterize the crossover time scales, it is analytically advantageous to introduce the time scales via $G_s(x,t)$ as it is a fundamental characteristic of glassy systems \cite{chaudhuri2007,giulioreview}, and other variables are related to it.

Several works have shown that the probability of particle displacements in many complex systems can generally lead to exponential tails \cite{chubynsky2014,barkai2020,wang2020}. As shown in the Appendix, Fig. \ref{vanhove_expo}, and many others \cite{szamel2006,chaudhuri2007,miotto2021,rusciano2022}, particle displacements in glassy liquids can also have exponential distribution at intermediate times. However, there is a crucial difference: compared to these complex systems, the exponential tail in glassy liquids is transient, and the distribution crosses over to Gaussian at long times \cite{chaudhuri2007,szamel2006,das2018}. To describe the intermediate and long-time behavior, we propose the following form:
\begin{equation}\label{formofvH}
G_s(x,t) = \frac{1}{t^\nu \Gamma(1+1/\mu)}\exp\left[-\left(\bar{D}\frac{|x|}{t^\nu}\right)^{\mu}\right],
\end{equation}
where we have used $|x|$ to emphasize that the distribution is symmetric, $\Gamma(\ldots)$ is the gamma function. We have set the constant $\bar{D}$, which is related to diffusivity, to unity. Equation (\ref{formofvH}) contains two independent parameters, $\mu$ and $\nu$; both are time-dependent. It is easy to see that $\nu$ determines the MSD, whereas $\mu$ governs the nature of the distribution. The sub-diffusive to diffusive crossover of MSD implies $\nu$ goes from a small value to 1/2 at long times. By contrast, as $G_s(x,t)$ goes from exponential to Gaussian behavior, we expect $\mu$ to vary from 1 to 2 at long times. For glassy systems, the times when $\nu$ goes to 1/2 and when $\mu$ goes to 2 are different: these are the two independent time scales given by $\tau_F$ and $\tau_G$, respectively. $\tau_G>\tau_F$ and all other crossover behaviors are related to one of them.

\subsection{Distinct mechanisms behind the two time-scales}
\label{twotime_mechanism}
For the simplicity of arguments, we confine ourselves to one dimension, extension to higher dimensions is straightforward. Two independent processes control the values of the exponents in Eq. (\ref{formofvH}). The scaling form, $x/t^\nu$, determines the behavior of MSD. By contrast, DH, via the subordination mechanism \cite{chechkin2017}, governs the distribution of particle displacement. We first concentrate on the sub-diffusion. It comes from the properties at the particulate level. At the intermediate time scale, each particle is inside the cage formed by its neighbors. The caging time increases as we approach $T_g$, and when the cage breaks, the particle jumps out of the cage \cite{berthier2011}. Cage breaking is an independent event, expected to have a distribution of times scale, $t_w$. Analytical calculations within trap models and simulation studies of glass-forming liquids \cite{scher1975,bertin2003,niblett2017} seem to suggest that this distribution is a power law,
\begin{equation} \label{waitingtimedist}
\Psi(t_w)\sim \frac{1}{t_w^{1+\delta}},
\end{equation}
with $0<\delta<1$, for which one obtains $\langle x(t)^2\rangle\sim D_\delta t^\delta$, where $D_\delta$ is the generalization of the ordinary diffusivity and has the dimension [Length]$^2$/[time]$^\delta$ \cite{metzler2014} and $\nu=\delta/2$. The sub-diffusive behavior can persist if the particles continue to find ever deeper traps. But, in a glassy system, we expect a crossover from sub-diffusive to diffusive behavior beyond the caging time scale. The distribution of particle displacements locally is Gaussian \cite{metzler2016,lampo2017}
\begin{equation}
G_s(x,t)=\frac{1}{\sqrt{4\pi D_\delta t^\delta}}\exp\left( -\frac{x^2}{4D_\delta t^\delta} \right).
\end{equation} 
However, it can become non-Gaussian if there is a distribution of $D_\delta$. For example, $G_s(x,t)$ can be exponential with an exponential distribution of $D_\delta$ \cite{chechkin2017,lampo2017}. In that case, we will have $\langle x(t)^2\rangle\sim \overline{D_\delta} t^\delta$, where $\overline{D_\delta}$ is the average diffusivity. Thus, sub-diffusive behavior can persist irrespective of the nature of $G_s(x,t)$.

On the other hand, the crossover from non-Gaussian to Gaussian behavior is governed by the medium-heterogeneity that changes over time. The overall particle displacement probability depends on the slow dynamics of the medium. The subordination mechanism \cite{chechkin2017}, proposed by Chechkin {\it et al.} for an annealed disorder, can be applied to glassy systems to explain the crossover of $G_s(x,t)$ from exponential to Gaussian behavior. In glassy liquids, there are two sources of disorder: thermal fluctuation and DH. The first leads to stochastic interaction between a homogeneous medium and a diffusing particle; it drives the stochastic particle motion in the medium. By contrast, the second leads to the rearrangement of the medium; it modulates the local interaction affecting the diffusion of the particle. The latter effect leads to diffusing diffusivity \cite{chubynsky2014,chechkin2017,lampo2017}: the diffusivity changes stochastically as the particle diffuses in the heterogeneous medium. We can define an `internal time' or path length $\tau$ as
\begin{equation}
\tau=\int_0^t D(t') \mathrm{d} t',
\end{equation}
where Ref. \cite{chechkin2017} expressed $D(t)$ via a Ornstein-Uhlenbeck process $Y(t)$: $D(t)=Y^2(t)$. 
Note $\tau$ has the dimension of [Length]$^2$. The probability density function for $\tau$ is $T(\tau,t)$. If the propagator $P(x,\tau)$ for the homogeneous medium is known, then the propagator for the subordinated process in the heterogeneous medium is obtained by averaging $P(x,\tau)$ with $T(\tau,t)$ for all possible $\tau$. The propagator for the homogeneous medium is Gaussian,
\begin{equation}
P(x,\tau)=\frac{1}{\sqrt{4\pi \tau}}\exp\left(-\frac{x^2}{4\tau}\right).
\end{equation}
Then the distribution of particle displacement, i.e., the propagator for the subordinated process, is obtained as
\begin{equation}\label{heteroP}
G_s(x,t)=\int_0^\infty T(\tau,t)P(x,\tau)\mathrm{d}\tau.
\end{equation}
Defining the Fourier transform at wave vector $k$ as
\begin{align}
\hat{P}(k,t)=\int_{-\infty}^\infty e^{ikx}P(x,t)\mathrm{d} x=e^{-k^2\tau},
\end{align}
we obtain $\hat{G}_s(k,t)=\tilde{T}(k^2,t)$, where $\tilde{T}(k^2,t)$ is the Laplace transform of $T(\tau,t)$. The expression of $\tilde{T}(k^2,t)$ is known in the literature \cite{chechkin2017,dankel1991}:
\begin{align}
\tilde{T}(k^2,t)=\frac{e^{t/2}}{\left[\frac{1}{2}\left(a_k+\frac{1}{a_k}\right)\sinh(t a_k)+\cosh(ta_k)\right]^{1/2}}
\end{align}
where $a_k=\sqrt{1+2k^2}$. Thus, the probability density, $\tilde{T}(k^2,t)$, is an explicit function of $t$. The expansion of $\tilde{T}(k^2,t)$ at small $t$ and large $t$ are different. Using these simplified expansions in Eq. (\ref{heteroP}), it is straightforward to see that $G_s(x,t)$ will be exponential for small $t$ and becomes Gaussian at large $t$ \cite{chechkin2017}. Thus, the mechanisms leading to the sub-diffusive MSD and non-Gaussian $G_s(x,t)$ are mutually independent. Our analyses of the simulation data seem to support this scenario for glassy systems.

\subsection{$Q(t)$ becomes power-law and MSD becomes diffusive at $\tau_F$}
\label{secQtGs}
We now discuss the behavior of the MSD and $Q(t)$ when $\nu=1/2$, but $\mu$ is arbitrary and specifically different from 2. A quick look at Eq. (\ref{formofvH}) reveals that $G_s(x,t)$ is non-Gaussian in this time-scale, $\tau_F$.
The MSD in $1d$ is 
\begin{equation}\label{msd_diff}
\text{MSD}(t) = \int_{-\infty}^\infty x^2 G_s(x,t)dx.
\end{equation}
Using Eq. (\ref{formofvH}) at $\tau_F$ for the $d$-dimensional definition of MSD (Eq. \ref{MSDoft}), we obtain MSD$(t)\propto t$ in any dimension.

Another crucial consequence of $\nu$ being 1/2 at $\tau_F$ is that $Q(t)$ becomes power-law. As detailed in the Appendix, using the definitions in the appendix, Eqs. (\ref{Qoft}) and (\ref{Gsoft}), we have
\begin{equation}
Q(t)=\int_0^a G_s(r,t)\dr,
\end{equation}
where $\dr$ is the volume element in dimension $d$. Using the above equation and the form of $G_s(x,t)$ at $\tau$, we obtain ({see Appendix Sec. IA for details), 
\begin{equation}\label{powerlawQt}
Q(t)\sim t^{-d/2}.
\end{equation}
Compared to an exponential, the power law is easier to detect at long times as its decay is slower. As shown in Fig. \ref{Qoftplot} and Fig. \ref{variousdim}, the power-law of $Q(t)$ is quite prominent in the simulation data for different dimensions. Moreover, the exponent of the power-law decay only depends on $d$ and not on $T$, in contrast to the SER exponent $\beta$. All the higher-order correlations, defined via the overlap function, should also follow power law beyond $\tau_F$. Thus, the necessary and sufficient condition for MSD to become diffusive and $Q(t)$ to be power law is $\nu=1/2$. On the other hand, the Gaussian nature of $G_s(x,t)$ at a time $\tau_G$, when $\mu=2$, has critical consequences that we discuss now.

\subsection{$F_s(k,t)$ becomes exponential at $\tau_G$}
\label{fs_exponential}
For the clarity of presentation, we show the calculation for $3d$ only (see Appendix Sec. IB for general dimension). It is convenient to define
\begin{equation}\label{Fsktiso}
F_s(k,t)=F_s(k_x,k_y,k_z,t)\big|_{k_x^2+k_y^2+k_z^2=k^2},
\end{equation}
where $k_i$'s are the wave vectors in directions $i=x,y,z$, and the condition enforces isotropy. Using the definitions in the appendix, Eqs. (\ref{Fskt}) and (\ref{Gsoft}), we have
\begin{align}\label{FsGsrelation}
F_s&(k_x,k_y,k_z,t)=\frac{1}{(2\pi)^{3/2}}\int e^{i\bk\cdot\br} G_s(\mathbf{r},t)\dr \nonumber\\
&=\frac{1}{{2\pi}^{3/2}}\int e^{i \bk\cdot\br}G_s(x,t)G_s(y,t)G_s(z,t) \mathrm{d} x\mathrm{d}y\mathrm{d}z,
\end{align}
where we have used the fact that particle displacements along various directions are independent. Then, we can write the integral in Eq. (\ref{FsGsrelation}) as products of three identical integrals, $F_s(k_x,k_y,k_z)=I(k_x)I(k_y)I(k_z)$, where
\begin{equation}\label{xint_Fs}
I(k_x)=\frac{1}{\sqrt{2\pi}}\int_{-\infty}^\infty e^{ik_xx}G_s(x,t){\rm d}x,
\end{equation}
and similarly for the other two integrals. Since we are interested at $t>\tau_G$, $\mu=2$ and $\nu=1/2$ (see Sec. \ref{twotimescales}). Then $G_s(x,t)\sim \exp(-x^2/t)/\sqrt{t}$, and as detailed in the Appendix, Sec. SIB, we obtain 
\begin{equation}\label{Fsktexponential}
F_s(k,t)=e^{-k^2t/4}.
\end{equation}
Thus, $F_s(k,t)$ becomes exponential and relaxation time varies as $\tau_\alpha\sim 1/k^2$ when the van-Hove function goes to Gaussian.

\subsection{Behavior of $Q(t)$ when $F_s(k,t)$ becomes exponential}
Finally, we discuss what happens to $Q(t)$ at $\tau_G$ when $G_s(x,t)$ becomes Gaussian and $F_s(k,t)$ becomes exponential. To address this question, we exploit the analytical relation between $Q(t)$ and $F_s(k,t)$:
\begin{equation}\label{QtFsktrel1}
Q(t)=\frac{1}{(2\pi)^{3/2}}\int_0^a\int_{-\infty}^{\infty}F_s(k,t)e^{-i\bk\cdot\br}{\mathrm d}\bk\dr,
\end{equation}
where ${\mathrm d}\bk$ and $\dr$ are the volume elements in Fourier and real spaces, respectively. Here we present the calculation for $3d$ only (see Appendix for other dimensions). First, let us consider the spatial integration. Using the isotropy of the system (since all directions are equivalent), we obtain
\begin{align}
\int_0^a e^{-i\bk\cdot\br}\dr&=\int_0^a\int_0^\pi e^{-ikr\cos\theta}2\pi r^2\sin\theta\mathrm{d}r\mathrm{d}\theta\nonumber\\
=\frac{4\pi}{k}&\int_0^a r\sin(kr) \mathrm{d}r=\frac{4\pi}{k^3}[\sin{(ka)}-ka\cos(ka)].
\end{align}
Using the above result in Eq. (\ref{QtFsktrel1}), we obtain, when $F_s(k,t)$ becomes exponential,
\begin{equation}\label{QtFsktrel2}
Q(t)=\sqrt{32\pi}\int_0^\infty e^{-k^2t/4}\frac{\sin{(ka)}-ka\cos(ka)}{k^3}k^2 \mathrm{d}k.
\end{equation}
Since we are interested in the long-time behavior alone, we take $t\to \infty$. In this limit, due to the exponential factor in the integrand, the small $k$ values will give the dominant contribution to the integral. The leading order contribution from the trigonometric part becomes $\sin{(ka)}-ka\cos(ka)\simeq (ka)^3/3$. Therefore, from Eq. (\ref{QtFsktrel2}), we obtain
\begin{equation}\label{Qt_a_cube}
Q(t)\simeq\frac{\sqrt{32\pi}a^3}{3}\int_0^\infty e^{-k^2t/4}k^2 \mathrm{d}k=\frac{8\sqrt{2}\pi a^3}{3}t^{-3/2}.
\end{equation}
We show in the Appendix (Fig. \ref{variousdim}) that $Q(t)\sim t^{-d/2}$ in dimension $d$. Note that $Q(t)$ already assumes the same power law at an earlier time, $\tau_F$. Thus, the change in $F_s(k,t)$ will not affect the behavior of $Q(t)$. The same arguments also apply to MSD. The crossovers in $Q(t)$ and MSD only comes from the change in $\nu$. The two times, $\tau_F$ and $\tau_G$, are controlled by different mechanisms and affect distinct observables. Now we show the simulation results supporting these analytical arguments.

\begin{figure}
	\includegraphics[width=8.6cm]{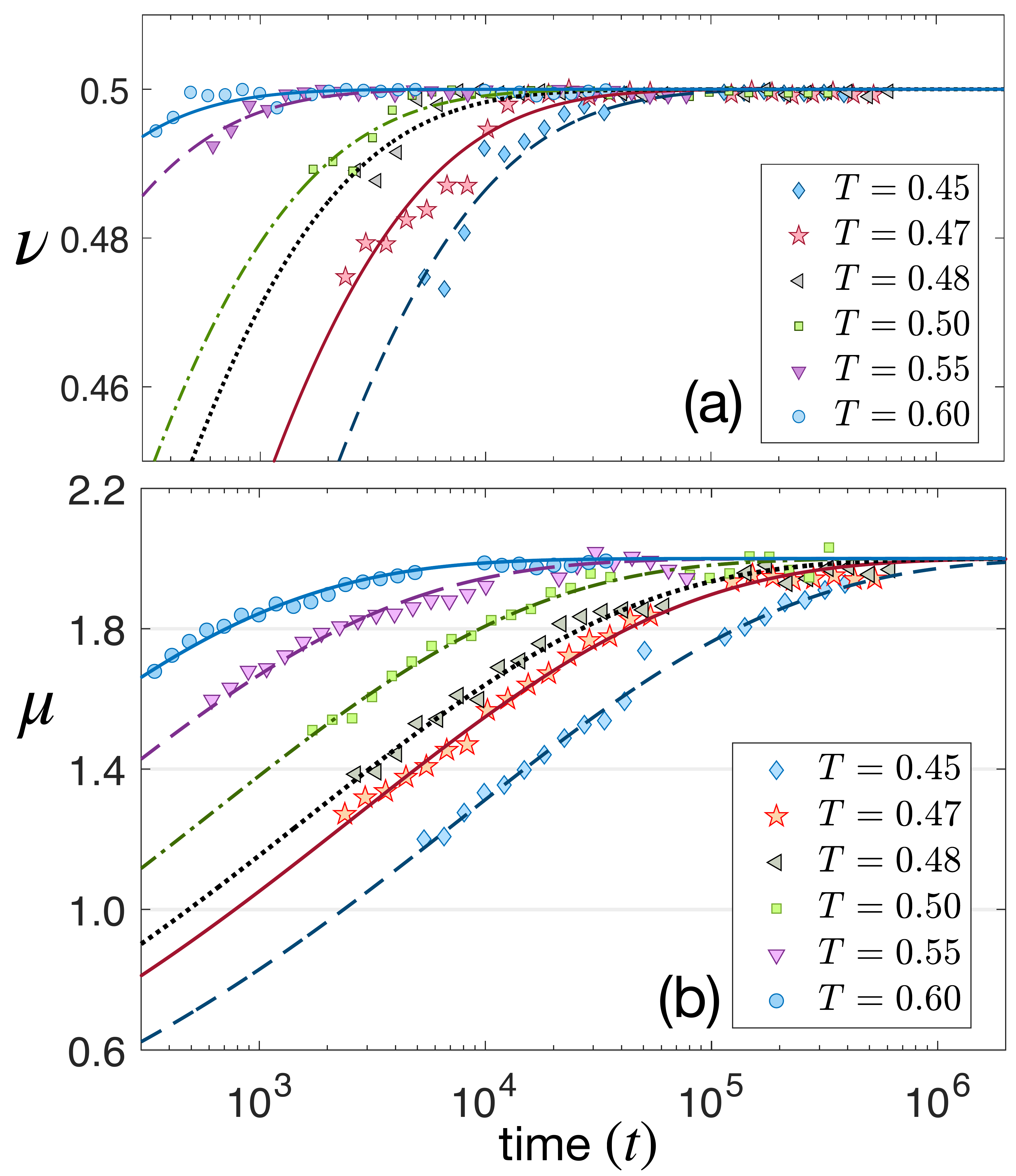}
	\caption{Existence of the two distinct time scales. We fit Eq. (\ref{formofvH}) with the simulation data of $G_s(x,t)$ at different times and obtain $\nu$ and $\mu$. $\nu$ and $\mu$ are shown as a function of time for various $T$ in (a) and (b), respectively. The lines represent fit to a stretched exponential form with the stretching exponent 0.3. The excellent fits confirm that $\nu=1/2$ and $\mu=2$ beyond specific time scales, defining $\tau_F$ and $\tau_G$ respectively.}
	\label{twotime_numu}
\end{figure}

\begin{figure*}
	\includegraphics[width=17.2cm]{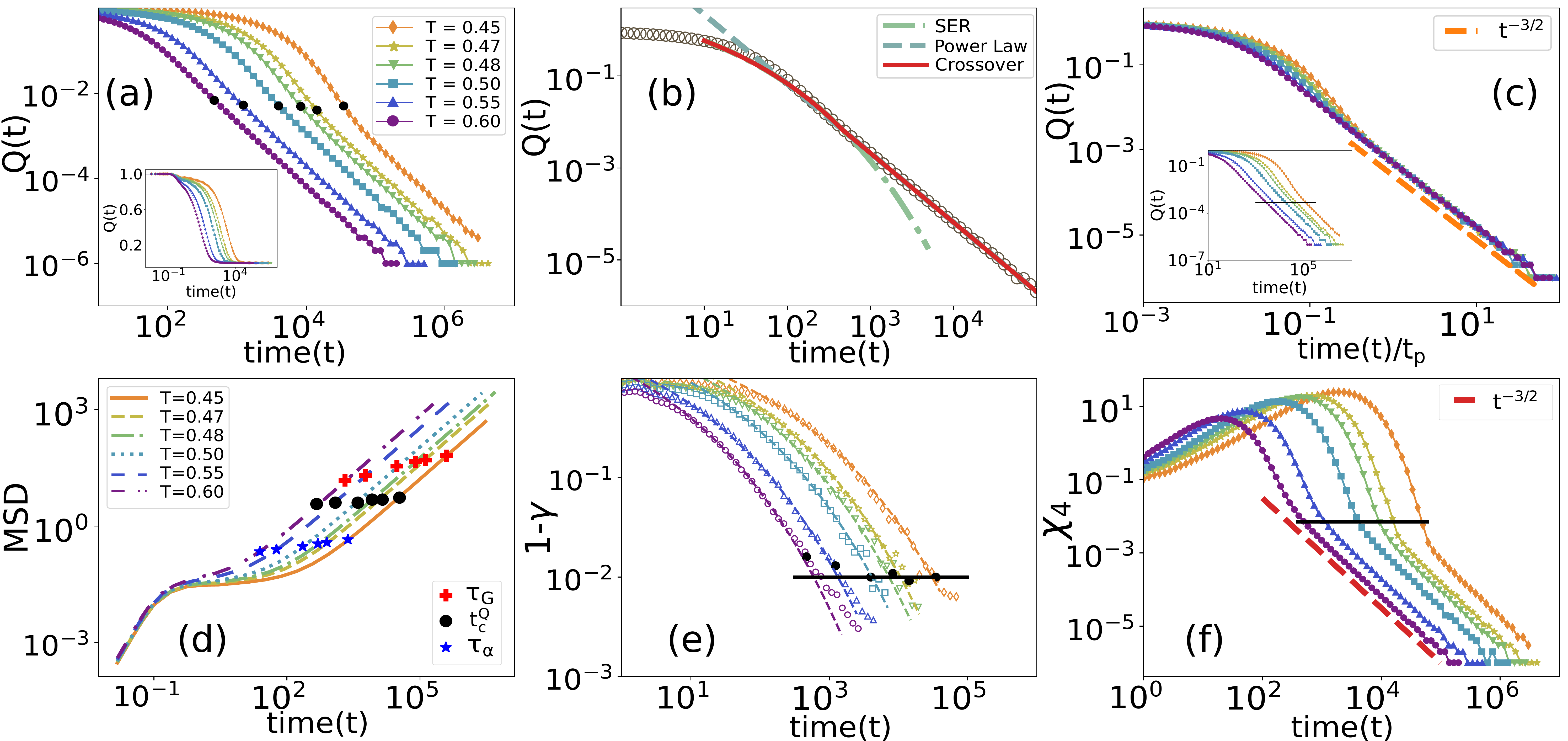}
	\caption{Crossover behaviors of $Q(t)$ and MSD. (a) Q(t) shows a crossover from SER to Power Law. {\bf Inset:} $Q(t)$ in the semi-log plot. (b) Separate fits of the SER and power-law forms to the intermediate and long-time data for $Q(t)$ at $T=0.60$. The solid line is fit with Eq. (\ref{wholerangefitting}). (c) We define a time scale $t_p$ in the power-law regime of $Q(t)$ such that $Q(t_p)=5\times10^{-4}$ (inset). Excellent data collapse in the long time when $Q(t)$ is plotted as a function of $t/t_p$ shows that the power law exponent only depends on $d$. (d) MSD as a function of $t$ in the log-log plot. The three time scales, $\tau_\alpha$, $\tau_F\equiv t_c^Q$, and $\tau_G$ are marked for comparison. (e) (1-$\gamma$) approaches zero as a stretched exponential; lines are fits and symbols are simulation data. We define $\gamma=0.99$ at $\tau_F$ (shown by the horizontal line). (f) We visually characterize the crossover time in $\chi_4(t)$, marked by the horizontal line. The dashed line is the plot of the function $\sim t^{-d/2}$.}
	\label{Qoftplot}
\end{figure*}

\section{Simulation results}

\subsection{The two timescales : $\tau_F$ and $\tau_G$}
\label{twotimes}
We now present simulation data supporting our analytical arguments. We show the simulation data for $3d$ in the main text and other dimensions ($d=2,4,5,6$, Figs. \ref{Q(t)_2d_chandan_analysis_suppl} - \ref{variousdim}) data in the Appendix. We first demonstrate the existence of the two distinct time scales, $\tau_F$ and $\tau_G$, defined via Eq. (\ref{formofvH}) when $\nu=1/2$ and $\mu=2$, respectively. We fit Eq. (\ref{formofvH}) with the data for $G_s(x,t)$ at different times (see Fig. \ref{vanhove_expo} for the fits) and obtain the values of $\nu$ and $\mu$ as shown in Fig. \ref{twotime_numu} for six different temperatures. Clearly, $\nu$ becomes 1/2 much before $\mu$ goes to 2. We fit the data of both $\nu$ and $\mu$ with the stretched exponential form, $f(x)=A(1-\exp[-(t/\tau)^\delta])$; we fix $A=0.5$ and $2$ for the data of $\nu$ and $\mu$ respectively as these are their saturation values. Furthermore, we were able to fit the simulation data of $\nu$ and $\mu$ with a constant $\delta=0.3$ and $\tau$ as a single fitting parameter. We show the fits in Figs. \ref{twotime_numu}(a) and (b) with the lines. The excellent fits with a single fitting parameter of the data with the stretched exponential form, as opposed to a power law, confirm that these two times are well-defined.
Since these are crossover time scales, obtaining their precise values from the data is challenging. Therefore, looking at them from different angles is crucial to gain insights. We now look at the effects and consequences of $\tau_F$ and $\tau_G$ on the other observables.

\subsection{Crossover behaviors of $Q(t)$ and MSD at $\tau_F$}
As analytically shown in Sec. \ref{secQtGs}, $Q(t)$ and MSD show crossover behaviors at $\tau_F$. We first look at the crossover of $Q(t)$ from SER to a power law. Computation of $Q(t)$ at long times is relatively easy in simulations due to the lack of fluctuations. We show the simulation data in spatial dimension three in Fig. \ref{Qoftplot} (a); the power-law decay is evident in the log-log plot. The initial form is a stretched exponential and then a power law. Although the crossover is unmistakable, the crossover time, $t_c^Q$, is difficult to estimate. For a reliable estimate of $t_c^Q$, we first fit the short-time and long-time parts of $Q(t)$ with  SER and power-law, respectively. We next define a function combining these two functional forms as follows
\begin{equation}\label{wholerangefitting}
Q(t) \sim h(t)\exp[-(t/\tau_D)^{\beta_Q}] + (1-h(t))t^{-d/2}
\end{equation}
where $h(t) = \exp[-(t/t_c^Q)^{n}]$, $n \geq 1$ is a real number. We obtain the relaxation time $\tau_D$ and the stretching exponent $\beta_Q$ of the SER form from the fit of the short-time data. $h(t)$ in Eq. (\ref{wholerangefitting}) extrapolates between the two regimes with the crossover time $t_c^Q$: the SER part dominates when $t << t_c^Q$ and the power law part dominates when $t >> t_c^Q$. The value of $n$ depends on the nature of the crossover; we kept it as a free parameter and found it to be generally greater than 1. Figure \ref{Qoftplot}(b) shows a particular fit of Eq. (\ref{wholerangefitting}) with the data. We indicate the values of $t_c^Q$ at different $T$ in Fig. \ref{Qoftplot}(a). Figure \ref{Qoftplot}(c) shows $Q(t)$ as a function of $t/t_p$, where we chose $t_p$ in the power-law regime via $Q(t_p)=5\times10^{-4}$; the excellent data collapse in the long times confirms that the power-law exponent depends only on $d$.

We now analyze the crossover behavior in MSD, shown in Fig. \ref{Qoftplot}(d) for various $T$. 
Let us consider the long-time behavior of MSD as MSD$(t)\sim t^\gamma$; the diffusive behavior sets in when $\gamma=1$. In practice, obtaining the precise time when $\gamma$ becomes unity is challenging \cite{berthiercomment}. Therefore, we study the approach of $(1-\gamma)$ towards zero to determine the crossover time. Figure \ref{Qoftplot}(e) shows the behavior of ($1-\gamma$) as a function of time for various $T$. If $(1-\gamma)$ approaches zero algebraically, then there is no intrinsic time scale, and $\tau_F$ becomes meaningless \cite{berthiercomment}. On the other hand, if the approach is some form of exponential, even if stretched, there is an intrinsic time-scale associated with the approach of $\gamma$ to unity. The dashed lines in Fig. \ref{Qoftplot}(e) show that the data of $(1-\gamma)$ fit well with a stretched exponential, i.e., we can define a time scale for the  diffusive nature of MSD. The difficulty of extracting this time scale from the simulation data still remains: it essentially becomes defining a tolerance value for $(1-\gamma)$. We have checked that all the timescales determined via values of $(1-\gamma)$ from 0.1 to 0.01 are proportional to each other (Fig. \ref{vanhove_expo} in the Appendix). This proportionality is not surprising when the cut-off is in the diffusive regime. The approach from sub-diffusive to diffusive is weak, but from our analytical arguments, we know that $t_c^Q=\tau_F$. Using this knowledge, we find $(1-\gamma)\sim 0.01$ at $\tau_F$ (Fig. \ref{Qoftplot}e). This definition is consistent with earlier works \cite{das2018,szamel2006}. The four-point correlation function, $\chi_4(t)$ that is the variance of $Q(t)$ (Eq. \ref{chi4eq}), also decays as $t^{-d/2}$ at long times (Fig. \ref{Qoftplot}f). We visually identify the crossover time to this power law decay and show by the horizontal line in Fig. \ref{Qoftplot}(f).

\subsection{Behavior at $\tau_G$}
What are the effects of the time scale $\tau_G$ when $\mu$ becomes 2, and $G_s(x,t)$ becomes Gaussian (Eq. \ref{formofvH})? We showed in Fig. \ref{twotime_numu}(b) that $\mu$ approaches 2 in a stretched exponential fashion; this implies the same behavior for the approach of $G_s(x,t)$ to the Gaussian behavior. However, in contrast, it has been suggested in the literature that this approach may be algebraic \cite{berthiercomment}, implying the inaccessibility of $\tau_G$. Therefore, we first analyze this aspect in more detail. The two most convenient and widely-used procedures to quantify this are via the Binder cumulant, $B(t)$, and the non-Gaussian parameter, $\alpha_2(t)$. $B(t) = -1 + \langle \Delta x^4(t)\rangle/3\langle \Delta x^2(t)\rangle^2$, where $\Delta x(t)$ is the displacement of a particle in a specific direction in time $t$ and the averages are over different time origins and ensembles \cite{binder1981}. Similarly, $\alpha_2(t)=3 \langle \Delta r^4(t)\rangle/5[\langle \Delta r^2(t)\rangle]^2-1$, where $\Delta r(t)$ is the displacement in time $t$ \cite{rahman1964}. $B(t)$ and $\alpha_2(t)$ are zero when the particle displacements are Gaussian. Thus, the approach of $B(t)$ and $\alpha_2(t)$ towards zero quantifies the evolution of $G_s(x,t)$ towards Gaussian behavior.

\begin{figure}
	\includegraphics[width=8.6cm]{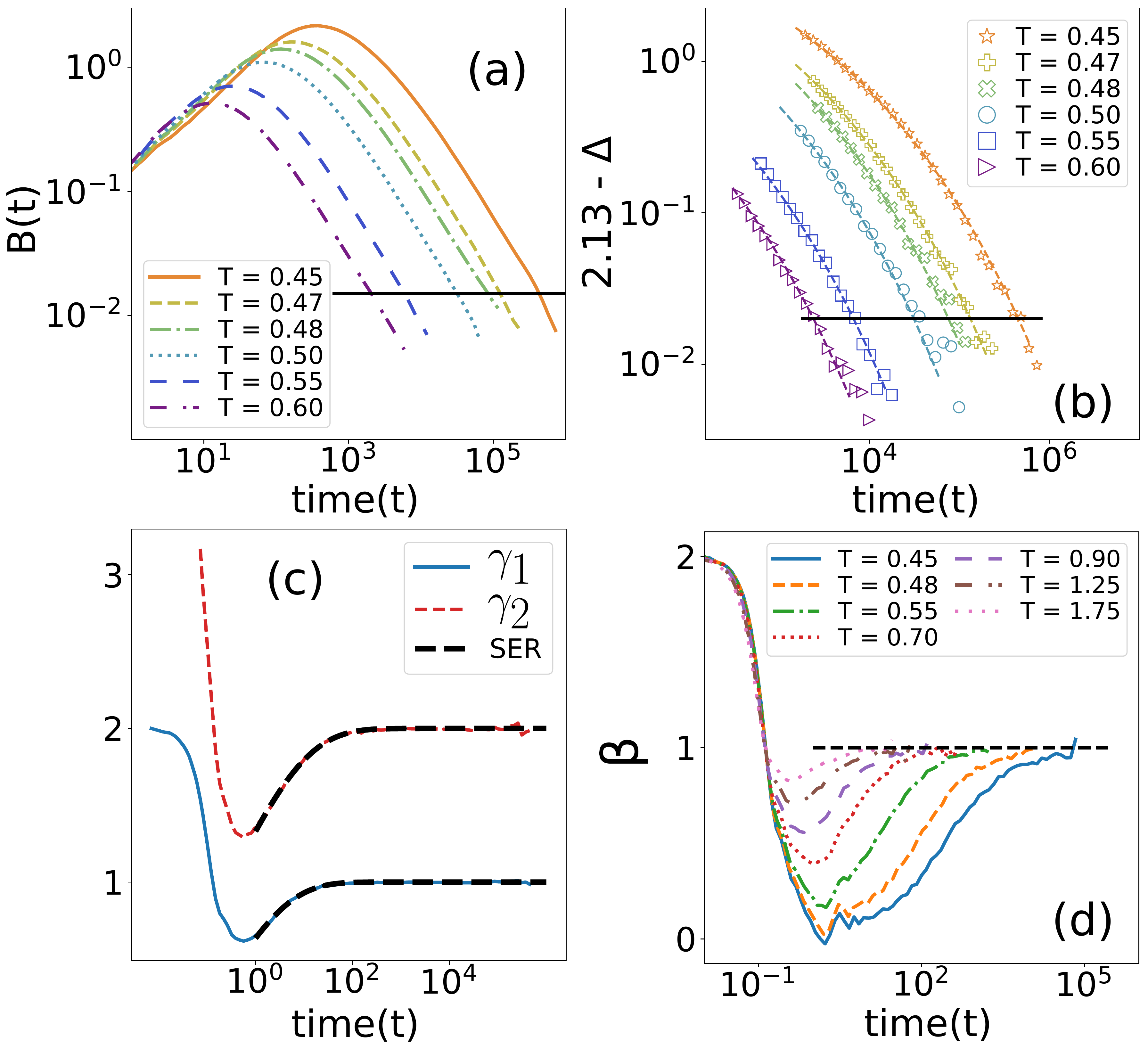}
	\caption{Behavior at $\tau_G$. (a) The decay of the Binder cumulant, $B(t)$, towards zero. The horizontal lines at 0.015 indicate when $G_s(x,t)$ becomes Gaussian. (b) $\Delta(t)=2.13$ for a Gaussian variable. $2.13-\Delta$ decays towards zero as a stretched exponential; dashed lines are fits, and symbols are the simulation data. (c) Analyzing the slopes $\gamma_1$ and $\gamma_2$ of $\ln \langle x^2(t)\rangle$ and $\ln\langle x^4(t)\rangle$ with respect to $\ln t$ provides an alternate procedure to test how $G_s(x,t)$ reaches the Gaussian behavior (see text). (d) We numerically evaluate the stretching exponent $\beta$ of $F_s(k,t)\sim \exp[-(t/\tau)^\beta]$ for $k=2$. $\beta$ goes from a small value at intermediate times towards unity at long times.}
	\label{tauGbehavior}
\end{figure}

The behaviors of $B(t)$ and $\alpha_2(t)$ at different $T$ are consistent with stretched exponential with a small stretching exponent (Fig. \ref{tauGbehavior}(a) and Fig. \ref{binder_fit_suppl}). When the stretching exponent is small, distinguishing a stretched exponential from a power law within a small range of data is challenging. 
We have checked that the data of both $B(t)$ and $\alpha_2(t)$ are compatible with the stretched exponential form with the same stretching exponent, 0.12 (Appendix Fig. \ref{binder_fit_suppl}); since the stretching exponent value is small, the function looks nearly a power law. Therefore, we looked at alternative tests for the approach of $G_s(x,t)$ towards Gaussian behavior. 
We looked into the peak height $\Delta(t)$ of $P(\log_{10} r,t)$, where $P(\log_{10}r,t)=(\ln 10) 4\pi r^3G_s(r,t)$. $\Delta(t)=2.13$ when $G_s(r,t)$ is Gaussian (Fig. \ref{p_log_r_T_0.60}). Figure \ref{tauGbehavior}(b) shows that $\Delta(t)$ also approaches 2.13 as a stretched-exponential.

In addition, we characterized the slopes of $\langle x^2(t)\rangle$ and $\langle x^4(t)\rangle$. When $x(t)$ has a Gaussian distribution, it is easy to see that $\langle x^2(t)\rangle\sim t$ and $\langle x^4(t)\rangle\sim t^2$. 
We define $\gamma_1=\partial \ln \langle x^2(t)\rangle/\partial \ln t$ and $\gamma_2=\partial \ln \langle x^4(t)\rangle/\partial \ln t$; thus, $\gamma_1=1$ and $\gamma_2=2$ for a Gaussian process.
Figure \ref{tauGbehavior}(c) shows that both $\gamma_1$ and $\gamma_2$ reach their respective values as stretched exponentials. Therefore, $B(t)$ will also approach zero with the same functional form. All these tests are consistent with the result that the approach of $G_s(x,t)$ towards the Gaussian nature is a stretched exponential; therefore, the time scale $\tau_G$ is well-defined \cite{ruscianoreply}. 
As before, we obtain $\tau_G$ from simulation data by defining a tolerance value: $B(\tau_G)=0.015$ and $\alpha_2(\tau_G)=0.015$. These definitions are not unique but guided by $t_c^Q$. As discussed in Sec. \ref{twotime_mechanism}, DH is the source of the non-Gaussian nature of $G_s(x,t)$ in glassy liquids. The DH in glasses is self-induced. In contrast, DH in many complex and biological systems \cite{wang2009,guan2014,skaug2013} is exogenous. Although the Fickian yet non-Gaussian behavior in these systems is similar to that in a glassy liquids \cite{rusciano2022,chechkin2017,chubynsky2014}, a direct comparison is imprecise.

Finally, we present the data supporting the analytical result of Sec. \ref{fs_exponential} that $F_s(k,t)$ crosses over from stretched exponential to an exponential form at $\tau_G$. We write $F_s(k,t)\sim \exp[-(t/\tau)^\beta]$ with $\tau$ being some relaxation time and $\beta$ the stretching exponent that depends on time. Therefore, the derivative of $\log[-\log\{F_s(k,t)\}]$ with respect to $\log t$ gives $\beta$. We numerically evaluate this derivative $\beta$ at different times. Figure \ref{tauGbehavior} (d) shows $\beta$ as a function of time for $k=2.0$. We find that at long times, $\beta$ becomes one, i.e., $F_s(k,t)$ becomes exponential. As detailed in the Appendix (Fig. \ref{binder_fit_suppl}c), $\beta$ approaches unity as a compressed exponential. Therefore, we can define a tolerance value and obtain $\tau_G$; via $F_s(k,t)$, we get $\tau_G$ when $\beta=0.99$.

\begin{figure}
	\includegraphics[width=8.6cm]{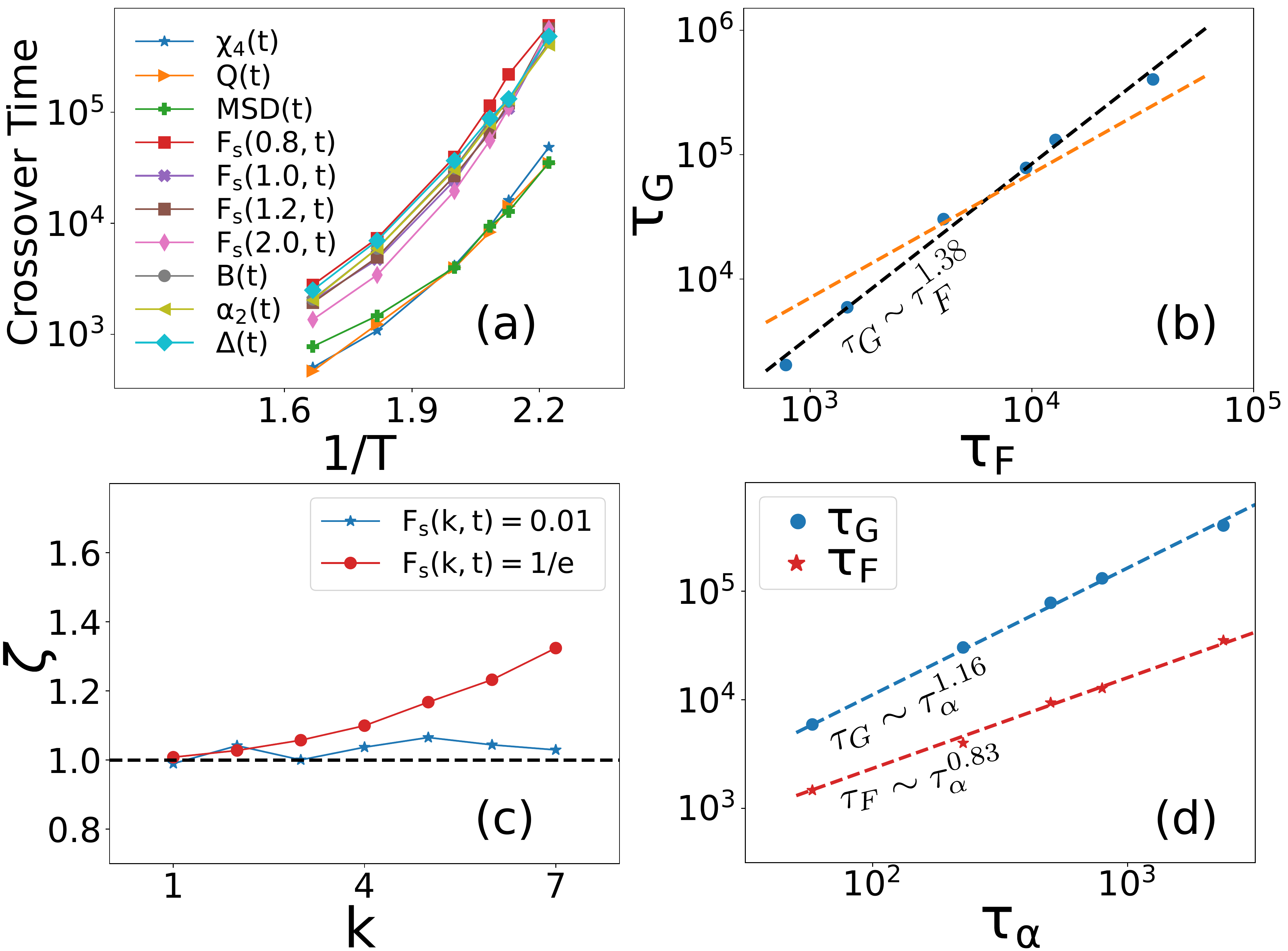}
	\caption{The emerging picture of glassiness. (a) The crossover times of various glassy characteristics are related to either $\tau_F$ or $\tau_G$. (b) The plot of $\tau_G$ vs $\tau_F$ in the log-log scale shows the slope is not unity, implying they are not proportional. The orange dashed line shows the line with slope unity. (c) We chose a set of parameters for which the breakdown of SE relation becomes stronger for larger $k$ when the relaxation time is defined via $F_s(k,\tau_\alpha)=1/e$. However, the relation is restored beyond $\tau_G$. (d) Variation of $\tau_G$ and $\tau_F$, as a function of $\tau_\alpha$, follow a power law with exponents 1.16 and 0.83, respectively.}
	\label{crossover_times}
\end{figure}

\subsection{The emerging picture involving the two time scales}
We now discuss the emerging picture, which is a consequence of having only two distinct time scales, $\tau_F$ and $\tau_G$. We have shown that the parameters $\nu$ and $\mu$ of Eq. (\ref{formofvH}) reach 1/2 and 2 in a stretched exponential fashion. Therefore, the time scales are well-defined. Furthermore, the crossover behaviors for different observables in a glassy system are related to one of them. Figure \ref{crossover_times}(a) shows that the crossover times of MSD, $Q(t)$, and $\chi_4(t)$ are similar, whereas the times related to $\alpha_2$, $B(T)$, $\Delta(t)$, and $F_s(k,t)$ are proportional to each other. A suitable choice of the cut-off parameters can make the different times overlap on the two distinct time scales. The value here is motivated by the pronounced signature of the crossover in $Q(t)$. The time scale in the first set of parameters is $\tau_F$, and the second is $\tau_G$. 
We plot these two times in Fig. \ref{crossover_times}(b). $\tau_G$ is {\it not} proportional to $\tau_F$; this shows that $\tau_F$ and $\tau_G$ are distinct and, as discussed in Sec. \ref{twotime_mechanism}, have different mechanisms controlling them. 
Thus, MSD becomes diffusive, and $Q(t)$ and $\chi_4(t)$ decay with a power law much before DH dies out in a glassy system; the behavior of these variables is governed by the caging that survives for a time scale $\tau_F$. On the other hand, the DH persists till a time $\tau_G$ and controls the behaviors of $G_s(x,t)$, $B(t)$, $\alpha_2(t)$, $F_s(k,t)$, etc. Thus, DH is a crucial aspect of glassy systems: it shows up in diverse glassy characteristics, has a self-life, and dies out after $\tau_G$. Beyond $\tau_G$, a glassy liquid resembles an ordinary liquid.

Let us now test this emerging picture. The SE relation \cite{einsteinpaper} breaks down in a glassy system due to the presence of DH \cite{sillescu1999,dhbook,shila2013,parmar2017,shell2005}. The SE relation states that diffusivity varies inversely with relaxation time. The violation of this relation has a complex nature: it holds at a small $k$ (long length scale) but breaks down at a large $k$ (short length scale) \cite{parmar2017}. 
Writing $D\sim \tau^{-\zeta}$, violation of the SE relation implies $\zeta\neq 1$. This definition is suitable for a system-independent investigation. When probed at long-length scales, distinct parts of the system look identical. The DH is prominent only when the probing length is smaller or of the order of the DH lengths. Our results imply that DH dies out at all $k$, and the system becomes homogeneous beyond $\tau_G$. To test this, we chose a set of $k$ for which the SE relation breaks down (Fig. \ref{crossover_times}c) when we probe the system at times of the order of $\tau_\alpha$ (Fig. \ref{d_vs_tau} in the Appendix). However, if we define the relaxation time as $\tau_G$ or higher, the SE relation should hold again. Figure \ref{crossover_times}(c) shows if we define the relaxation time via $F_s(k,\tau)=0.01$, the relation becomes valid again at all values of $k$. A similar result was also reported in Ref. \cite{parmar2017}, though the reason was not evident. The restoration of the SE relation has also been reported in the past \cite{kawasaki2017} with varying definitions of relaxations. 
As emphasized earlier, the violation of the SE relation is related to the DH \cite{kawasaki2017,parmar2017,shell2005}. However, our work reveals that the diffusive dynamics in glassy liquids is controlled by an entirely distinct mechanism that leads to the other set of time scales related to $\tau_F$.

Restoration of the SE relation at longer times confirms a fascinating phenomenology about DH. The traditional way of defining the relaxation time $\tau_\alpha$ via the relation $F_s(k,\tau_\alpha)=1/e$ is adequate when all the particles have similar characteristics. However, this definition predominantly includes the fast particles. True relaxation of the system also requires relaxing the slow particles; when this happens, the decay of $F_s(k,t)$ becomes exponential, similar to an ordinary liquid. Reference \cite{stillinger1988} argued that slow-moving particles are more relevant for glassy dynamics as the fast-moving particles are unlikely to be strongly affected by the change in $T$. Subsequently, many works have argued along similar lines \cite{donati1998,kumar2006}. The slow-moving particles also seem to play a crucial role in the aging behavior of glassy systems \cite{douglass2022}. Besides, $F_s(k,t)$ is exponential at small $k$ because one is sampling a large length scale compared to that of DH. However, at long times, all possible nature of heterogeneity passes through the system at all lengths; the time scale for this to happen is $\tau_G$. Beyond this time, the system appears homogeneous, SE relation becomes valid again, and $F_s(k,t)$ decays exponentially at all $k$. This picture is consistent with recent experimental results \cite{niss2020} and has consequences for theories of glassy dynamics.

Considering that $\tau_\alpha$ predominantly signifies the relaxation of the fast particles, one aspect of $\tau_F$ and $\tau_G$ deserves attention. As Fig. \ref{crossover_times}(d) shows, $\tau_F$ and $\tau_G$ vary with $\tau_\alpha$ as power laws: $\tau_G\sim \tau_\alpha^a$ and $\tau_F\sim \tau_\alpha^b$. Figure \ref{crossover_times}(d) reveals two aspects of our data: (1) both $\tau_F$ and $\tau_G$ are greater than $\tau_\alpha$ in the range of our simulations. (2) our data suggest that $a>1$ and $b<1$. The second aspect implies $\tau_G$ remains greater than $\tau_\alpha$ at all $T$. However, since $b<1$, beyond a particular $T$, $\tau_F$ becomes smaller than $\tau_\alpha$. This result represents a crossover of the nature of the caging. The slow particles govern caging at relatively high $T$, whereas the fast particles dominate it at lower $T$. The implication of this result is significant: particle caging becomes irrelevant for glassy dynamics, contrary to some recent proposals \cite{massimo2023,li2020}. Conversely, the power-law relation of $\tau_F$ with $\tau_\alpha$ may break down at lower $T$. Resolution of this aspect can provide crucial insights into caging and its role in glassy dynamics.

\begin{figure}
\includegraphics[width=8.6cm]{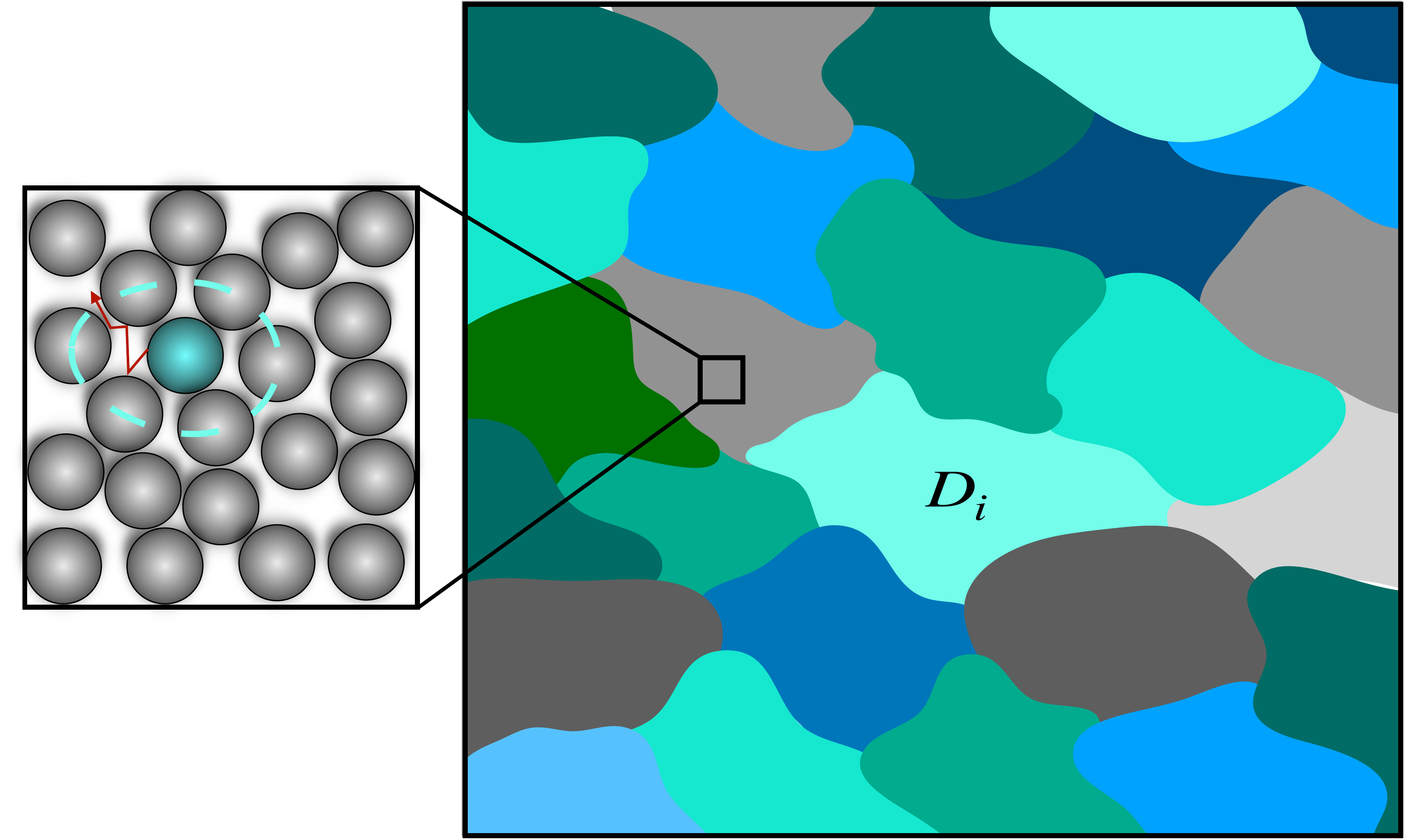}
\caption{The caging property is related to the particulate length scale, much shorter than the DH lengths, where the subordination mechanism is applicable. The latter associates the varying diffusivities within each region to particle displacement $r$ at time $t$. When a particle moves in a homogeneous medium, the distribution of $r$ is Gaussian. However, in a heterogeneous medium, it is a function of the path length or ‘internal time’ $\tau$, which is the integration of the stochastic diffusivities in its path in time $t$. This mechanism implies that we obtain the distribution in the heterogeneous medium by averaging that in the homogeneous medium over the probability density of $\tau$. This subordination leads to non-Gaussian behavior at short times and Gaussian at long times.}
\label{subordinate}
\end{figure}

\section{Discussion and Conclusion}
\label{disc}

To conclude, we have focused on the nature of the defining traits of a glassy liquid. Specifically, if various glassy features are distinct or manifestations of something more fundamental. Our results suggest that the diverse characteristics are manifestations of two primary mechanisms. The first is the caging of particles at an intermediate time scale $\tau_F$. Beyond $\tau_F$, MSD becomes diffusive, and $Q(t)$ and the corresponding $\chi_4(t)$ decay as a power law. The second is the DH that vanishes at a time scale $\tau_G$, with $\tau_G>\tau_F$. Beyond $\tau_G$, the slow-moving particles of the system also relax. Then $G_s(x,t)$ becomes Gaussian, $B(t)$ and $\alpha_2(t)$ become zero, the decay of $F_s(k,t)$ becomes exponential, and SE relation is restored at all $k$ again. Thus, glassy liquids resemble an ordinary liquid beyond $\tau_G$. We have tested these results in simulations of different models in various dimensions. Our work provides a coherent picture of a glassy system and what needs to be explained to understand glassiness: one must elucidate the origins and implications of caged particle motions and dynamical heterogeneity; everything else is a simple consequence of these two.

The basis of our analysis of the various glassy characteristics is their crossover time scales. A precise evaluation of a crossover time is generically challenging. However, we have shown that the crossover in $Q(t)$ is quite prominent. We have defined an interpolating function between the SER and power law forms and objectively estimated the crossover time scale. This crossover time facilitates the analysis of the other time scales. $Q(t)$ and $F_s(k,t)$ are assumed to contain similar information. For example, the relaxation times $\tau$ and $\tau_\alpha$, defined via $Q(\tau)=F_s(k,\tau_\alpha)=1/e$, are equivalent. Note that the definition of $Q(t)$ has a scale, $a$, associated with it (Eq. \ref{Qoft}), and the wavevector $k$ is explicit in $F_s(k,t)$. Thus, the relaxation times are functions of $a$ and $k$. As shown in the Appendix Fig. \ref{tau_q_tau_k_T_0.47}, $\tau$ and $\tau_\alpha$ are similar when we take $a=7/4k$. However, a detailed understanding of the explicit behaviors of the autocorrelation functions, $Q(t)$ and $F_s(k,t)$, is lacking. Both of them decay via SER at short times, and they both show crossover to some other forms.  However, they cross over at distinct time scales, $\tau_F$ and $\tau_G$, and to different functions, power law and exponential.

Our results show that a comprehensive analysis of all the crucial glassy features within a unified framework leads to several surprising results and a coherent picture of glassy liquids. One of the characteristic features of glassiness is the stretched exponential relaxation due to the spatially heterogeneous nature of the dynamics. Thus, characterizing and theoretically explaining the source of this stretching exponent $\beta$ is a critical test for the theories. We emphasize that one should be careful while computing $\beta$ from the data due to the crossover of relaxation functions. Since the long-time data for $F_s(k,t)$ is quite noisy, the exponential part is nearly invisible at low temperatures. Alternatively, the computation of $Q(t)$ at long times is more manageable due to the lack of fluctuations. This aspect explains the increasing use of $Q(t)$ to characterize glassy dynamics in recent simulations. As we showed in our simulation data, the power-law regime in $Q(t)$ is readily visible. A stretched exponential form will fit power-law data with a small stretching exponent. Thus, one must be careful while analyzing the simulation data; exclude the power-law regime while evaluating $\beta$. Otherwise, the value of $\beta$ will be wrong.

We have discussed that there exist two distinct mechanisms: caging and DH. The first is the particulate-scale property, whereas the second is the property at a larger length scale. The sub-diffusive and non-Gaussian system properties are independent of each other. Whereas the distribution of cage-breaking time can explain the sub-diffusive behavior, the subordination concept explains the crossover from non-Gaussian to Gaussian behavior. We have schematically elucidated these two mechanisms in Fig. \ref{subordinate}. What happens at $\tau_\alpha$? 
The canonical definition of relaxation time is $F_s(k,\tau_\alpha)=1/e$ or $Q(\tau_\alpha)=1/e$. Since $F_s(k,t)$ and $Q(t)$ do not distinguish between the fast and the slow particles, $\tau_\alpha$ primarily represents the relaxation of the fast-moving particles. However, glassiness seems to be predominantly controlled by slow particles \cite{stillinger1988,donati1998,kumar2006,douglass2022}. In the regime of our simulations, $\tau_\alpha$ is smaller than both $\tau_F$. However, we can have $\tau_F<\tau_\alpha$ at low enough $T$ when $\tau_\alpha$ becomes quite large. Then fast-moving particles will dominate $\tau_F$ at low $T$. This scenario raises questions on the role of caging in glassy dynamics, contrasting some recent proposals \cite{li2020,massimo2023}.

The results presented here have severe implications for theories of glassy dynamics and impose stricter conditions. For example, one of the most celebrated theories of glassy dynamics, the mode-coupling theory (MCT) \cite{goetzebook,spdas2004}, gives an equation of motion for $F_s(k,t)$. MCT has also been extended for DH \cite{IMCT}. It provides a stretched exponential form for $F_s(k,t)$ and captures the correct trends for the DH. It is well-known that the theory breaks down at low enough $T$ where it predicts a non-ergodicity transition. Yet, there are scenarios where MCT works surprisingly well. However, our results suggest that MCT is incorrect even in the regime where it is assumed to work. We have shown that $F_s(k,t)$ shows a crossover from SER to exponential at long times, but MCT does not exhibit any such crossover. We have also discussed that DH relaxes at $\tau_G$. Then should we use $\tau_G$ instead of $\tau_\alpha$ within various theories to characterize relaxation time? We have shown that $\tau_G\propto \tau_\alpha^a$. The theories that posit relaxation dynamics as an activated event, such as the RFOT theory or the Adam-Gibbs-Di Marzio theory \cite{lubchenko2007,adamgibbs1965}, changing this definition will not have any effect as this only rescales the energy scale by a constant. On the other hand, this redefinition will be significant for critical theories such as MCT that predict power-law divergence. The exponents will differ compared to when $\tau_\alpha$ gives the relaxation time.

We have argued that $\tau_F$ and $\tau_G$ have different microscopic origins: cage breaking for $\tau_F$ and disappearance of dynamical heterogeneity for $\tau_G$. We emphasize that $\tau_G$ is {\it different} from the time scale of dynamical heterogeneity defined as the time $\tau_4$ when $\chi_4(t)$ exhibits a peak; $\tau_4$ is close to $\tau_\alpha$ and much smaller than $\tau_G$ for the temperatures considered here. This is important because there are reasons to expect that caging and dynamical heterogeneity at time scales comparable to $\tau_4$ are related.  Within mean-field theories \cite{franz2011} of the glass transition, $\chi_4$ diverges at the dynamical transition temperature $T_d$, which is usually interpreted as the critical temperature of MCT. There is a single free energy minimum for $T>T_d$, and fluctuations are supposed to be confined to the basin of this minimum representing the liquid state. One may argue that caging, interpreted as fluctuations near a single free-energy minimum, and dynamical heterogeneity (measured by $\chi_4$) are closely related In the mean-field limit of infinite dimension, local fluctuations give rise to non-Gaussianity, $\alpha_2$ diverges at the dynamical transition \cite{biroli2022}, and fluctuations are confined to single minima for $T<T_d$. But, in finite dimensions, activated events preempt this dynamical transition. Hence, the behavior found in these mean-field studies should only be relevant at time scales shorter than the time at which the plateau in the mean-square displacement ends The heterogeneity time scale $\tau_G$, considered here, is much longer than such cage-breaking time scales. Charbonneau {\it et al.} studied the dynamics of the Mari-Kurchan model \cite{mari2011} at finite spatial dimensions \cite{charbonneau2014}. They showed that cage-breaking events lead to a breakdown of the Stokes-Einstein relation between the diffusivity and the structural relaxation time $\tau_\alpha$ \cite{charbonneau2014}. These results at time scales comparable to $\tau_\alpha$ are consistent with those in Fig. 4(c), where we have shown that the relaxation time $\tau_\alpha$ obtained from $F_s(k,t=\tau_\alpha) =1/e$ exhibits a violation of the Stokes-Einstein relation. In addition, we find (Fig. 4c) that the relation gets restored when we use the much longer time scale $\tau_G$ as the relaxation time. The earlier studies mentioned above did not consider the mechanism(s) that govern this time scale $\tau_G$ and do not contradict our assertion that these mechanisms are distinct from those related to the much shorter time scale $\tau_F$ associated with caging.

Past results have shown that glassy liquids can be Fickian yet non-Gaussian. A similar scenario has been observed for many complex biological systems too. We have shown that this result alone is {\it not} surprising, as two distinct mechanisms - caging and DH - govern the mean and the distribution of particle displacements. They can be related, but other scenarios are also feasible. For example, caging is absent in many complex biological systems \cite{wang2009,guan2014,chubynsky2014,chechkin2017}, or sometimes it is always present \cite{skaug2013}. However, these scenarios are distinct from glassy liquids \cite{berthiercomment}. Caging and DH fade after finite time scales $\tau_F$ and $\tau_G$. They are the essence of a glassy system: how they appear and control the dynamical behaviors are what a theory of glass should strive to explain. The stricter constraints and the comprehensive picture emerging from the results presented in this work should help guide the development of such a theory.

\section{Acknowledgements}
We thank L. Berthier, S. Karmakar, Vishnu V. Krishnan, Manoj Kumar Nandi, Kabir Ramola, and S. Sastry for discussions. We acknowledge the support of the Department of Atomic Energy, Government of India, under Project Identification No. RTI 4007. SKN thanks SERB for grant via SRG/2021/002014.

\appendix

\label{materials}

\section{Various observables and analytical relations among them}

We define the variables that have been the core of our analytical arguments. The mean-squared displacement, MSD$(t)$, at time $t$ is
\begin{equation}\label{MSDoft}
	\text{MSD}(t)=\Big \langle\frac{1}{N} \sum_{i=1}^N [r_i(t+t_0)-r_i(t_0)]^2\Big \rangle,
\end{equation}
where $N$ is the total number of particles, $r_i(t)$ is the position of the $i$th particle at time $t$, and $\langle \ldots\rangle$ represents average over different ensembles as well as time origins $t_0$. The distribution of particle displacement in a particular spatial direction is characterized via the self part of the van-Hove function, $G_s(x,t)$, given as
\begin{equation}\label{Gsoft}
	G_s(x,t)=\Big \langle\frac{1}{N} \sum_{i=1}^N \delta[x-x_i(t+t_0)+x_i(t_0)]\Big \rangle,
\end{equation}
where $\delta[\ldots]$ is the delta function.
The self intermediate scattering function, $F_s(k,t)$, at an wave vector $k$, is
\begin{equation}\label{Fskt}
	F_{s}(k,t) = \frac{1}{N}\Big \langle \sum_{i = 1}^{N} e^{\dot{\iota} \mathbf{k}.(\mathbf{r}_{i}(t+t_0) - \mathbf{r}_{i}(t_0))} \Big \rangle.
\end{equation}

The overlap function, $Q(t)$, is 
\begin{equation}\label{Qoft}
	Q(t)= \langle \tilde{Q}(t)\rangle=\Big \langle\frac{1}{N}\sum_{i=1}^{N} W\Big(a - |\mathbf{r}_{i}(t+t_{0})-\mathbf{r}_{i}(t_{0})|\Big)\Big \rangle,
\end{equation}
where $W(x)$ is the Heaviside Step Function, 
\begin{equation*}
	W(x)=\begin{cases}
		1, \,\,\, \text{when } x>0\\
		0,\,\,\, \text{when } x<0,
	\end{cases}
\end{equation*}
The value of $a$ is usually chosen as $0.3$, where MSD shows a plateau. The precise value of $a$ does not affect the qualitative results, but the choice of $a$ around the MSD plateau helps better data analysis \cite{guiselin2020}. In this work, we have primarily used $a=0.3$. However, we show in Fig. \ref{qt_vs_parameter_a} the variation of the overlap function at constant $t$ with changing $a$, and in Fig. \ref{tau_q_tau_k_T_0.47}, the variation of $\tau$ as a function of $a$.
Note that we define $Q(t)$ via a step function; at any particular time, $Q(t)$ is the average of $N$ numbers comprising 0 or 1. Thus, $Q(t)$ can never be negative. On the other hand, $F_s(k,t)$ requires averaging the oscillating cosine or sine functions; therefore, it can also be negative. Thus, when the average values of both functions approach zero at long times, we expect higher fluctuations in $F_s(k,t)$ compared to $Q(t)$.
The four-point correlation function is defined as the variance of $Q(t)$ as
\begin{equation}\label{chi4eq}
	\chi_4(t)=N(\langle \tilde{Q}(t)^2\rangle-Q(t)^2).
\end{equation}

It is easy to see from the above definitions that 
\begin{equation}\label{relQG}
	Q(t) = \int_0^a G_s(r,t) \d\mathbf{r},
\end{equation}
where $\mathrm{d}\mathbf{r}$ is the volume element in dimension $d$. We now provide the details of the analytical results discussed in the main text.

\subsection{$Q(t)$ decays as $t^{-d/2}$ at long times in $d$ dimensions}
We will use Eq. (\ref{relQG}) to obtain the behavior of $Q(t)$ via $G_s(r,t)$. We have proposed a generic form for $G_s(r,t)$ in Eq. 1 in the main text. For a general dimension $d$, the functional form will remain the same,
\begin{equation}
	G_{s}\left( r,t\right)=Ce^{-\left( \dfrac{r}{t^{\nu}}\right) ^{\mu } }.
\end{equation}
However, the normalization constant $C$ will depend on $d$. We first obtain $C$ in $d$-dimension. Using the normalization condition of $G_s(r,t)$, we obtain
\begin{equation}
	\int_0^{\infty} S_d  r^{d-1} C e^{-\left(\frac{r}{t^\nu}\right)^\mu} \d r=1,
\end{equation}
where $S_d={2 \pi^{d/2}}/{\Gamma(d/2)}$ is the surface area of a $d$-dimensional hypersphere of unit radius. Defining $x = \left(\frac{r}{t^\nu}\right)^\mu$, we get
\begin{equation}
	C \frac{2\pi^{d/2}}{\mu \Gamma({d / 2})} t^{d\nu} \int_0^{\infty}  x^{\frac{d}{\mu}-1} e^{-x} \d x=1.
\end{equation}
Recognizing that the integral is the definition of gamma function, we have
\begin{equation}
	C \frac{2\pi^{d/2}}{\mu \Gamma({d / 2})} t^{d\nu} \Gamma(d/\mu)=1.
\end{equation}

Thus, we obtain the normalized $G_s(r,t)$ as
\begin{equation}\label{G_s_d_normalised}
	G_s(r,t) = \frac{\mu \Gamma({d / 2})}{t^{d\nu} \Gamma(d/\mu) 2\pi^{d/2}} e^{-(r/t^\nu)^\mu}
\end{equation}
Now, using the Eq. (\ref{G_s_d_normalised}), in Eq. (\ref{relQG}), we obtain
\begin{equation}
	Q(t) = \frac{\mu}{t^{d \nu}\Gamma(d/\mu)} \int_0^a e^{-\left(\frac{r}{t^\nu}\right)^\mu} r^{d-1} \d r 
\end{equation}
Since we are interested in the long time regime, when the MSD becomes diffusive, the leading order contribution for $\nu = 1/2$ becomes
\begin{equation}
	Q(t) \sim t^{-d/2} .
\end{equation}
We find that this result agrees with simulation data in dimensions $d=2$ to $6$ for various systems.

\subsection{$F_s(k,t)$ becomes exponential}
Let us first write down $F_s(k,t)$ as follows
\begin{equation}
	F_s(k,t)=F_s(k_1,k_2,....,k_d,t)\big|_{k_1^2+k_2^2+....+k_d^2=k^2},
\end{equation}
where the $d$-dimensional Fourier transform is defined as
\begin{equation}\label{discreteFs}
	F_s(k_1,k_2,....,k_d,t)=\frac{1}{(2\pi)^{d/2}}\int e^{i\mathbf{k}\cdot\mathbf{r}} G_s(\mathbf{r},t) \d\mathbf{r}.
\end{equation}
This implies
\begin{align}
	F_s(k_1,k_2,....,k_d,t)&= \nonumber\\
	\frac{1}{(2\pi)^{d/2}}\int e^{i\mathbf{k}\cdot\mathbf{r}} G_s&(x_1,t)G_s(x_2,t)...G_s(x_d,t) \d x_1\d x_2..d x_d.
\end{align}
We write the above relation as
$F_s(k_1,k_2,...,k_d)=I(k_1)I(k_2)....I(k_d)$,
where each of the integrals are 
\begin{equation}
	I(k_1)=\frac{1}{\sqrt{2\pi}}\int_{-\infty}^\infty e^{ik_1x_1}G_s(x_1,t){\rm d}x_1.
\end{equation}
At large times such that $t>\tau_G$, $\mu = 2$ and $\nu = 1/2$, hence we get $I(k_1)=\exp(-k_1^2t/4)$. Which leads to 
\begin{equation}
	F_s(k,t)=e^{-k^2t/4}
\end{equation}
where $k^2=k_1^2+k_2^2+ \ldots +k_d^2$ in $d$ dimensions.

\subsection{Behavior of $Q(t)$ when $F_s(k,t)$ is exponential}
From Eqs. (\ref{relQG}) and (\ref{discreteFs}), it is easy to see that
\begin{equation}\label{QFs}
	Q(t)=\frac{1}{(2\pi)^{d/2}}\int_0^a\int_{-\infty}^{\infty}F_s(k,t)e^{-i\mathbf{k}\cdot\mathbf{r}}{\mathrm d}\mathbf{k}{\mathrm d}\mathbf{r}.
\end{equation}
Let us first solve the space integral in the spherical coordinates for $d$ dimension, which have coordinates $$
\left\{r, \theta_1, \theta_2, \ldots, \theta_{d-2}, \phi\right\}
,$$where $\theta_i$ goes from $[0, \pi]$ and $\phi$ goes from $[0, 2\pi]$. The above integral, considering the isotropy of space, is
\begin{align}
	\int_0^a e^{-i\mathbf{k}\cdot\mathbf{r}}& {\mathrm d}\mathbf{r}=\int_0^a \int_0^\pi \ldots \int_0^\pi \int_0^{2\pi} e^{-i k r \cos \theta_{d-2}}r^{d-1} \times\nonumber\\
	&\sin {}^{d-2}\theta_1 \sin {}^{d-3}\theta_2 \ldots \sin {\theta_{d-2}} \d r \d \theta_1 \ldots \d \theta_{d-2} \d \phi.
\end{align}
The angular integrals can be represented in terms of $\Gamma$-function, 
\begin{equation}
	\int_0^\pi \sin ^{n} \theta d \theta = 2 \int_0^{\pi / 2} \sin ^n \theta d \theta
	= \frac{\sqrt{\pi} \Gamma(\frac{n+1}{2})}{\Gamma(\frac{n}{2}+1)}.
\end{equation}
Using the above relation for $\theta_1$ to $\theta_{d-3}$, we have 
\begin{align}
	\int_0^a e^{-i\mathbf{k}\cdot\mathbf{r}} {\mathrm d}\mathbf{r}&= 2 \pi\left(\frac{\pi^{\frac{d-3}{2}} \Gamma(3/2)}{\Gamma(d / 2)}\right) \times \nonumber\\
	& \int_0^a \int_0^\pi r^{d-1} e^{-i k r \cos \theta_{d-2}} \sin \theta_{d-2} \d \theta_{d-2} \d r.
\end{align}
Substituting $- \cos(\theta_{d-2}) = l$ and solving we get,
\begin{equation}
	\int_0^a e^{-i\mathbf{k}\cdot\mathbf{r}} {\mathrm d}\mathbf{r}= \dfrac{4\pi }{k}\left( \dfrac{\pi^{\dfrac{d-3}{2}}\Gamma(\dfrac{3}{2})}{\Gamma(\dfrac{d}{2})}\right) \int ^{a}_{0}\sin \left( kr\right) r^{d-2}\d r
\end{equation}
Since we are interested in the long-time, when $F_s(k,t)$ is exponential, we set $F_s(k,t)=e^{-k^2t/4}$ in Eq. (\ref{QFs}) and obtain
\begin{align}
	Q(t) = 2^{2-d/2}\left( \dfrac{\Gamma(\dfrac{3}{2})}{(\sqrt{\pi}) \Gamma(\dfrac{d}{2})}\right)&\int_0^a e^{-k^2t/4}\frac{\sin \left( kr\right)}{k}r^{d-2}\d r \times\nonumber\\
	&\int_0^\infty\frac{2 \pi^{d/2}}{\Gamma({d / 2})} k^{d-1} \d k
\end{align}
Since $t$ is very large, only small $k$ in the exponential will contribute. Then, we have
\begin{equation}
	Q(t) \simeq \dfrac{2^{3-d/2}\pi ^{\left( \dfrac{d-1}{2}\right) }\Gamma(\dfrac{3}{2})}{d\left( \Gamma(\dfrac{d}{2})\right) ^{2}}a^{d}\int ^{\infty }_{0}e^{-\dfrac{k^{2}t}{4}}k^{d-1}\d k
\end{equation}
which gives
\begin{equation}\label{Qofa}
	Q(t) = \dfrac{2^{d/2+2}\pi ^{\left( \dfrac{d-1}{2}\right) }\Gamma(\dfrac{3}{2})}{d\left( \Gamma(\dfrac{d}{2})\right) }a^{d}t^{-d/2}.
\end{equation}
Thus, $Q(t)\sim t^{-d/2}$ when $F_s(k,t)$ becomes exponential.

\section{Model and simulation details}
The simulation results presented in the main text are for the $3d$ binary Kob-Andersen Lennard-Jones mixture with number-ratio $80:20$ for $A$ and $B$ type particles. This system rarely crystallizes and is an excellent glass former \cite{kob1995}. The interaction potential is given by
\begin{equation}\label{eq:lj}
U_{ij} = 4\epsilon_{ij}\Big(\Big(\frac{\sigma_{ij}}{r}\Big)^{12}-\Big(\frac{\sigma_{ij}}{r}\Big)^{6}\Big), \,\,\, \text{for} r < r_c,
\end{equation}
and $0$ otherwise. $i$ and $j$ are particle indices $A$ or $B$. The parameters of the model are as follows: $\epsilon_{AA} = 1.0$, $\sigma_{AA} = 1.0$; $\epsilon_{AB} = 1.5$, $\sigma_{AB} = 0.8$ and $\epsilon_{BB} = 0.5$, $\sigma_{BB} = 0.88$. The cut-off radius for the interaction potential, $r_c = 2.5$. We kept the number density $\rho = 1.2$ fixed. For reference, $T_{MCT}$, where MCT predicts the non-ergodicity transition for this system, is around $0.435$. The temperature ranges explored in this work are in the regime of moderately supercooled; going to lower $T$ is challenging due to the large computation time requirement. However, we expect the qualitative results will remain valid even at lower $T$. We have used the time step $dt = 0.005$. The Molecular Dynamics simulations were performed on Large-scale Atomic/Molecular Massively Parallel Simulator (LAMMPS) \cite{LAMMPS}. 

For the 2$d$ model, we used the same binary Kob-Andersen Lennard-Jones model, as described in the main text with density $\rho=1.2$, however, the number ratio of the two types particles is 65:35. 

For the higher dimensions, we have investigated a $50:50$ binary mixture of particles that interact with a harmonic potential given by \cite{durian1995foam,berthier2009compressing}:
\begin{eqnarray}
\nonumber
V_{\alpha \beta}(r) &=& \epsilon_{\alpha \beta}\left(1-\frac{r}{\sigma_{\alpha \beta}}\right)^2, \hspace{0.24cm}\text{when } r_{\alpha \beta} \leq \sigma_{\alpha \beta}\\ 
&=&0 ,\hspace{2.6cm} \text{when }r_{\alpha \beta } > \sigma_{\alpha \beta}
\end{eqnarray}
where $\alpha, \;  \beta$ $\in$ (A,B), indicates the type of particle. The two types of particle differ in their sizes, with $\sigma_{BB} =1.4 \sigma_{AA}$ and $\sigma_{AB}= (\sigma_{AA}+\sigma_{BB})/2$, but with the interaction strengths being the same for all pairs. In reporting results for this system, we use reduced units, with units of length, energy and time scales being $\sigma_{AA}$, $\epsilon_{AA}$  and  $\sqrt \frac{\sigma^2_{AA}m_{AA}}{\epsilon_{AA}} $ respectively.
We present results for $d=4$ fixing the density at $1.3\phi_J$, where $\phi_J$ is the jamming density. We have used $\phi_J$=$0.467$ using estimates by Charbonneau \emph{et al.} \cite{charbonneau2011glass}. 
The number density, $\rho$ is related to the volume fraction $\phi$ for the binary mixture in  the following way 
\begin{equation}
\phi =  \rho 2^{-d}\frac{\pi^{d/2}}{\Gamma(1+\frac{d}{2})}((c_A\sigma^d_{AA} + c_B\sigma^d_{BB}) 
\end{equation}   
where $\rho = N/V$, with $N$ being the number of particles, and $V$ the volume, and the fractions $c_A = c_B = 1/2$. The corresponding number densities are following: $0.8132$.
The system size is fixed at $2000$ particles, which is large enough that the linear dimension $L$ is $> 2 \sigma_{BB}$ in all dimensions. Molecular dynamics (MD) simulations are performed in a hyper-cubic box with periodic boundary conditions in the constant number, volume, and temperature (NVT) ensemble. The integration time step was fixed at $dt=0.01$. Temperatures are kept constant using the  Brown and Clarke \cite{brown1984comparison} algorithm. The data, presented here, have run lengths of around $100\tau$ (where $\tau$ is the relaxation time, defined below). We present results that  are averaged over $100$ independent samples.


\section{Additional simulation results}

\subsection{Variation of $Q(t)$ as a function of $a$, fixed $t$}
Equation (19) in the main text, we have shown that $Q(t)\sim a^d$ in the power-law regime. We have tested it in $2d$ and $3d$ and present the data for $3d$ alone in Fig. \ref{qt_vs_parameter_a}.
We take the data for $T = 0.47$, $0.48$, and $0.50$ and varied $a$. Figure \ref{qt_vs_parameter_a} shows $Q(t)$ as a function of $a$ at a specific value of $t=8\times10^4$. We find that $Q(a)\sim a^3$ is in agreement with the analytical result.

\begin{figure}
	\includegraphics[width=8.6cm]{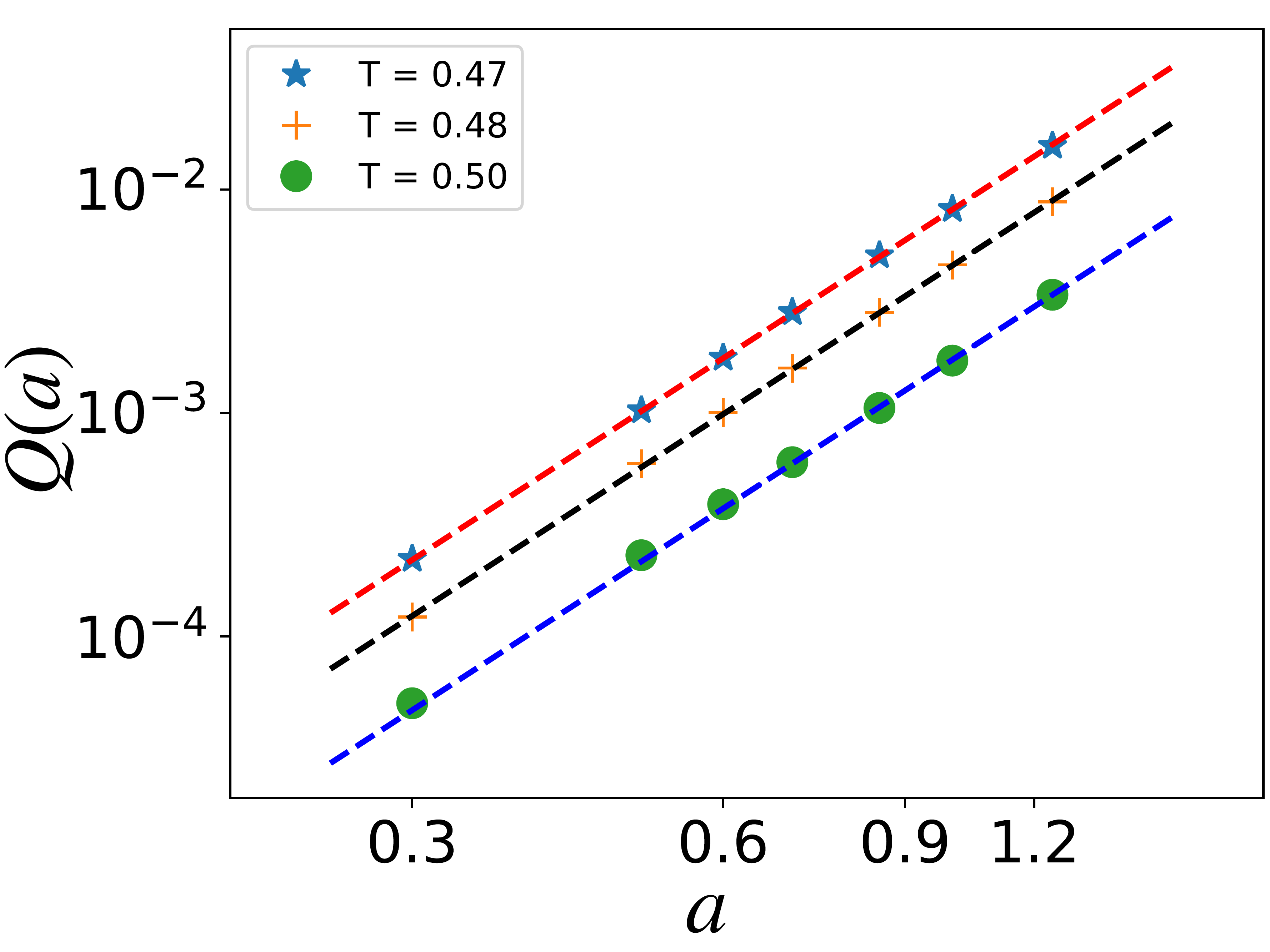}
	\caption{The variation of the overlap function with the Parameter $a$ for $T = 0.47, 0.48$, and $0.5$ in 3d in the power-law regime. The power law fit gives the exponent close to $3$, consistent with the analytical relation (Eq. \ref{Qofa}).}
	\label{qt_vs_parameter_a}
\end{figure}

\subsection{Particle displacements and fit of simulation data with the proposed general form}
As discussed in the main text, the probability of particle displacements in glassy systems show exponential tails in glassy systems. Figures \ref{vanhove_expo} (a) and (b) show that the van-Hove function in our simulations has exponential tails at intermediate times.
We have proposed a general form for the probability of particle displacements in the main text, Eq. (1). Figure \ref{vanhove_expo} (c) shows the fits of the simulation data with this form at various times. We obtain the values of $\mu$ and $\nu$ from these fits and show these values in Table \ref{vanhove_expo}(d).

\begin{figure}
	\includegraphics[width=8.6cm]{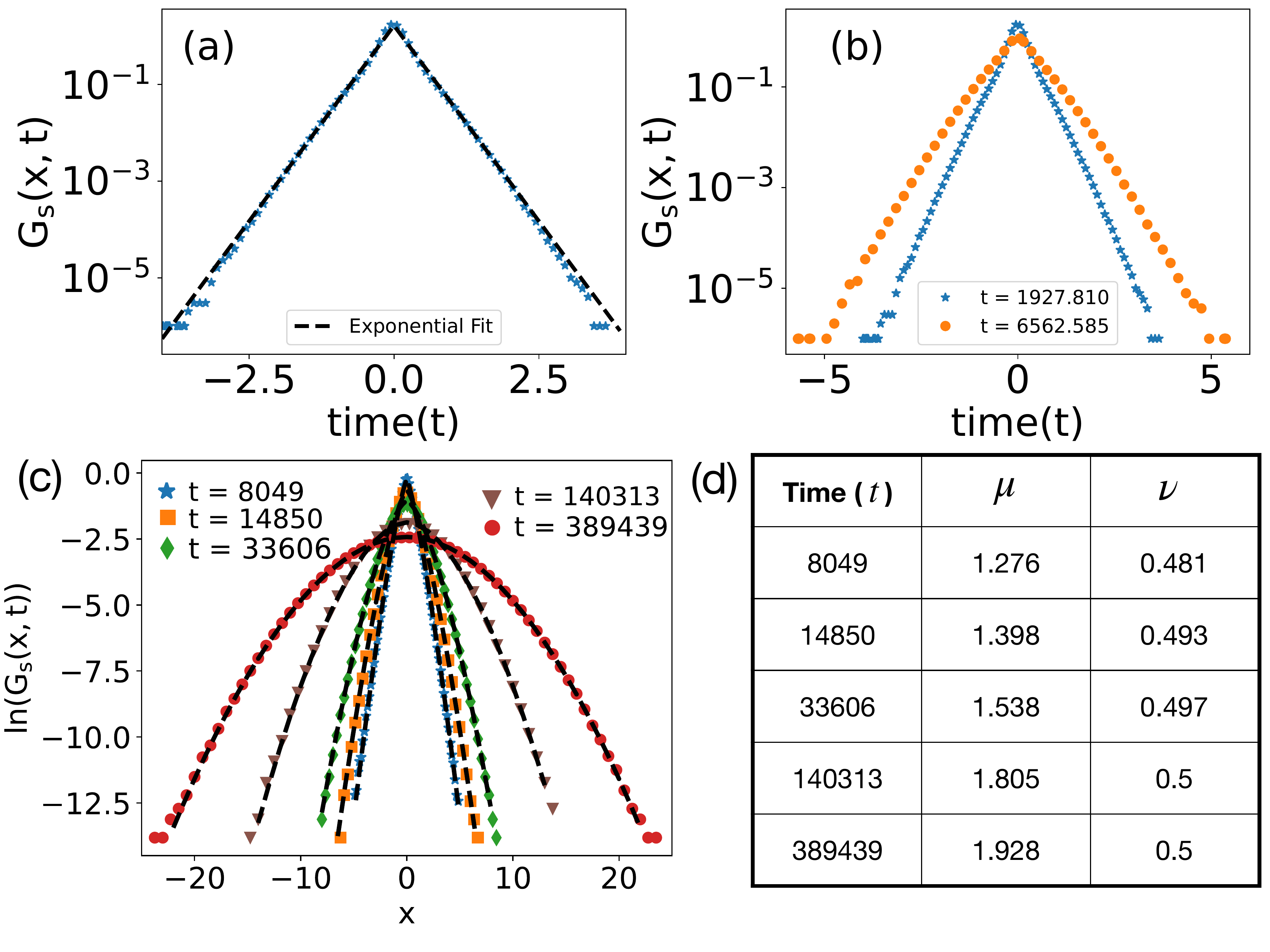}
	\caption{(a) The exponential tails of the van-Hove function, $G_s(x,t)$, at intermediate times, the dashed line is fit to the exponential form. These data are for $t=1927$. (b) $G_s(x,t)$ becomes progressively more Gaussian as time progresses, implying the data at higher time is more Gaussian. (c) Fits of the simulation data with the proposed form of $G_s(x,t)$, Eq. \ref{formofvH}, give values of $\mu$ and $\nu$. We show the fits to the data at various times. All the data in this figure are for a system with $T = 0.45$. (d) The corresponding values of $\mu$ and $\nu$ obtained from the fits in (c) are listed.}
	\label{vanhove_expo}
\end{figure}

\begin{figure}
	\includegraphics[width=8.6cm]{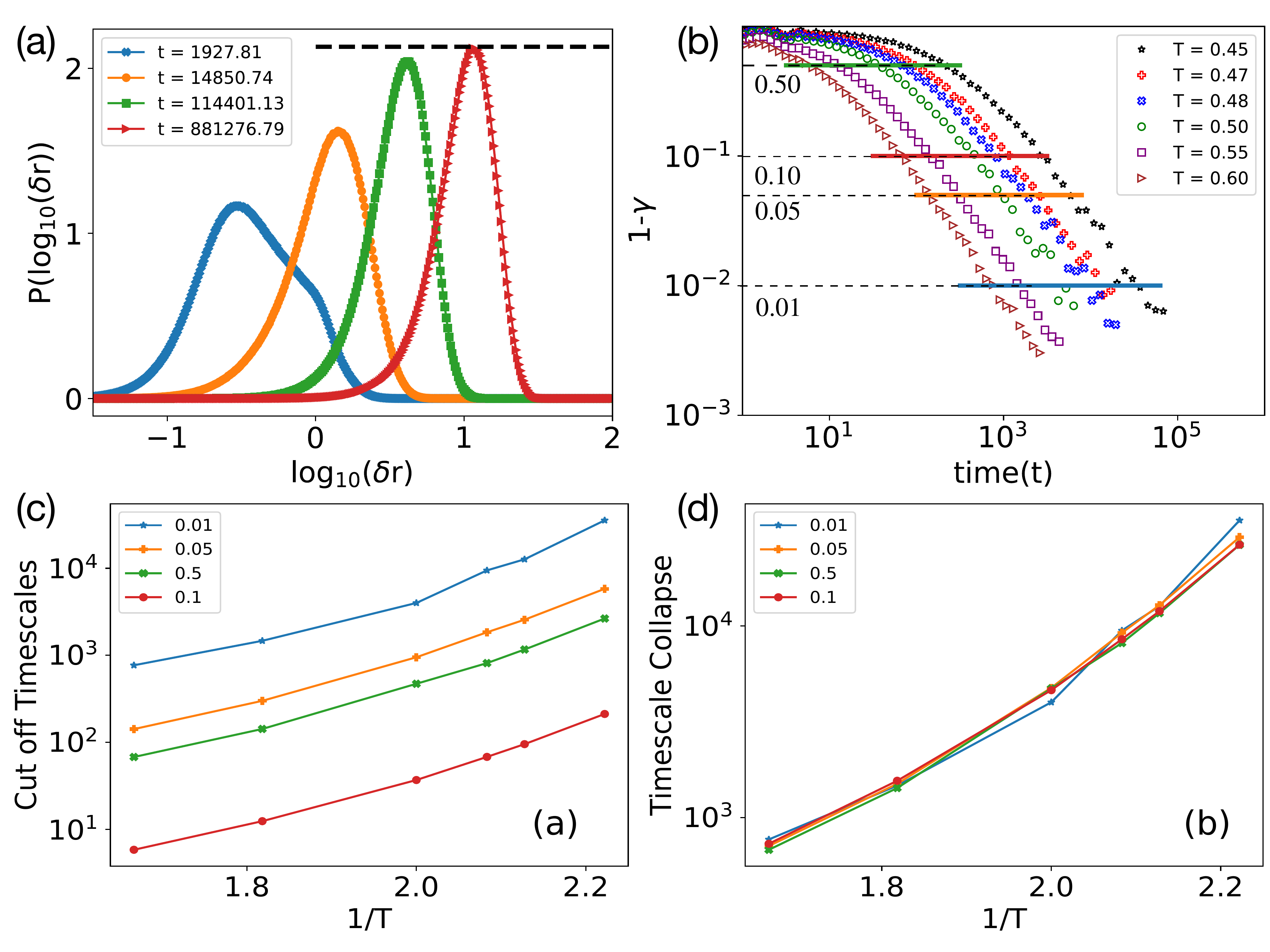}
	\caption{(a) $P({\rm log_{10}}(\delta r),t)$ for T = 0.45 at different times. The dashed black lines is at a value of 2.13. (b) (1-$\gamma$) variation with time at different $T$. The horizontal lines specify different cut-off values (marked in the figure) when MSD becomes diffusive. (c) The timescales corresponding to different cut-offs of $\gamma$ are proportional to each other. (d) Data-collapse of various timescales when we scale them to a single value at a particular $T$. }
	\label{p_log_r_T_0.60}
\end{figure}

\subsection{Single particle displacements}
The single particle displacements, $P({\rm log_{10}}(\delta r),t)$, at time $t$ is related to $G_s(r,t)$ via the following equation :
\begin{equation}
P({\rm log_{10}}(\delta r),t) = ({\rm ln}10)  4\pi \delta r^3 G_s(\delta r,t).
\end{equation}
At long times when $G_s(r,t)$ becomes Gaussian, the peak value of $P({\rm log_{10}}(\delta r),t)$ is $\Delta(t)=2.13$. Figure \ref{p_log_r_T_0.60}(a) shows the behaviour of $P({\rm log_{10}}(\delta r),t)$ at four different $t$. To characterize the behavior of $\Delta(t)$, we plot $ 2.13-\Delta(t)$ as a function of $t$ in Fig. \ref{tauGbehavior}(b). 

At shorter times, $P({\rm log_{10}}(\delta r),t)$ shows two peaks. The second peak (see the plot corresponding to $t = 1927.81$ above) grows and the first peak vanishes as time increases. Therefore, we focused on the second peak alone and presented the data in Fig. \ref{tauGbehavior}(b).

\subsection{MSD timescales for different Cut offs}
We have characterized the behavior of MSD as $\sim t^\gamma$. MSD is sub-diffusive when $\gamma<1$ and diffusive when $\gamma=1$. Figure \ref{p_log_r_T_0.60}(b) shows the variation of ($1-\gamma$) as a function of $t$. The diffusive behavior is equivalent to defining a cut-off value for $\gamma$. We have chosen four different cut-offs, shown by the lines.

Figure \ref{p_log_r_T_0.60}(c) shows the cut-off time scales as a function of $1/T$ for the various cut-offs. Figure \ref{p_log_r_T_0.60}(d) shows we can collapse these time scales on a single curve. This data collapse implies that these definitions are equivalent. Our definition of this cut-off, when $\gamma$ becomes 1, is guided by the crossover time of $Q(t)$.

\begin{figure*}
	\includegraphics[width=14.2cm]{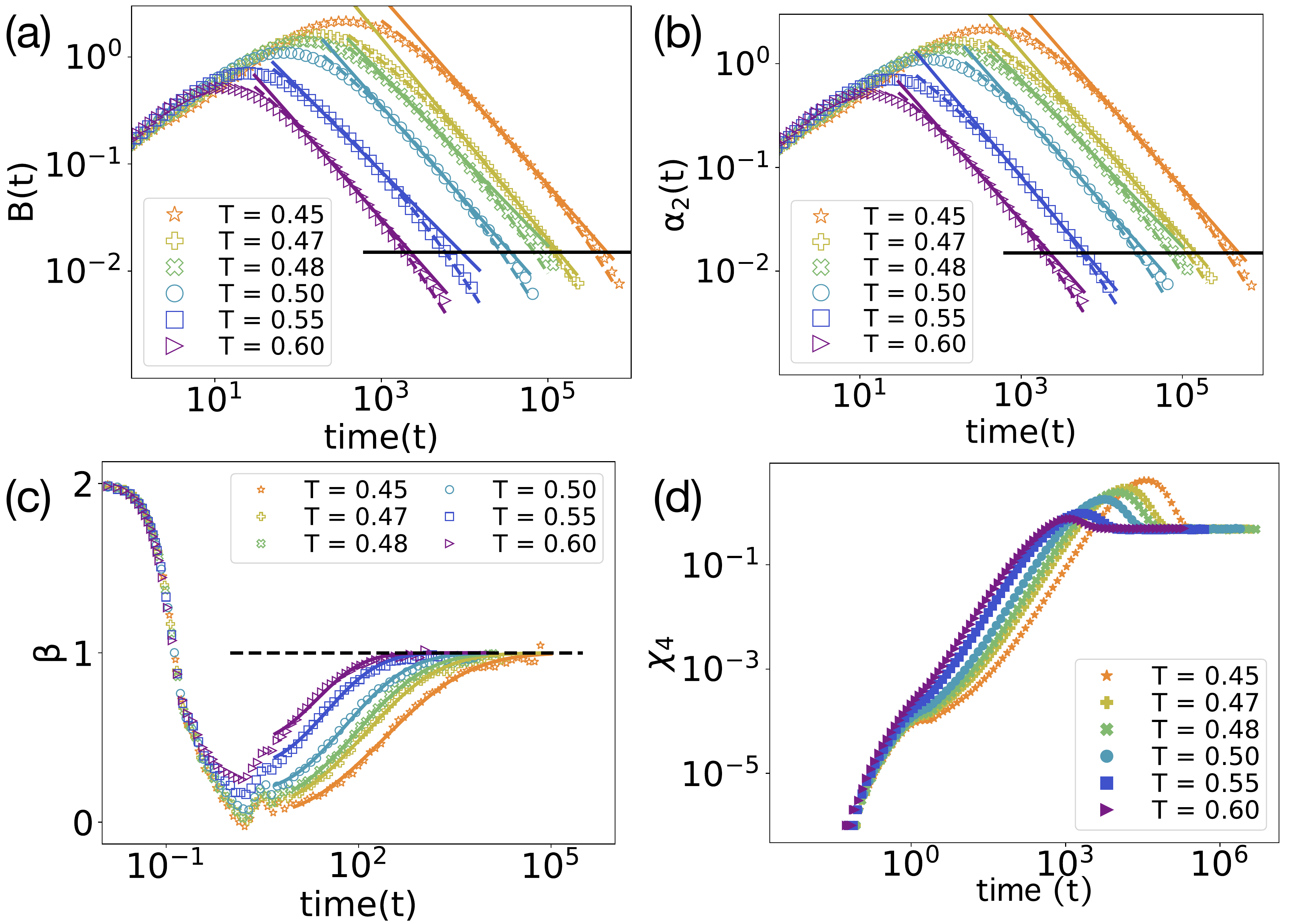}
	\caption{(a) $B(t)$ shows a stretched exponential behaviour. We fit the equation $ae^{-(t/b)^c}$ (for the stretched exponential behaviour), with the data and obtain the stretching exponent $c=0.12$ whereas we fit $d(t/e)^{-f}$ for the power law behaviour. (b) $\alpha_2(t)$ also has a stretched exponential behavior with the same stretching exponent as for $B(t)$. The dashed lines represent the fits for stretched exponential whereas straight lines represent the fits for power law. In both these plots, the data seems to be better represented by a stretched-exponential than a power law. (c) The stretching exponent ($\beta$) of $F_s(k,t)$ with $k=2.0$ shows a compressed exponential approach to 1. The exponent of the compressed exponential is close to 2.8. The lines are the fits for the compressed exponential behaviour. (d) $\chi_4(k,t)$ defined for $F_s(k,t)$ does not decay to zero at long times. We have plotted $\chi_4(k,t)$ with the value of $k = 0.8$.
	}
	\label{binder_fit_suppl}
\end{figure*}

\begin{figure*}
	\includegraphics[width=15.5cm]{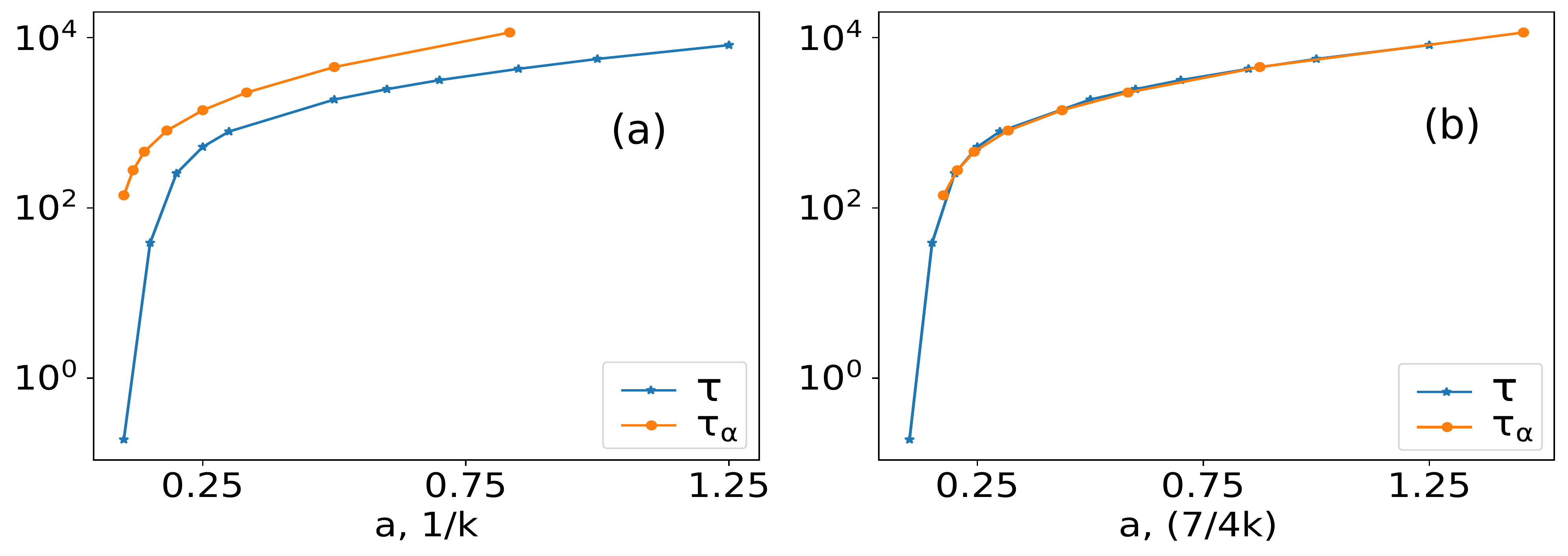}
	\caption{{\bf Left}: $\tau_\alpha$ defined via $F_s(k,\tau_\alpha)=1/e$ and $\tau$, defined via $Q(\tau)=1/e$ are plotted as functions of $1/k$ and $a$, respectively. {\bf Right}: The data collapse to a single curve if we scale $1/k$ to $7/4k$. We have used $T = 0.47$ for this plot.}
	\label{tau_q_tau_k_T_0.47}
\end{figure*}

\subsection{Stretched Exponential Behaviour of $B(t)$ and $\alpha_2(t)$}
As discussed in the main text, we can quantify the non-Gaussian nature of $G_s(r,t)$ via Binder's cumulant, $B(t)$, and non-Gaussian parameter, $\alpha_2(t)$. They are defined such that they are zero for Gaussian variables. We show these parameters in Figs. \ref{binder_fit_suppl}(a) and \ref{binder_fit_suppl}(b), respectively. It has been argued in the literature that these parameters go to zero as a power law. In that case, since there is no characteristic time in power law, it is not possible to define a time scale at which $G_s(r,t)$ becomes Gaussian.

However, we find that they are more consistent with a stretched-exponential decay with a tiny stretching exponent, 0.12. We show the fits with the stretched exponential by the dashed lines in Fig. \ref{binder_fit_suppl} (a) and (b).

As described in the main text we can write $F_s(k,t)\sim \exp[-(t/\tau_\alpha)^\beta]$ with $\tau_\alpha$ being a relaxation time and treating $\beta$ as a function of time. Then, the derivative of $\log[-\log\{F_s(k,t)\}]$ with respect to $\log t$ gives $\beta$. Figure \ref{binder_fit_suppl}(c) shows the behavior of $\beta$ with time: it goes to 1 as a compressed exponential with an exponent $\sim$ 2.8.

\begin{figure*}
	\includegraphics[width=16.6cm]{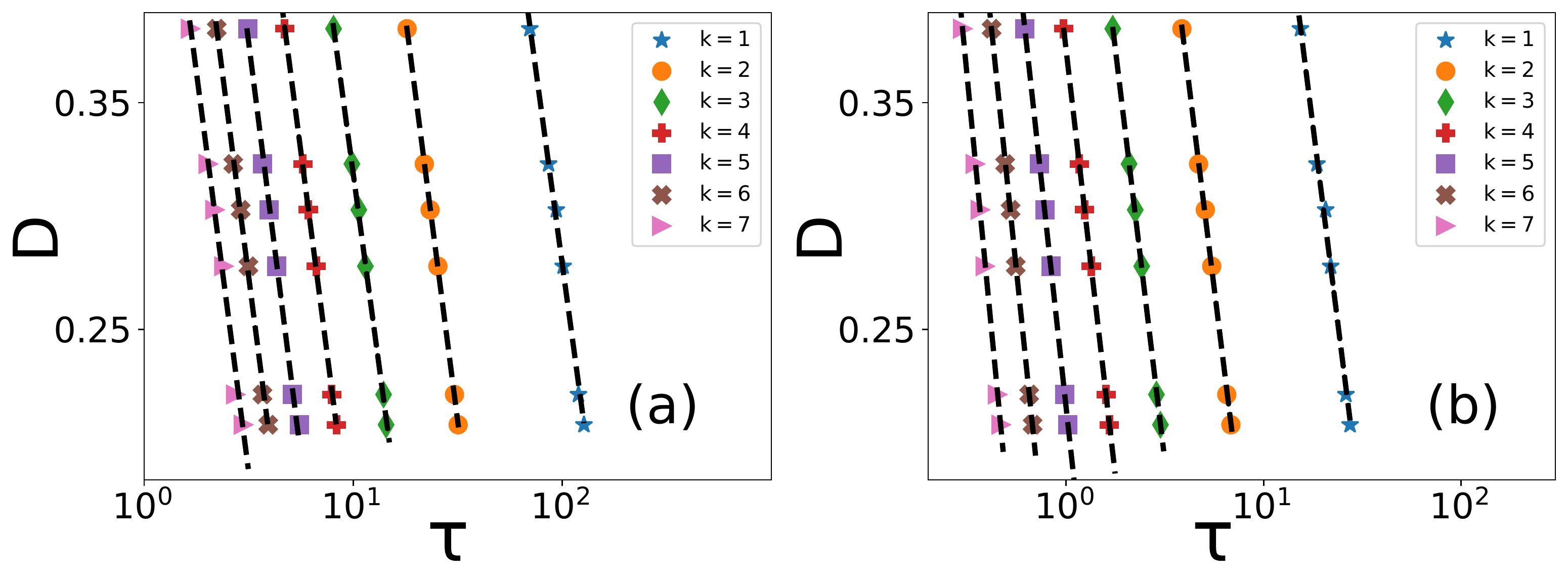}
	\caption{$D$ vs $\tau$ for the temperature range $T = 1.75$ to 2.50 for different values of $k$. Different curves correspond to distinct $k$, whereas various $\tau$ in a particular curve comes from varying $T$. (a) corresponds to the data of $D$ vs $\tau$ when $F_s(k,\tau) = 0.01$ and (b) corresponds to the data when $F_s(k,\tau) = 1/e$. The dashed black lines are the fits with power law behaviour ($D \sim\tau^{-\zeta}$). We obtain the values of $\zeta$ from these fits and show them in the main text, Fig. \ref{crossover_times}(d).}
	\label{d_vs_tau}
\end{figure*} 

\subsection{The four-point correlation function defined via $F_s(k,t)$}
The four-point correlation function, defined via $F_s(k,t)$ is
\begin{equation}
\chi_4(k,t)=N(\langle \tilde{f_s}(k,t)^2\rangle-F_s(k,t)^2),
\end{equation} 
where $\langle \tilde{f_s}(k,t)\rangle = F_s(k,t)$. $\chi_4(k,t)$ goes to a non-zero constant at long times (Fig. \ref{binder_fit_suppl}d).

\subsection{Comparisons of relaxation times defined via $F_s(k,t)$ and $Q(t)$}

We can define relaxation times, $\tau_\alpha$ and $\tau$ from both $F_s(k,t)$ and $Q(t)$ when they become $1/e$. Since $F_s(k,t)$ is an explicit function of $k$, the relaxation time will also be a function of $k$, i.e., $\tau_\alpha=\tau_\alpha(k)$. On the other hand, it is clear from Eq. (\ref{Qoft}) that $Q(t)$ is a function of $a$, thus $\tau=\tau(a)$. How are these two definitions related? We show $\tau_\alpha(k)$ as a function of $1/k$ and $Q(t)$ as a function of $a$ in Fig. \ref{tau_q_tau_k_T_0.47}(a). We see that they have similar behavior. As we show in Fig. \ref{tau_q_tau_k_T_0.47}(b), the data can be collapsed into a single curve when we scale $1/k$ by 4/7. Thus, the two definitions of relaxation time are equivalent.

\subsection{Diffusion Constant vs. Relaxation Time}
As mentioned in Sec. IID of the main that D $\sim$ $\tau^{-\zeta}$ and in the Fig. \ref{crossover_times}(d), $\zeta$ is plotted with $k$, here we specify the parameters we used for Fig. \ref{crossover_times}(d). As shown in Fig. \ref{d_vs_tau}, we fit the data of $D$ as a function of $\tau$ with a power law to obtain $\zeta$. Figure \ref{d_vs_tau}(a) shows the fits when we define $\tau$ as $F_s(k,\tau)=0.01$ and Fig. \ref{d_vs_tau}(b) shows the fits when we define $\tau$ as $F_s(k,\tau)=1/e$. We have used different $T$ range to have various $\tau$ shown in the figure (Fig. \ref{d_vs_tau}).

\begin{figure}
	\includegraphics[width=8.6cm]{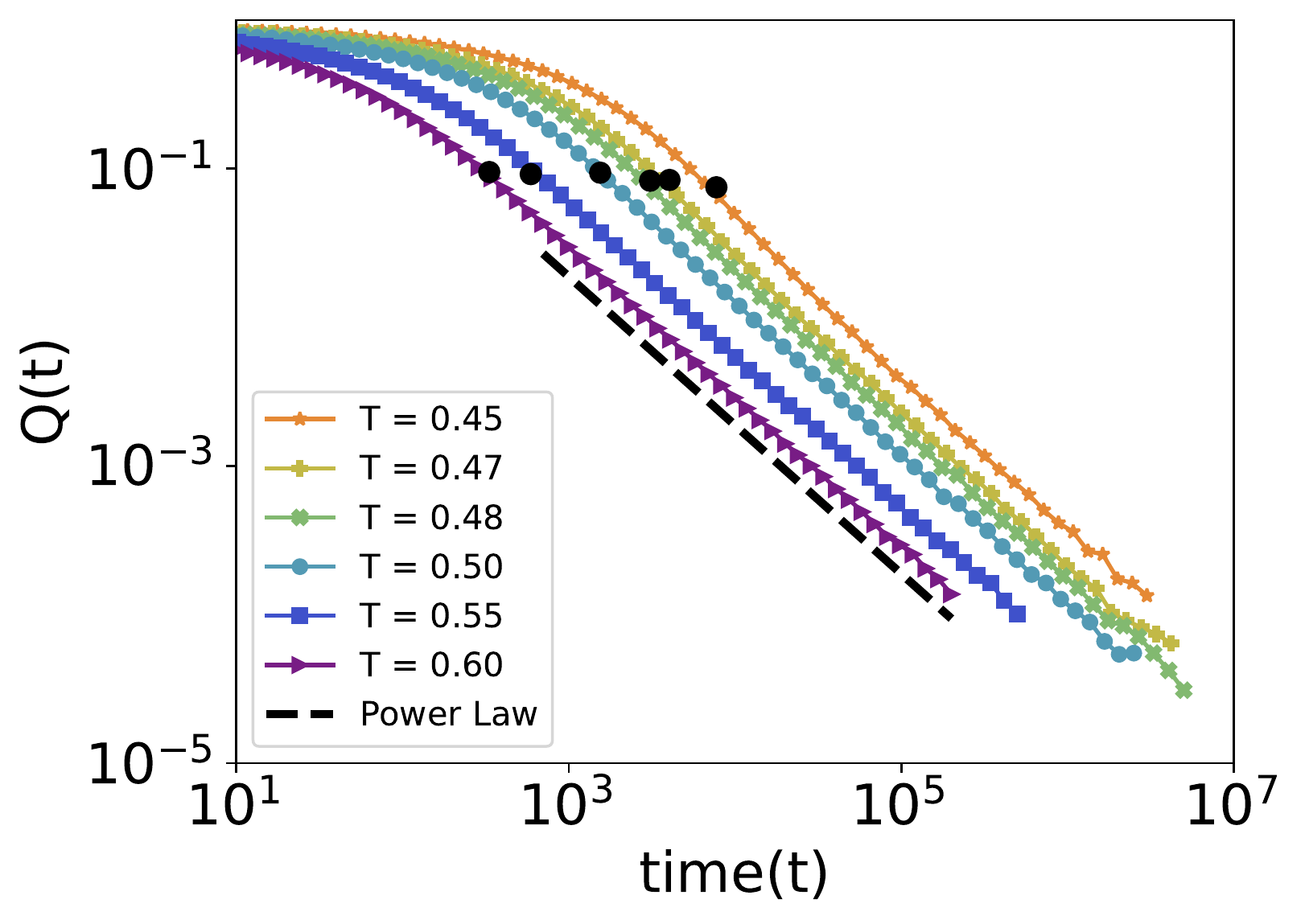}
	\caption{$Q(t)$ in $2d$ has a similar behaviour to that in $3d$. The power law exponent is 1.0 in $2d$, consistent with the analytical result $Q(t)\sim t^{-d/2}$ at long times. The crossover times are shown as black dots for different temperatures.}
	\label{Q(t)_2d_chandan_analysis_suppl}
\end{figure}

\subsection{Analysis of Q(t) form $2d$ simulation}
Q(t) shows a power law decay with power law exponent equal 1.0 in $2d$. This is consistent with the expression that $Q(t)$ goes as a power law, $t^{-d/2}$ ( Fig. \ref{Q(t)_2d_chandan_analysis_suppl}). The crossover times obtained via a similar analysis outlined in the main text are also indicated.

\begin{figure*}
	\centering
	\includegraphics[width=16cm]{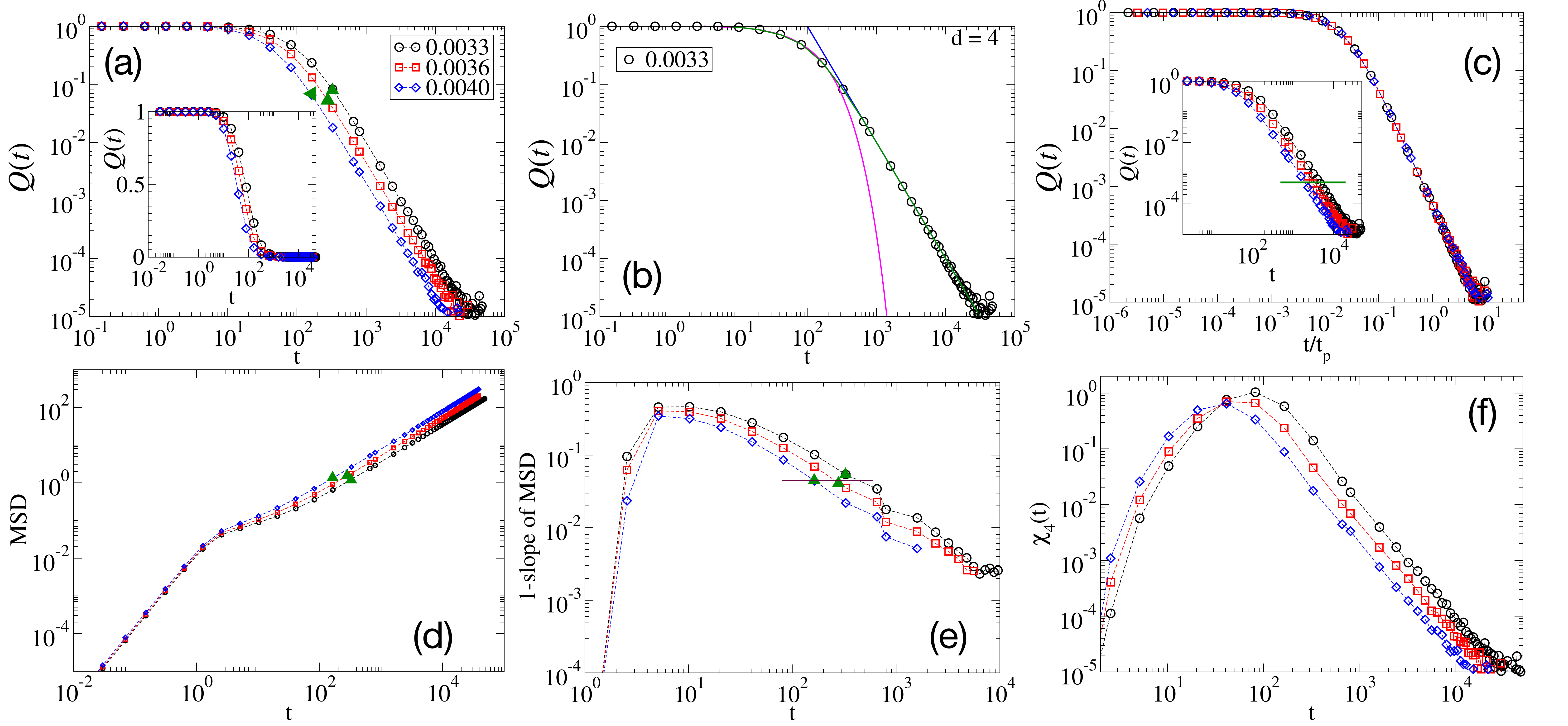}
	\caption{Simulation data for a system in $4d$. (a) $Q(t)$ is plotted as a function of time for different temperatures. The green triangles denote the points where we expect the crossover from SER to power law. (b) $Q(t)$ is shown as a function of time for a specific temperature, $T = 0.0033$. Magenta lines show a stretched exponential fit, whereas the blue line corresponds to a power law fit. (c) $Q(t)$ is shown against time scaled with a time $t_p$. In the inset, the green line shows the value of $t_p$. This corresponds to the figure 2(c) in the main text, but shown for a system in $4d$ here. (d) MSD as a function of time. The green triangle marks the crossover point.
		(e) $1-$slope of MSD vs. time. The plot demonstrates how to extract the time scale for diffusion (corresponding to Fig. 2(e) in the main text).
		(f) $\chi_4(t)$ is shown against time for the same temperatures as indicated in the legend in (a). It shows that at long times, $\chi_4(t)$ follows a power law with an exponent $d/2$.  } 
	\label{4dfig}
\end{figure*}

\subsection{Results in 4d: Q(t), MSD, and $\alpha_2(t)$}
We now show the simulation results in other dimensions. Figure \ref{4dfig} shows the results for the system in spatial dimension four.

\begin{figure}
	\centering
	\includegraphics[width=0.5\textwidth]{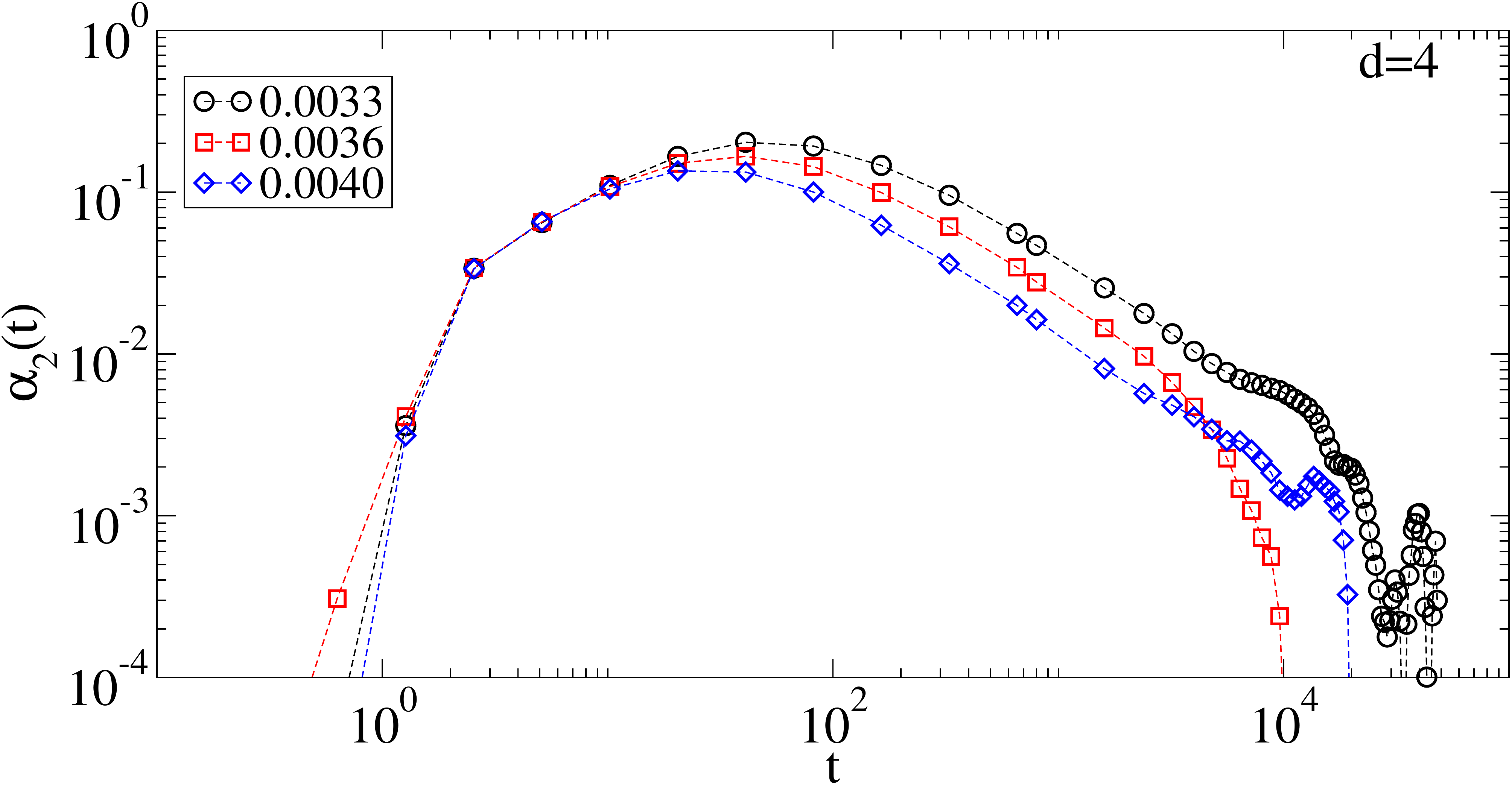}
	\caption{$\alpha_2(t)$ is plotted against time for dimension $d=4$, showing that  it decays to zero at long time. 
}
	\label{}
\end{figure}

\subsection{Power law dependence of $Q(t)$ in higher dimensions}
We finally show the power law dependence of $Q(t)$ at various dimensions $d=3-6$. Figure \ref{variousdim} shows that the power-law nature of the long-time decay of $Q(t)$ is quite prominent in all these dimensions.

\begin{figure}
	\centering
	\includegraphics[width=8.6cm]{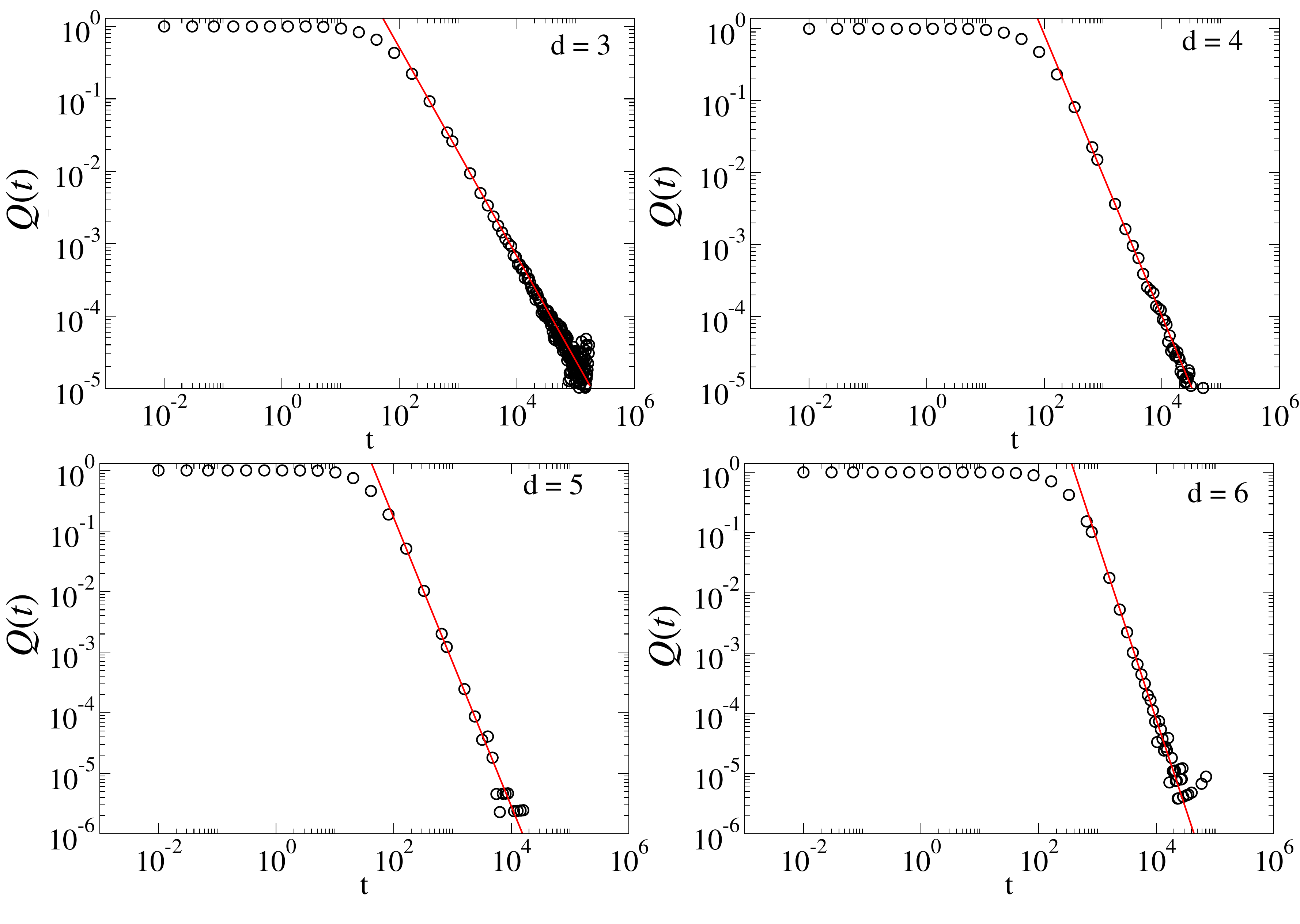}
	\caption{$Q(t)$ is plotted as a function of time for a fixed temperature in different dimensions, $d=3-6$. The red line corresponds to a power law fit with an exponent $-d/2$. These plots demonstrate the validity of the power law in different dimensions and confirm the value of the exponent. }
	\label{variousdim}
\end{figure}

\section{Data Availability}
All the data is included in the paper.


\begin{thebibliography}{88}%
\makeatletter
\providecommand \@ifxundefined [1]{%
 \@ifx{#1\undefined}
}%
\providecommand \@ifnum [1]{%
 \ifnum #1\expandafter \@firstoftwo
 \else \expandafter \@secondoftwo
 \fi
}%
\providecommand \@ifx [1]{%
 \ifx #1\expandafter \@firstoftwo
 \else \expandafter \@secondoftwo
 \fi
}%
\providecommand \natexlab [1]{#1}%
\providecommand \enquote  [1]{``#1''}%
\providecommand \bibnamefont  [1]{#1}%
\providecommand \bibfnamefont [1]{#1}%
\providecommand \citenamefont [1]{#1}%
\providecommand \href@noop [0]{\@secondoftwo}%
\providecommand \href [0]{\begingroup \@sanitize@url \@href}%
\providecommand \@href[1]{\@@startlink{#1}\@@href}%
\providecommand \@@href[1]{\endgroup#1\@@endlink}%
\providecommand \@sanitize@url [0]{\catcode `\\12\catcode `\$12\catcode
  `\&12\catcode `\#12\catcode `\^12\catcode `\_12\catcode `\%12\relax}%
\providecommand \@@startlink[1]{}%
\providecommand \@@endlink[0]{}%
\providecommand \url  [0]{\begingroup\@sanitize@url \@url }%
\providecommand \@url [1]{\endgroup\@href {#1}{\urlprefix }}%
\providecommand \urlprefix  [0]{URL }%
\providecommand \Eprint [0]{\href }%
\providecommand \doibase [0]{https://doi.org/}%
\providecommand \selectlanguage [0]{\@gobble}%
\providecommand \bibinfo  [0]{\@secondoftwo}%
\providecommand \bibfield  [0]{\@secondoftwo}%
\providecommand \translation [1]{[#1]}%
\providecommand \BibitemOpen [0]{}%
\providecommand \bibitemStop [0]{}%
\providecommand \bibitemNoStop [0]{.\EOS\space}%
\providecommand \EOS [0]{\spacefactor3000\relax}%
\providecommand \BibitemShut  [1]{\csname bibitem#1\endcsname}%
\let\auto@bib@innerbib\@empty
\bibitem [{\citenamefont {Berthier}\ and\ \citenamefont
  {Biroli}(2011)}]{giulioreview}%
  \BibitemOpen
  \bibfield  {author} {\bibinfo {author} {\bibfnamefont {L.}~\bibnamefont
  {Berthier}}\ and\ \bibinfo {author} {\bibfnamefont {G.}~\bibnamefont
  {Biroli}},\ }\bibfield  {title} {\bibinfo {title} {Theoretical perspective on
  the glass transition and amorphous materials},\ }\href
  {https://doi.org/10.1103/RevModPhys.83.587} {\bibfield  {journal} {\bibinfo
  {journal} {Rev. Mod. Phys.}\ }\textbf {\bibinfo {volume} {83}},\ \bibinfo
  {pages} {587} (\bibinfo {year} {2011})}\BibitemShut {NoStop}%
\bibitem [{\citenamefont {Cicerone}\ and\ \citenamefont
  {Ediger}(1996)}]{cicerone1996}%
  \BibitemOpen
  \bibfield  {author} {\bibinfo {author} {\bibfnamefont {M.~T.}\ \bibnamefont
  {Cicerone}}\ and\ \bibinfo {author} {\bibfnamefont {M.~D.}\ \bibnamefont
  {Ediger}},\ }\bibfield  {title} {\bibinfo {title} {Enhanced translation of
  probe molecules in supercooled o-terphenyl: Signature of spatially
  heterogeneous dynamics?},\ }\href {https://doi.org/10.1063/1.471433}
  {\bibfield  {journal} {\bibinfo  {journal} {J. Chem. Phys.}\ }\textbf
  {\bibinfo {volume} {104}},\ \bibinfo {pages} {7210} (\bibinfo {year}
  {1996})}\BibitemShut {NoStop}%
\bibitem [{\citenamefont {Sillescu}(1999)}]{sillescu1999}%
  \BibitemOpen
  \bibfield  {author} {\bibinfo {author} {\bibfnamefont {H.}~\bibnamefont
  {Sillescu}},\ }\bibfield  {title} {\bibinfo {title} {Heterogeneity at the
  glass transition: a review},\ }\href
  {https://doi.org/10.1016/S0022-3093(98)00831-X} {\bibfield  {journal}
  {\bibinfo  {journal} {J. Non-crystalline solids}\ }\textbf {\bibinfo {volume}
  {243}},\ \bibinfo {pages} {81} (\bibinfo {year} {1999})}\BibitemShut
  {NoStop}%
\bibitem [{\citenamefont {Williams}\ and\ \citenamefont
  {Watts}(1970)}]{williams1970}%
  \BibitemOpen
  \bibfield  {author} {\bibinfo {author} {\bibfnamefont {G.}~\bibnamefont
  {Williams}}\ and\ \bibinfo {author} {\bibfnamefont {D.~C.}\ \bibnamefont
  {Watts}},\ }\bibfield  {title} {\bibinfo {title} {Non-symmetrical dielectric
  relaxation behaviour arising from a simple empirical decay function},\ }\href
  {https://doi.org/10.1039/TF9706600080} {\bibfield  {journal} {\bibinfo
  {journal} {Trans. Faraday Soc.}\ }\textbf {\bibinfo {volume} {66}},\ \bibinfo
  {pages} {80} (\bibinfo {year} {1970})}\BibitemShut {NoStop}%
\bibitem [{\citenamefont {Szamel}\ and\ \citenamefont
  {Flenner}(2006)}]{szamel2006}%
  \BibitemOpen
  \bibfield  {author} {\bibinfo {author} {\bibfnamefont {G.}~\bibnamefont
  {Szamel}}\ and\ \bibinfo {author} {\bibfnamefont {E.}~\bibnamefont
  {Flenner}},\ }\bibfield  {title} {\bibinfo {title} {Time scale for the onset
  of fickian diffusion in supercooled liquids},\ }\href
  {https://doi.org/10.1103/PhysRevE.73.011504} {\bibfield  {journal} {\bibinfo
  {journal} {Phys. Rev. E}\ }\textbf {\bibinfo {volume} {73}},\ \bibinfo
  {pages} {011504} (\bibinfo {year} {2006})}\BibitemShut {NoStop}%
\bibitem [{\citenamefont {Chaudhuri}\ \emph {et~al.}(2007)\citenamefont
  {Chaudhuri}, \citenamefont {Berthier},\ and\ \citenamefont
  {Kob}}]{chaudhuri2007}%
  \BibitemOpen
  \bibfield  {author} {\bibinfo {author} {\bibfnamefont {P.}~\bibnamefont
  {Chaudhuri}}, \bibinfo {author} {\bibfnamefont {L.}~\bibnamefont
  {Berthier}},\ and\ \bibinfo {author} {\bibfnamefont {W.}~\bibnamefont
  {Kob}},\ }\bibfield  {title} {\bibinfo {title} {Universal nature of particle
  displacements close to glass and jamming transitions},\ }\href
  {https://doi.org/10.1103/PhysRevLett.99.060604} {\bibfield  {journal}
  {\bibinfo  {journal} {Phys. Rev. Lett.}\ }\textbf {\bibinfo {volume} {99}},\
  \bibinfo {pages} {060604} (\bibinfo {year} {2007})}\BibitemShut {NoStop}%
\bibitem [{sci(2005)}]{science2005}%
  \BibitemOpen
  \bibfield  {title} {\bibinfo {title} {What don't we know?},\ }\href@noop {}
  {\bibfield  {journal} {\bibinfo  {journal} {Science}\ }\textbf {\bibinfo
  {volume} {309}},\ \bibinfo {pages} {83} (\bibinfo {year} {2005})}\BibitemShut
  {NoStop}%
\bibitem [{\citenamefont {Hecksher}\ \emph {et~al.}(2008)\citenamefont
  {Hecksher}, \citenamefont {Nielsen}, \citenamefont {Olsen},\ and\
  \citenamefont {Dyre}}]{hecksher2008}%
  \BibitemOpen
  \bibfield  {author} {\bibinfo {author} {\bibfnamefont {T.}~\bibnamefont
  {Hecksher}}, \bibinfo {author} {\bibfnamefont {A.~I.}\ \bibnamefont
  {Nielsen}}, \bibinfo {author} {\bibfnamefont {N.~B.}\ \bibnamefont {Olsen}},\
  and\ \bibinfo {author} {\bibfnamefont {J.~C.}\ \bibnamefont {Dyre}},\
  }\bibfield  {title} {\bibinfo {title} {Little evidence for dynamic
  divergences in ultraviscous molecular liquids},\ }\href
  {https://doi.org/10.1038/nphys1033} {\bibfield  {journal} {\bibinfo
  {journal} {Nat. Phys.}\ }\textbf {\bibinfo {volume} {4}},\ \bibinfo {pages}
  {737} (\bibinfo {year} {2008})}\BibitemShut {NoStop}%
\bibitem [{\citenamefont {Berthier}\ \emph {et~al.}(2011)\citenamefont
  {Berthier}, \citenamefont {Biroli}, \citenamefont {Bouchaud},\ and\
  \citenamefont {Jack}}]{dhbook}%
  \BibitemOpen
  \bibfield  {author} {\bibinfo {author} {\bibfnamefont {L.}~\bibnamefont
  {Berthier}}, \bibinfo {author} {\bibfnamefont {G.}~\bibnamefont {Biroli}},
  \bibinfo {author} {\bibfnamefont {J.}~\bibnamefont {Bouchaud}},\ and\
  \bibinfo {author} {\bibfnamefont {R.~L.}\ \bibnamefont {Jack}},\ }\href@noop
  {} {\emph {\bibinfo {title} {Dynamical Heterogeneities in Glasses, Colloids,
  and Granular Media}}}\ (\bibinfo  {publisher} {Oxford University Press},\
  \bibinfo {year} {2011})\BibitemShut {NoStop}%
\bibitem [{\citenamefont {Miotto}\ \emph {et~al.}(2021)\citenamefont {Miotto},
  \citenamefont {Pigolotti}, \citenamefont {Chechkin},\ and\ \citenamefont
  {Rold{\'{a}}n-Vargas}}]{miotto2021}%
  \BibitemOpen
  \bibfield  {author} {\bibinfo {author} {\bibfnamefont {J.~M.}\ \bibnamefont
  {Miotto}}, \bibinfo {author} {\bibfnamefont {S.}~\bibnamefont {Pigolotti}},
  \bibinfo {author} {\bibfnamefont {A.~V.}\ \bibnamefont {Chechkin}},\ and\
  \bibinfo {author} {\bibfnamefont {S.}~\bibnamefont {Rold{\'{a}}n-Vargas}},\
  }\bibfield  {title} {\bibinfo {title} {Length scales in brownian yet
  non-gaussian dynamics},\ }\href {https://doi.org/10.1103/PhysRevX.11.031002}
  {\bibfield  {journal} {\bibinfo  {journal} {Phys. Rev. X}\ }\textbf {\bibinfo
  {volume} {11}},\ \bibinfo {pages} {031002} (\bibinfo {year}
  {2021})}\BibitemShut {NoStop}%
\bibitem [{\citenamefont {Rusciano}\ \emph
  {et~al.}(2022{\natexlab{a}})\citenamefont {Rusciano}, \citenamefont
  {Pastore},\ and\ \citenamefont {Greco}}]{rusciano2022}%
  \BibitemOpen
  \bibfield  {author} {\bibinfo {author} {\bibfnamefont {F.}~\bibnamefont
  {Rusciano}}, \bibinfo {author} {\bibfnamefont {R.}~\bibnamefont {Pastore}},\
  and\ \bibinfo {author} {\bibfnamefont {F.}~\bibnamefont {Greco}},\ }\bibfield
   {title} {\bibinfo {title} {Fickian non-gaussian diffusion in glass-forming
  liquids},\ }\href {https://doi.org/10.1103/PhysRevLett.128.168001} {\bibfield
   {journal} {\bibinfo  {journal} {Phys. Rev. Lett.}\ }\textbf {\bibinfo
  {volume} {128}},\ \bibinfo {pages} {168001} (\bibinfo {year}
  {2022}{\natexlab{a}})}\BibitemShut {NoStop}%
\bibitem [{\citenamefont {Angelini}\ \emph {et~al.}(2011)\citenamefont
  {Angelini}, \citenamefont {Hannezo}, \citenamefont {Trepat}, \citenamefont
  {Marquez}, \citenamefont {Fredberg},\ and\ \citenamefont
  {Weitz}}]{angelini2011}%
  \BibitemOpen
  \bibfield  {author} {\bibinfo {author} {\bibfnamefont {T.~E.}\ \bibnamefont
  {Angelini}}, \bibinfo {author} {\bibfnamefont {E.}~\bibnamefont {Hannezo}},
  \bibinfo {author} {\bibfnamefont {X.}~\bibnamefont {Trepat}}, \bibinfo
  {author} {\bibfnamefont {M.}~\bibnamefont {Marquez}}, \bibinfo {author}
  {\bibfnamefont {J.~J.}\ \bibnamefont {Fredberg}},\ and\ \bibinfo {author}
  {\bibfnamefont {D.~A.}\ \bibnamefont {Weitz}},\ }\bibfield  {title} {\bibinfo
  {title} {Glass-like dynamics of collective cell migration},\ }\href
  {https://doi.org/10.1073/pnas.1010059108} {\bibfield  {journal} {\bibinfo
  {journal} {Proc. Natl. Acad. Sci. (USA)}\ }\textbf {\bibinfo {volume}
  {108}},\ \bibinfo {pages} {4717} (\bibinfo {year} {2011})}\BibitemShut
  {NoStop}%
\bibitem [{\citenamefont {Garcia}\ \emph {et~al.}(2015)\citenamefont {Garcia},
  \citenamefont {Hannezo}, \citenamefont {Elgeti}, \citenamefont {Joanny},
  \citenamefont {Silberzan},\ and\ \citenamefont {Gov}}]{garcia2015}%
  \BibitemOpen
  \bibfield  {author} {\bibinfo {author} {\bibfnamefont {S.}~\bibnamefont
  {Garcia}}, \bibinfo {author} {\bibfnamefont {E.}~\bibnamefont {Hannezo}},
  \bibinfo {author} {\bibfnamefont {J.}~\bibnamefont {Elgeti}}, \bibinfo
  {author} {\bibfnamefont {J.~F.}\ \bibnamefont {Joanny}}, \bibinfo {author}
  {\bibfnamefont {P.}~\bibnamefont {Silberzan}},\ and\ \bibinfo {author}
  {\bibfnamefont {N.~S.}\ \bibnamefont {Gov}},\ }\bibfield  {title} {\bibinfo
  {title} {Physics of active jamming during collective cellular motion in a
  monolayer},\ }\href {https://doi.org/10.1073/pnas.1510973112} {\bibfield
  {journal} {\bibinfo  {journal} {Proc. Natl. Acad. Sci. (USA)}\ }\textbf
  {\bibinfo {volume} {112}},\ \bibinfo {pages} {15314} (\bibinfo {year}
  {2015})}\BibitemShut {NoStop}%
\bibitem [{\citenamefont {Sadhukhan}\ and\ \citenamefont
  {Nandi}(2021)}]{souvik2021}%
  \BibitemOpen
  \bibfield  {author} {\bibinfo {author} {\bibfnamefont {S.}~\bibnamefont
  {Sadhukhan}}\ and\ \bibinfo {author} {\bibfnamefont {S.~K.}\ \bibnamefont
  {Nandi}},\ }\bibfield  {title} {\bibinfo {title} {Theory and simulation for
  equilibrium glassy dynamics in cellular potts model of confluent biological
  tissue},\ }\href {https://doi.org/10.1103/PhysRevE.103.062403} {\bibfield
  {journal} {\bibinfo  {journal} {Phys. Rev. E}\ }\textbf {\bibinfo {volume}
  {103}},\ \bibinfo {pages} {062403} (\bibinfo {year} {2021})}\BibitemShut
  {NoStop}%
\bibitem [{\citenamefont {Sadhukhan}\ and\ \citenamefont
  {Nandi}(2022)}]{sadhukhan2022}%
  \BibitemOpen
  \bibfield  {author} {\bibinfo {author} {\bibfnamefont {S.}~\bibnamefont
  {Sadhukhan}}\ and\ \bibinfo {author} {\bibfnamefont {S.~K.}\ \bibnamefont
  {Nandi}},\ }\bibfield  {title} {\bibinfo {title} {On the origin of universal
  cell shape variability in confluent epithelial monolayers},\ }\href
  {https://doi.org/10.7554/elife.76406} {\bibfield  {journal} {\bibinfo
  {journal} {eLife}\ }\textbf {\bibinfo {volume} {11}},\ \bibinfo {pages}
  {e76406} (\bibinfo {year} {2022})}\BibitemShut {NoStop}%
\bibitem [{\citenamefont {Parry}\ \emph {et~al.}(2014)\citenamefont {Parry},
  \citenamefont {Surovtsev}, \citenamefont {Cabeen}, \citenamefont {O'Hern},
  \citenamefont {Dufresne},\ and\ \citenamefont {Jacobs-Wagner}}]{parry2014}%
  \BibitemOpen
  \bibfield  {author} {\bibinfo {author} {\bibfnamefont {B.~R.}\ \bibnamefont
  {Parry}}, \bibinfo {author} {\bibfnamefont {I.~V.}\ \bibnamefont
  {Surovtsev}}, \bibinfo {author} {\bibfnamefont {M.~T.}\ \bibnamefont
  {Cabeen}}, \bibinfo {author} {\bibfnamefont {C.~S.}\ \bibnamefont {O'Hern}},
  \bibinfo {author} {\bibfnamefont {E.~R.}\ \bibnamefont {Dufresne}},\ and\
  \bibinfo {author} {\bibfnamefont {C.}~\bibnamefont {Jacobs-Wagner}},\
  }\bibfield  {title} {\bibinfo {title} {The bacterial cytoplasm has glass-like
  properties and is fluidized by metabolic activity},\ }\href
  {https://doi.org/10.1016/j.cell.2013.11.028} {\bibfield  {journal} {\bibinfo
  {journal} {Cell}\ }\textbf {\bibinfo {volume} {156}},\ \bibinfo {pages} {183}
  (\bibinfo {year} {2014})}\BibitemShut {NoStop}%
\bibitem [{\citenamefont {Zhou}\ \emph {et~al.}(2009)\citenamefont {Zhou},
  \citenamefont {Trepat}, \citenamefont {Park}, \citenamefont {Lenormand},
  \citenamefont {Oliver}, \citenamefont {Mijailovich}, \citenamefont {Hardin},
  \citenamefont {Weitz}, \citenamefont {Butler},\ and\ \citenamefont
  {Fredberg}}]{zhou2009}%
  \BibitemOpen
  \bibfield  {author} {\bibinfo {author} {\bibfnamefont {E.~H.}\ \bibnamefont
  {Zhou}}, \bibinfo {author} {\bibfnamefont {X.}~\bibnamefont {Trepat}},
  \bibinfo {author} {\bibfnamefont {C.~Y.}\ \bibnamefont {Park}}, \bibinfo
  {author} {\bibfnamefont {G.}~\bibnamefont {Lenormand}}, \bibinfo {author}
  {\bibfnamefont {M.~N.}\ \bibnamefont {Oliver}}, \bibinfo {author}
  {\bibfnamefont {S.~M.}\ \bibnamefont {Mijailovich}}, \bibinfo {author}
  {\bibfnamefont {C.}~\bibnamefont {Hardin}}, \bibinfo {author} {\bibfnamefont
  {D.~A.}\ \bibnamefont {Weitz}}, \bibinfo {author} {\bibfnamefont {J.~P.}\
  \bibnamefont {Butler}},\ and\ \bibinfo {author} {\bibfnamefont {J.~J.}\
  \bibnamefont {Fredberg}},\ }\bibfield  {title} {\bibinfo {title} {Universal
  behavior of the osmotically compressed cell and its analogy to the colloidal
  glass transition},\ }\href {https://doi.org/10.1073pnas.0901462106}
  {\bibfield  {journal} {\bibinfo  {journal} {Proc. Natl. Acad. Sci. (USA)}\
  }\textbf {\bibinfo {volume} {106}},\ \bibinfo {pages} {10632} (\bibinfo
  {year} {2009})}\BibitemShut {NoStop}%
\bibitem [{\citenamefont {Fabry}\ \emph {et~al.}(2001)\citenamefont {Fabry},
  \citenamefont {Maksym}, \citenamefont {Butler}, \citenamefont {Glogauer},
  \citenamefont {Navajas},\ and\ \citenamefont {Fredberg}}]{fabry2001}%
  \BibitemOpen
  \bibfield  {author} {\bibinfo {author} {\bibfnamefont {B.}~\bibnamefont
  {Fabry}}, \bibinfo {author} {\bibfnamefont {G.~N.}\ \bibnamefont {Maksym}},
  \bibinfo {author} {\bibfnamefont {J.~P.}\ \bibnamefont {Butler}}, \bibinfo
  {author} {\bibfnamefont {M.}~\bibnamefont {Glogauer}}, \bibinfo {author}
  {\bibfnamefont {D.}~\bibnamefont {Navajas}},\ and\ \bibinfo {author}
  {\bibfnamefont {J.~J.}\ \bibnamefont {Fredberg}},\ }\bibfield  {title}
  {\bibinfo {title} {Scaling the microrheology of living cells},\ }\href
  {https://doi.org/10.1103/PhysRevLett.87.148102} {\bibfield  {journal}
  {\bibinfo  {journal} {Phys. Rev. Lett.}\ }\textbf {\bibinfo {volume} {87}},\
  \bibinfo {pages} {148102} (\bibinfo {year} {2001})}\BibitemShut {NoStop}%
\bibitem [{\citenamefont {Streitberger}\ \emph {et~al.}(2019)\citenamefont
  {Streitberger}, \citenamefont {Lilaj}, \citenamefont {Schrank}, \citenamefont
  {Braun}, \citenamefont {Hoffmann}, \citenamefont {Reiss-Zimmermann},
  \citenamefont {K{\"a}s},\ and\ \citenamefont {Sack}}]{streitberger2019}%
  \BibitemOpen
  \bibfield  {author} {\bibinfo {author} {\bibfnamefont {K.-J.}\ \bibnamefont
  {Streitberger}}, \bibinfo {author} {\bibfnamefont {L.}~\bibnamefont {Lilaj}},
  \bibinfo {author} {\bibfnamefont {F.}~\bibnamefont {Schrank}}, \bibinfo
  {author} {\bibfnamefont {J.}~\bibnamefont {Braun}}, \bibinfo {author}
  {\bibfnamefont {K.-T.}\ \bibnamefont {Hoffmann}}, \bibinfo {author}
  {\bibfnamefont {M.}~\bibnamefont {Reiss-Zimmermann}}, \bibinfo {author}
  {\bibfnamefont {J.~A.}\ \bibnamefont {K{\"a}s}},\ and\ \bibinfo {author}
  {\bibfnamefont {I.}~\bibnamefont {Sack}},\ }\bibfield  {title} {\bibinfo
  {title} {How tissue fluidity influences brain tumor progression},\ }\href
  {https://doi.org/10.1073/pnas.1913511116} {\bibfield  {journal} {\bibinfo
  {journal} {Proceedings of the National Academy of Sciences}\ }\textbf
  {\bibinfo {volume} {117}},\ \bibinfo {pages} {128} (\bibinfo {year}
  {2019})}\BibitemShut {NoStop}%
\bibitem [{\citenamefont {Biroli}\ \emph {et~al.}(2002)\citenamefont {Biroli},
  \citenamefont {Cocco},\ and\ \citenamefont {Monasson}}]{biroli2002}%
  \BibitemOpen
  \bibfield  {author} {\bibinfo {author} {\bibfnamefont {G.}~\bibnamefont
  {Biroli}}, \bibinfo {author} {\bibfnamefont {S.}~\bibnamefont {Cocco}},\ and\
  \bibinfo {author} {\bibfnamefont {R.}~\bibnamefont {Monasson}},\ }\bibfield
  {title} {\bibinfo {title} {Phase transitions and complexity in computer
  science: an overview of the statistical physics approach to the random
  satisfiability problem},\ }\href
  {https://doi.org/10.1016/s0378-4371(02)00516-2} {\bibfield  {journal}
  {\bibinfo  {journal} {Physica A: Statistical Mechanics and its Applications}\
  }\textbf {\bibinfo {volume} {306}},\ \bibinfo {pages} {381} (\bibinfo {year}
  {2002})}\BibitemShut {NoStop}%
\bibitem [{\citenamefont {Wolynes}\ \emph {et~al.}(1995)\citenamefont
  {Wolynes}, \citenamefont {Onuchic},\ and\ \citenamefont
  {Thirumalai}}]{wolynes1995}%
  \BibitemOpen
  \bibfield  {author} {\bibinfo {author} {\bibfnamefont {P.~G.}\ \bibnamefont
  {Wolynes}}, \bibinfo {author} {\bibfnamefont {J.~N.}\ \bibnamefont
  {Onuchic}},\ and\ \bibinfo {author} {\bibfnamefont {D.}~\bibnamefont
  {Thirumalai}},\ }\bibfield  {title} {\bibinfo {title} {Navigating the folding
  routes},\ }\href {https://doi.org/10.1126/science.7886447} {\bibfield
  {journal} {\bibinfo  {journal} {Science}\ }\textbf {\bibinfo {volume}
  {267}},\ \bibinfo {pages} {1619} (\bibinfo {year} {1995})}\BibitemShut
  {NoStop}%
\bibitem [{\citenamefont {Dauchot}\ \emph {et~al.}(2005)\citenamefont
  {Dauchot}, \citenamefont {Marty},\ and\ \citenamefont
  {Biroli}}]{dauchot2005}%
  \BibitemOpen
  \bibfield  {author} {\bibinfo {author} {\bibfnamefont {O.}~\bibnamefont
  {Dauchot}}, \bibinfo {author} {\bibfnamefont {G.}~\bibnamefont {Marty}},\
  and\ \bibinfo {author} {\bibfnamefont {G.}~\bibnamefont {Biroli}},\
  }\bibfield  {title} {\bibinfo {title} {Dynamical heterogeneity close to the
  jamming transition in a sheared granular material},\ }\href
  {https://doi.org/10.1103/PhysRevLett.95.265701} {\bibfield  {journal}
  {\bibinfo  {journal} {Phys. Rev. Lett.}\ }\textbf {\bibinfo {volume} {95}},\
  \bibinfo {pages} {265701} (\bibinfo {year} {2005})}\BibitemShut {NoStop}%
\bibitem [{\citenamefont {J.Deseigne}\ \emph {et~al.}(2010)\citenamefont
  {J.Deseigne}, \citenamefont {Dauchot},\ and\ \citenamefont
  {Chat{\'{e}}}}]{deseigne2010}%
  \BibitemOpen
  \bibfield  {author} {\bibinfo {author} {\bibnamefont {J.Deseigne}}, \bibinfo
  {author} {\bibfnamefont {O.}~\bibnamefont {Dauchot}},\ and\ \bibinfo {author}
  {\bibfnamefont {H.}~\bibnamefont {Chat{\'{e}}}},\ }\bibfield  {title}
  {\bibinfo {title} {Collective motion of vibrated polar disks},\ }\href
  {https://doi.org/10.1103/PhysRevLett.105.098001} {\bibfield  {journal}
  {\bibinfo  {journal} {Phys. Rev. Lett.}\ }\textbf {\bibinfo {volume} {105}},\
  \bibinfo {pages} {135702} (\bibinfo {year} {2010})}\BibitemShut {NoStop}%
\bibitem [{\citenamefont {Dreyfus}\ \emph {et~al.}(2005)\citenamefont
  {Dreyfus}, \citenamefont {Baudry}, \citenamefont {Roper}, \citenamefont
  {Fermigier}, \citenamefont {Stone},\ and\ \citenamefont
  {Bibette}}]{dreyfus2005}%
  \BibitemOpen
  \bibfield  {author} {\bibinfo {author} {\bibfnamefont {R.}~\bibnamefont
  {Dreyfus}}, \bibinfo {author} {\bibfnamefont {J.}~\bibnamefont {Baudry}},
  \bibinfo {author} {\bibfnamefont {M.~L.}\ \bibnamefont {Roper}}, \bibinfo
  {author} {\bibfnamefont {M.}~\bibnamefont {Fermigier}}, \bibinfo {author}
  {\bibfnamefont {H.~A.}\ \bibnamefont {Stone}},\ and\ \bibinfo {author}
  {\bibfnamefont {J.}~\bibnamefont {Bibette}},\ }\bibfield  {title} {\bibinfo
  {title} {Microscopic artificial swimmers},\ }\href
  {https://doi.org/10.1038/nature04090} {\bibfield  {journal} {\bibinfo
  {journal} {Nature}\ }\textbf {\bibinfo {volume} {437}},\ \bibinfo {pages}
  {862} (\bibinfo {year} {2005})}\BibitemShut {NoStop}%
\bibitem [{\citenamefont {Palacci}\ \emph {et~al.}(2010)\citenamefont
  {Palacci}, \citenamefont {Cottin-Bizonne}, \citenamefont {Ybert},\ and\
  \citenamefont {Bocquet}}]{palacci2010}%
  \BibitemOpen
  \bibfield  {author} {\bibinfo {author} {\bibfnamefont {J.}~\bibnamefont
  {Palacci}}, \bibinfo {author} {\bibfnamefont {C.}~\bibnamefont
  {Cottin-Bizonne}}, \bibinfo {author} {\bibfnamefont {C.}~\bibnamefont
  {Ybert}},\ and\ \bibinfo {author} {\bibfnamefont {L.}~\bibnamefont
  {Bocquet}},\ }\bibfield  {title} {\bibinfo {title} {Sedimentation and
  effective temperature of active colloidal suspensions},\ }\href
  {https://doi.org/10.1103/PhysRevLett.105.088304} {\bibfield  {journal}
  {\bibinfo  {journal} {Phys. Rev. Lett.}\ }\textbf {\bibinfo {volume} {105}},\
  \bibinfo {pages} {088304} (\bibinfo {year} {2010})}\BibitemShut {NoStop}%
\bibitem [{\citenamefont {Ni}\ \emph {et~al.}(2013)\citenamefont {Ni},
  \citenamefont {Stuart},\ and\ \citenamefont {Dijkstra}}]{ni2013}%
  \BibitemOpen
  \bibfield  {author} {\bibinfo {author} {\bibfnamefont {R.}~\bibnamefont
  {Ni}}, \bibinfo {author} {\bibfnamefont {M.~A.~C.}\ \bibnamefont {Stuart}},\
  and\ \bibinfo {author} {\bibfnamefont {M.}~\bibnamefont {Dijkstra}},\
  }\bibfield  {title} {\bibinfo {title} {Pushing the glass transition towards
  random close packing using self-propelled hard spheres},\ }\href
  {https://doi.org/10.1038/ncomms3704} {\bibfield  {journal} {\bibinfo
  {journal} {Nat. Commun}\ }\textbf {\bibinfo {volume} {4}},\ \bibinfo {pages}
  {2704} (\bibinfo {year} {2013})}\BibitemShut {NoStop}%
\bibitem [{\citenamefont {Berthier}(2014)}]{berthier2014}%
  \BibitemOpen
  \bibfield  {author} {\bibinfo {author} {\bibfnamefont {L.}~\bibnamefont
  {Berthier}},\ }\bibfield  {title} {\bibinfo {title} {Nonequilibrium glassy
  dynamics of self-propelled hard disks},\ }\href
  {https://doi.org/10.1103/PhysRevLett.112.220602} {\bibfield  {journal}
  {\bibinfo  {journal} {Phys. Rev. Lett.}\ }\textbf {\bibinfo {volume} {112}},\
  \bibinfo {pages} {220602} (\bibinfo {year} {2014})}\BibitemShut {NoStop}%
\bibitem [{\citenamefont {Kohlrausch}(1854)}]{kohlrausch1854}%
  \BibitemOpen
  \bibfield  {author} {\bibinfo {author} {\bibfnamefont {R.}~\bibnamefont
  {Kohlrausch}},\ }\bibfield  {title} {\bibinfo {title} {Theorie des
  elektrischen rückstandes in der leidener flasche},\ }\href
  {https://doi.org/10.1002/andp.18541670203} {\bibfield  {journal} {\bibinfo
  {journal} {Annalen der Physik}\ }\textbf {\bibinfo {volume} {167}},\ \bibinfo
  {pages} {179} (\bibinfo {year} {1854})}\BibitemShut {NoStop}%
\bibitem [{\citenamefont {Einstein}(1956)}]{einsteinpaper}%
  \BibitemOpen
  \bibfield  {author} {\bibinfo {author} {\bibfnamefont {A.}~\bibnamefont
  {Einstein}},\ }\href@noop {} {\emph {\bibinfo {title} {Investigations on the
  theory of the Brownian Movement}}}\ (\bibinfo  {publisher} {Dover
  Publications, INC., New York},\ \bibinfo {year} {1956})\BibitemShut {NoStop}%
\bibitem [{\citenamefont {Hansen}\ and\ \citenamefont
  {McDonald}(2018)}]{hansenmcdonald}%
  \BibitemOpen
  \bibfield  {author} {\bibinfo {author} {\bibfnamefont {J.}~\bibnamefont
  {Hansen}}\ and\ \bibinfo {author} {\bibfnamefont {I.~R.}\ \bibnamefont
  {McDonald}},\ }\href@noop {} {\emph {\bibinfo {title} {Theory of Simple
  Liquids}}},\ \bibinfo {edition} {3rd}\ ed.\ (\bibinfo  {publisher} {Elsevier,
  New York},\ \bibinfo {year} {2018})\BibitemShut {NoStop}%
\bibitem [{\citenamefont {Parmar}\ \emph {et~al.}(2017)\citenamefont {Parmar},
  \citenamefont {Sengupta},\ and\ \citenamefont {Sastry}}]{parmar2017}%
  \BibitemOpen
  \bibfield  {author} {\bibinfo {author} {\bibfnamefont {A.~D.~S.}\
  \bibnamefont {Parmar}}, \bibinfo {author} {\bibfnamefont {S.}~\bibnamefont
  {Sengupta}},\ and\ \bibinfo {author} {\bibfnamefont {S.}~\bibnamefont
  {Sastry}},\ }\bibfield  {title} {\bibinfo {title} {Length-scale dependence of
  the stokes-einstein and adam-gibbs relations in model glass formers},\ }\href
  {https://doi.org/10.1103/PhysRevLett.119.056001} {\bibfield  {journal}
  {\bibinfo  {journal} {Phys. Rev. Lett.}\ }\textbf {\bibinfo {volume} {119}},\
  \bibinfo {pages} {056001} (\bibinfo {year} {2017})}\BibitemShut {NoStop}%
\bibitem [{\citenamefont {Sengupta}\ \emph {et~al.}(2013)\citenamefont
  {Sengupta}, \citenamefont {Karmakar}, \citenamefont {Dasgupta},\ and\
  \citenamefont {Sastry}}]{shila2013}%
  \BibitemOpen
  \bibfield  {author} {\bibinfo {author} {\bibfnamefont {S.}~\bibnamefont
  {Sengupta}}, \bibinfo {author} {\bibfnamefont {S.}~\bibnamefont {Karmakar}},
  \bibinfo {author} {\bibfnamefont {C.}~\bibnamefont {Dasgupta}},\ and\
  \bibinfo {author} {\bibfnamefont {S.}~\bibnamefont {Sastry}},\ }\bibfield
  {title} {\bibinfo {title} {Breakdown of the stokes-einstein relation in two,
  three, and four dimensions},\ }\href {https://doi.org/10.1063/1.4792356}
  {\bibfield  {journal} {\bibinfo  {journal} {J. Chem. Phys.}\ }\textbf
  {\bibinfo {volume} {138}},\ \bibinfo {pages} {12A548} (\bibinfo {year}
  {2013})}\BibitemShut {NoStop}%
\bibitem [{\citenamefont {Shell}\ \emph {et~al.}(2005)\citenamefont {Shell},
  \citenamefont {Debenedetti},\ and\ \citenamefont {Stillinger}}]{shell2005}%
  \BibitemOpen
  \bibfield  {author} {\bibinfo {author} {\bibfnamefont {M.~S.}\ \bibnamefont
  {Shell}}, \bibinfo {author} {\bibfnamefont {P.~G.}\ \bibnamefont
  {Debenedetti}},\ and\ \bibinfo {author} {\bibfnamefont {F.~H.}\ \bibnamefont
  {Stillinger}},\ }\bibfield  {title} {\bibinfo {title} {Dynamic heterogeneity
  and non-gaussian behaviour in a model supercooled liquid},\ }\href
  {https://doi.org/10.1088/0953-8984/17/49/002} {\bibfield  {journal} {\bibinfo
   {journal} {J. Phys.: Condens. Matter}\ }\textbf {\bibinfo {volume} {17}},\
  \bibinfo {pages} {S4035} (\bibinfo {year} {2005})}\BibitemShut {NoStop}%
\bibitem [{\citenamefont {Ediger}(2000)}]{ediger2000}%
  \BibitemOpen
  \bibfield  {author} {\bibinfo {author} {\bibfnamefont {M.~D.}\ \bibnamefont
  {Ediger}},\ }\bibfield  {title} {\bibinfo {title} {Spatially heterogeneous
  dynamics in supercooled liquids},\ }\href
  {https://doi.org/10.1146/annurev.physchem.51.1.99} {\bibfield  {journal}
  {\bibinfo  {journal} {Annu. Rev. Phys. Chem.}\ }\textbf {\bibinfo {volume}
  {51}},\ \bibinfo {pages} {99} (\bibinfo {year} {2000})}\BibitemShut {NoStop}%
\bibitem [{\citenamefont {Berthier}(2011)}]{berthier2011}%
  \BibitemOpen
  \bibfield  {author} {\bibinfo {author} {\bibfnamefont {L.}~\bibnamefont
  {Berthier}},\ }\bibfield  {title} {\bibinfo {title} {Dynamic heterogeneity in
  amorphous materials},\ }\href {https://doi.org/10.1103/Physics.4.42}
  {\bibfield  {journal} {\bibinfo  {journal} {Physics}\ }\textbf {\bibinfo
  {volume} {4}},\ \bibinfo {pages} {42} (\bibinfo {year} {2011})}\BibitemShut
  {NoStop}%
\bibitem [{\citenamefont {Yamamoto}\ and\ \citenamefont
  {Onuki}(1998)}]{yamamoto1998}%
  \BibitemOpen
  \bibfield  {author} {\bibinfo {author} {\bibfnamefont {R.}~\bibnamefont
  {Yamamoto}}\ and\ \bibinfo {author} {\bibfnamefont {A.}~\bibnamefont
  {Onuki}},\ }\bibfield  {title} {\bibinfo {title} {Heterogeneous diffusion in
  highly supercooled liquids},\ }\href
  {https://doi.org/10.1103/PhysRevLett.81.4915} {\bibfield  {journal} {\bibinfo
   {journal} {Phys. Rev. Lett.}\ }\textbf {\bibinfo {volume} {81}},\ \bibinfo
  {pages} {4915} (\bibinfo {year} {1998})}\BibitemShut {NoStop}%
\bibitem [{\citenamefont {Franz}\ and\ \citenamefont
  {Parisi}(2000)}]{franz2000}%
  \BibitemOpen
  \bibfield  {author} {\bibinfo {author} {\bibfnamefont {S.}~\bibnamefont
  {Franz}}\ and\ \bibinfo {author} {\bibfnamefont {G.}~\bibnamefont {Parisi}},\
  }\bibfield  {title} {\bibinfo {title} {On non-linear susceptibility in
  supercooled liquids},\ }\href {https://doi.org/10.1088/0953-8984/12/29/305}
  {\bibfield  {journal} {\bibinfo  {journal} {J. Phys.: Condens. Matter}\
  }\textbf {\bibinfo {volume} {12}},\ \bibinfo {pages} {6335} (\bibinfo {year}
  {2000})}\BibitemShut {NoStop}%
\bibitem [{\citenamefont {Weeks}\ \emph {et~al.}(2000)\citenamefont {Weeks},
  \citenamefont {Crocker}, \citenamefont {Levitt}, \citenamefont {Schofield},\
  and\ \citenamefont {Weitz}}]{weeks2000}%
  \BibitemOpen
  \bibfield  {author} {\bibinfo {author} {\bibfnamefont {E.~R.}\ \bibnamefont
  {Weeks}}, \bibinfo {author} {\bibfnamefont {J.~C.}\ \bibnamefont {Crocker}},
  \bibinfo {author} {\bibfnamefont {A.~C.}\ \bibnamefont {Levitt}}, \bibinfo
  {author} {\bibfnamefont {A.}~\bibnamefont {Schofield}},\ and\ \bibinfo
  {author} {\bibfnamefont {D.~A.}\ \bibnamefont {Weitz}},\ }\bibfield  {title}
  {\bibinfo {title} {Three-dimensional direct imaging of structural relaxation
  near the colloidal glass transition},\ }\href
  {https://doi.org/10.1126/science.287.5453.627} {\bibfield  {journal}
  {\bibinfo  {journal} {Science}\ }\textbf {\bibinfo {volume} {287}},\ \bibinfo
  {pages} {627} (\bibinfo {year} {2000})}\BibitemShut {NoStop}%
\bibitem [{\citenamefont {Karmakar}\ \emph {et~al.}(2015)\citenamefont
  {Karmakar}, \citenamefont {Dasgupta},\ and\ \citenamefont
  {Sastry}}]{smarajit2015}%
  \BibitemOpen
  \bibfield  {author} {\bibinfo {author} {\bibfnamefont {S.}~\bibnamefont
  {Karmakar}}, \bibinfo {author} {\bibfnamefont {C.}~\bibnamefont {Dasgupta}},\
  and\ \bibinfo {author} {\bibfnamefont {S.}~\bibnamefont {Sastry}},\
  }\bibfield  {title} {\bibinfo {title} {Length scales in glass-forming liquids
  and related systems: a review},\ }\href
  {https://doi.org/10.1088/0034-4885/79/1/016601} {\bibfield  {journal}
  {\bibinfo  {journal} {Rep. Prog. Phys.}\ }\textbf {\bibinfo {volume} {79}},\
  \bibinfo {pages} {016601} (\bibinfo {year} {2015})}\BibitemShut {NoStop}%
\bibitem [{\citenamefont {Guan}\ \emph {et~al.}(2014)\citenamefont {Guan},
  \citenamefont {Wang},\ and\ \citenamefont {Granick}}]{guan2014}%
  \BibitemOpen
  \bibfield  {author} {\bibinfo {author} {\bibfnamefont {J.}~\bibnamefont
  {Guan}}, \bibinfo {author} {\bibfnamefont {B.}~\bibnamefont {Wang}},\ and\
  \bibinfo {author} {\bibfnamefont {S.}~\bibnamefont {Granick}},\ }\bibfield
  {title} {\bibinfo {title} {Even hard-sphere colloidal suspensions display
  fickian yet non-gaussian diffusion},\ }\href
  {https://doi.org/10.1021/nn405476t} {\bibfield  {journal} {\bibinfo
  {journal} {ACS Nano}\ }\textbf {\bibinfo {volume} {8}},\ \bibinfo {pages}
  {3331} (\bibinfo {year} {2014})}\BibitemShut {NoStop}%
\bibitem [{\citenamefont {Wang}\ \emph {et~al.}(2009)\citenamefont {Wang},
  \citenamefont {Anthony}, \citenamefont {Bae},\ and\ \citenamefont
  {Granick}}]{wang2009}%
  \BibitemOpen
  \bibfield  {author} {\bibinfo {author} {\bibfnamefont {B.}~\bibnamefont
  {Wang}}, \bibinfo {author} {\bibfnamefont {S.~M.}\ \bibnamefont {Anthony}},
  \bibinfo {author} {\bibfnamefont {S.~C.}\ \bibnamefont {Bae}},\ and\ \bibinfo
  {author} {\bibfnamefont {S.}~\bibnamefont {Granick}},\ }\bibfield  {title}
  {\bibinfo {title} {Anomalous yet brownian},\ }\href
  {https://doi.org/10.1073/pnas.0903554106} {\bibfield  {journal} {\bibinfo
  {journal} {Proc. Natl. Acad. Sci. (USA)}\ }\textbf {\bibinfo {volume}
  {106}},\ \bibinfo {pages} {15160} (\bibinfo {year} {2009})}\BibitemShut
  {NoStop}%
\bibitem [{\citenamefont {Chechkin}\ \emph {et~al.}(2017)\citenamefont
  {Chechkin}, \citenamefont {Seno}, \citenamefont {Metzler},\ and\
  \citenamefont {Sokolov}}]{chechkin2017}%
  \BibitemOpen
  \bibfield  {author} {\bibinfo {author} {\bibfnamefont {A.~V.}\ \bibnamefont
  {Chechkin}}, \bibinfo {author} {\bibfnamefont {F.}~\bibnamefont {Seno}},
  \bibinfo {author} {\bibfnamefont {R.}~\bibnamefont {Metzler}},\ and\ \bibinfo
  {author} {\bibfnamefont {I.~M.}\ \bibnamefont {Sokolov}},\ }\bibfield
  {title} {\bibinfo {title} {Brownian yet non-gaussian diffusion: From
  superstatistics to subordination of diffusing diffusivities},\ }\href
  {https://doi.org/10.1103/PhysRevX.7.021002} {\bibfield  {journal} {\bibinfo
  {journal} {Phys. Rev. X}\ }\textbf {\bibinfo {volume} {7}},\ \bibinfo {pages}
  {021002} (\bibinfo {year} {2017})}\BibitemShut {NoStop}%
\bibitem [{\citenamefont {Chubynsky}\ and\ \citenamefont
  {Slater}(2014)}]{chubynsky2014}%
  \BibitemOpen
  \bibfield  {author} {\bibinfo {author} {\bibfnamefont {M.~V.}\ \bibnamefont
  {Chubynsky}}\ and\ \bibinfo {author} {\bibfnamefont {G.~W.}\ \bibnamefont
  {Slater}},\ }\bibfield  {title} {\bibinfo {title} {Diffusing diffusivity: A
  model for anomalous, yet brownian, diffusion},\ }\href
  {https://doi.org/10.1103/PhysRevLett.113.098302} {\bibfield  {journal}
  {\bibinfo  {journal} {Phys. Rev. Lett.}\ }\textbf {\bibinfo {volume} {113}},\
  \bibinfo {pages} {098302} (\bibinfo {year} {2014})}\BibitemShut {NoStop}%
\bibitem [{\citenamefont {Metzler}\ \emph {et~al.}(2014)\citenamefont
  {Metzler}, \citenamefont {Jeon}, \citenamefont {Cherstvy},\ and\
  \citenamefont {Barkai}}]{metzler2014}%
  \BibitemOpen
  \bibfield  {author} {\bibinfo {author} {\bibfnamefont {R.}~\bibnamefont
  {Metzler}}, \bibinfo {author} {\bibfnamefont {J.-H.}\ \bibnamefont {Jeon}},
  \bibinfo {author} {\bibfnamefont {A.~G.}\ \bibnamefont {Cherstvy}},\ and\
  \bibinfo {author} {\bibfnamefont {E.}~\bibnamefont {Barkai}},\ }\bibfield
  {title} {\bibinfo {title} {Anomalous diffusion models and their properties:
  non-stationarity, non-ergodicity, and ageing at the centenary of single
  particle tracking},\ }\href {https://doi.org/10.1039/c4cp03465a} {\bibfield
  {journal} {\bibinfo  {journal} {Soft Matter}\ }\textbf {\bibinfo {volume}
  {16}},\ \bibinfo {pages} {24128} (\bibinfo {year} {2014})}\BibitemShut
  {NoStop}%
\bibitem [{\citenamefont {Metzler}(2016)}]{metzler2016}%
  \BibitemOpen
  \bibfield  {author} {\bibinfo {author} {\bibfnamefont {R.}~\bibnamefont
  {Metzler}},\ }\bibfield  {title} {\bibinfo {title} {Gaussianity fair: The
  riddle of anomalous yet non-gaussian diffusion},\ }\href
  {https://doi.org/10.1016/j.bpj.2016.12.019} {\bibfield  {journal} {\bibinfo
  {journal} {Biophys. J.}\ }\textbf {\bibinfo {volume} {112}},\ \bibinfo
  {pages} {413} (\bibinfo {year} {2016})}\BibitemShut {NoStop}%
\bibitem [{\citenamefont {Jain}\ and\ \citenamefont
  {Sebastian}(2016)}]{Jain2016}%
  \BibitemOpen
  \bibfield  {author} {\bibinfo {author} {\bibfnamefont {R.}~\bibnamefont
  {Jain}}\ and\ \bibinfo {author} {\bibfnamefont {K.~L.}\ \bibnamefont
  {Sebastian}},\ }\bibfield  {title} {\bibinfo {title} {Diffusion in a crowded,
  rearranging environment},\ }\href {https://doi.org/10.1021/acs.jpcb.6b01527}
  {\bibfield  {journal} {\bibinfo  {journal} {The Journal of Physical Chemistry
  B}\ }\textbf {\bibinfo {volume} {120}},\ \bibinfo {pages} {3988} (\bibinfo
  {year} {2016})}\BibitemShut {NoStop}%
\bibitem [{\citenamefont {Berthier}\ \emph {et~al.}(2022)\citenamefont
  {Berthier}, \citenamefont {Flenner},\ and\ \citenamefont
  {Szamel}}]{berthiercomment}%
  \BibitemOpen
  \bibfield  {author} {\bibinfo {author} {\bibfnamefont {L.}~\bibnamefont
  {Berthier}}, \bibinfo {author} {\bibfnamefont {E.}~\bibnamefont {Flenner}},\
  and\ \bibinfo {author} {\bibfnamefont {G.}~\bibnamefont {Szamel}},\
  }\bibfield  {title} {\bibinfo {title} {Comment on ‘fickian non-gaussian
  diffusion in glass-forming liquids’},\ }\href
  {https://arxiv.org/abs/2210.07119v2} {\bibfield  {journal} {\bibinfo
  {journal} {arXiv}\ ,\ \bibinfo {pages} {arXiv:2210.07119}} (\bibinfo {year}
  {2022})}\BibitemShut {NoStop}%
\bibitem [{\citenamefont {Das}\ \emph {et~al.}(2018)\citenamefont {Das},
  \citenamefont {Dasgupta},\ and\ \citenamefont {Karmakar}}]{das2018}%
  \BibitemOpen
  \bibfield  {author} {\bibinfo {author} {\bibfnamefont {R.}~\bibnamefont
  {Das}}, \bibinfo {author} {\bibfnamefont {C.}~\bibnamefont {Dasgupta}},\ and\
  \bibinfo {author} {\bibfnamefont {S.}~\bibnamefont {Karmakar}},\ }\bibfield
  {title} {\bibinfo {title} {Time scales of fickian diffusion and the lifetime
  of dynamic heterogeneity},\ }\href {https://doi.org/10.3389/fphy.2020.00210}
  {\bibfield  {journal} {\bibinfo  {journal} {Front. in Phys.}\ }\textbf
  {\bibinfo {volume} {8}},\ \bibinfo {pages} {210} (\bibinfo {year}
  {2018})}\BibitemShut {NoStop}%
\bibitem [{\citenamefont {Rusciano}\ \emph {et~al.}(2023)\citenamefont
  {Rusciano}, \citenamefont {Pastore},\ and\ \citenamefont
  {Greco}}]{rusciano2023}%
  \BibitemOpen
  \bibfield  {author} {\bibinfo {author} {\bibfnamefont {F.}~\bibnamefont
  {Rusciano}}, \bibinfo {author} {\bibfnamefont {R.}~\bibnamefont {Pastore}},\
  and\ \bibinfo {author} {\bibfnamefont {F.}~\bibnamefont {Greco}},\ }\bibfield
   {title} {\bibinfo {title} {Universal evolution of fickian non-gaussian
  diffusion in two- and three-dimensional glass-forming liquids},\ }\href
  {https://doi.org/10.3390/ijms24097871} {\bibfield  {journal} {\bibinfo
  {journal} {Int. J. Mol. Sci.}\ }\textbf {\bibinfo {volume} {24}},\ \bibinfo
  {pages} {7871} (\bibinfo {year} {2023})}\BibitemShut {NoStop}%
\bibitem [{\citenamefont {Skaug}\ \emph {et~al.}(2013)\citenamefont {Skaug},
  \citenamefont {Mabry},\ and\ \citenamefont {Schwartz}}]{skaug2013}%
  \BibitemOpen
  \bibfield  {author} {\bibinfo {author} {\bibfnamefont {M.~J.}\ \bibnamefont
  {Skaug}}, \bibinfo {author} {\bibfnamefont {J.}~\bibnamefont {Mabry}},\ and\
  \bibinfo {author} {\bibfnamefont {D.~K.}\ \bibnamefont {Schwartz}},\
  }\bibfield  {title} {\bibinfo {title} {Intermittent molecular hopping at the
  solid-liquid interface},\ }\href
  {https://doi.org/10.1103/PhysRevLett.110.256101} {\bibfield  {journal}
  {\bibinfo  {journal} {Phys. Rev. Lett.}\ }\textbf {\bibinfo {volume} {110}},\
  \bibinfo {pages} {256101} (\bibinfo {year} {2013})}\BibitemShut {NoStop}%
\bibitem [{\citenamefont {Acharya}\ \emph {et~al.}(2017)\citenamefont
  {Acharya}, \citenamefont {Nandi},\ and\ \citenamefont
  {Bhattacharyya}}]{acharya2017}%
  \BibitemOpen
  \bibfield  {author} {\bibinfo {author} {\bibfnamefont {S.}~\bibnamefont
  {Acharya}}, \bibinfo {author} {\bibfnamefont {U.~K.}\ \bibnamefont {Nandi}},\
  and\ \bibinfo {author} {\bibfnamefont {S.~M.}\ \bibnamefont
  {Bhattacharyya}},\ }\bibfield  {title} {\bibinfo {title} {Fickian yet
  non-gaussian behaviour: A dominant role of the intermittent dynamics},\
  }\href {https://doi.org/10.1063/1.4979338} {\bibfield  {journal} {\bibinfo
  {journal} {J. Chem. Phys.}\ }\textbf {\bibinfo {volume} {146}},\ \bibinfo
  {pages} {134504} (\bibinfo {year} {2017})}\BibitemShut {NoStop}%
\bibitem [{\citenamefont {Rusciano}\ \emph
  {et~al.}(2022{\natexlab{b}})\citenamefont {Rusciano}, \citenamefont
  {Pastore},\ and\ \citenamefont {Greco}}]{ruscianoreply}%
  \BibitemOpen
  \bibfield  {author} {\bibinfo {author} {\bibfnamefont {F.}~\bibnamefont
  {Rusciano}}, \bibinfo {author} {\bibfnamefont {R.}~\bibnamefont {Pastore}},\
  and\ \bibinfo {author} {\bibfnamefont {F.}~\bibnamefont {Greco}},\ }\bibfield
   {title} {\bibinfo {title} {Reply to 'comment on `fickian non-gaussian
  diffusion in glass-forming liquids' '},\ }\href
  {https://doi.org/10.48550/arXiv.2212.09679} {\bibfield  {journal} {\bibinfo
  {journal} {arXiv}\ ,\ \bibinfo {pages} {arXiv:2212.09679}} (\bibinfo {year}
  {2022}{\natexlab{b}})}\BibitemShut {NoStop}%
\bibitem [{\citenamefont {Saltzman}\ \emph {et~al.}(2008)\citenamefont
  {Saltzman}, \citenamefont {Yatsenko},\ and\ \citenamefont
  {Schweizer}}]{saltzman2008}%
  \BibitemOpen
  \bibfield  {author} {\bibinfo {author} {\bibfnamefont {E.~J.}\ \bibnamefont
  {Saltzman}}, \bibinfo {author} {\bibfnamefont {G.}~\bibnamefont {Yatsenko}},\
  and\ \bibinfo {author} {\bibfnamefont {K.~S.}\ \bibnamefont {Schweizer}},\
  }\bibfield  {title} {\bibinfo {title} {Anomalous diffusion, structural
  relaxation and shear thinning in glassy hard sphere fluids},\ }\href
  {https://doi.org/10.1088/0953-8984/20/24/244129} {\bibfield  {journal}
  {\bibinfo  {journal} {J. Phys.: Condens. Matter}\ }\textbf {\bibinfo {volume}
  {20}},\ \bibinfo {pages} {244129} (\bibinfo {year} {2008})}\BibitemShut
  {NoStop}%
\bibitem [{\citenamefont {Kim}\ and\ \citenamefont {Saito}(2009)}]{kim2009}%
  \BibitemOpen
  \bibfield  {author} {\bibinfo {author} {\bibfnamefont {K.}~\bibnamefont
  {Kim}}\ and\ \bibinfo {author} {\bibfnamefont {S.}~\bibnamefont {Saito}},\
  }\bibfield  {title} {\bibinfo {title} {Multiple time scales hidden in
  heterogeneous dynamics of glass-forming liquids},\ }\href
  {https://doi.org/10.1103/PhysRevE.79.060501} {\bibfield  {journal} {\bibinfo
  {journal} {Phys. Rev. E}\ }\textbf {\bibinfo {volume} {79}},\ \bibinfo
  {pages} {060501(R)} (\bibinfo {year} {2009})}\BibitemShut {NoStop}%
\bibitem [{\citenamefont {Mezard}\ \emph {et~al.}(1987)\citenamefont {Mezard},
  \citenamefont {Parisi},\ and\ \citenamefont {Virasoro}}]{spinglassbook}%
  \BibitemOpen
  \bibfield  {author} {\bibinfo {author} {\bibfnamefont {M.}~\bibnamefont
  {Mezard}}, \bibinfo {author} {\bibfnamefont {G.}~\bibnamefont {Parisi}},\
  and\ \bibinfo {author} {\bibfnamefont {M.~A.}\ \bibnamefont {Virasoro}},\
  }\href@noop {} {\emph {\bibinfo {title} {Spin Glass Theory And Beyond: An
  Introduction To The Replica Method And Its Applications}}}\ (\bibinfo
  {publisher} {World Scientific Publishing Company},\ \bibinfo {year}
  {1987})\BibitemShut {NoStop}%
\bibitem [{\citenamefont {Guiselin}\ \emph {et~al.}(2020)\citenamefont
  {Guiselin}, \citenamefont {Tarjus},\ and\ \citenamefont
  {Berthier}}]{guiselin2020}%
  \BibitemOpen
  \bibfield  {author} {\bibinfo {author} {\bibfnamefont {B.}~\bibnamefont
  {Guiselin}}, \bibinfo {author} {\bibfnamefont {G.}~\bibnamefont {Tarjus}},\
  and\ \bibinfo {author} {\bibfnamefont {L.}~\bibnamefont {Berthier}},\
  }\bibfield  {title} {\bibinfo {title} {On the overlap between configurations
  in glassy liquids},\ }\href {https://doi.org/10.1063/5.0022614} {\bibfield
  {journal} {\bibinfo  {journal} {J. Chem. Phys.}\ }\textbf {\bibinfo {volume}
  {153}},\ \bibinfo {pages} {224502} (\bibinfo {year} {2020})}\BibitemShut
  {NoStop}%
\bibitem [{\citenamefont {Barkai}\ and\ \citenamefont
  {Burov}(2020)}]{barkai2020}%
  \BibitemOpen
  \bibfield  {author} {\bibinfo {author} {\bibfnamefont {E.}~\bibnamefont
  {Barkai}}\ and\ \bibinfo {author} {\bibfnamefont {S.}~\bibnamefont {Burov}},\
  }\bibfield  {title} {\bibinfo {title} {Packets of diffusing particles exhibit
  universal exponential tails},\ }\href
  {https://doi.org/10.1103/PhysRevLett.124.060603} {\bibfield  {journal}
  {\bibinfo  {journal} {Phys. Rev. Lett.}\ }\textbf {\bibinfo {volume} {124}},\
  \bibinfo {pages} {060603} (\bibinfo {year} {2020})}\BibitemShut {NoStop}%
\bibitem [{\citenamefont {Wang}\ \emph {et~al.}(2020)\citenamefont {Wang},
  \citenamefont {Barkai},\ and\ \citenamefont {Burov}}]{wang2020}%
  \BibitemOpen
  \bibfield  {author} {\bibinfo {author} {\bibfnamefont {W.}~\bibnamefont
  {Wang}}, \bibinfo {author} {\bibfnamefont {E.}~\bibnamefont {Barkai}},\ and\
  \bibinfo {author} {\bibfnamefont {S.}~\bibnamefont {Burov}},\ }\bibfield
  {title} {\bibinfo {title} {Large deviations for continuous time random
  walks},\ }\href {https://doi.org/10.3390/e22060697} {\bibfield  {journal}
  {\bibinfo  {journal} {Entropy}\ }\textbf {\bibinfo {volume} {22}},\ \bibinfo
  {pages} {697} (\bibinfo {year} {2020})}\BibitemShut {NoStop}%
\bibitem [{\citenamefont {Scher}\ and\ \citenamefont
  {Montroll}(1975)}]{scher1975}%
  \BibitemOpen
  \bibfield  {author} {\bibinfo {author} {\bibfnamefont {H.}~\bibnamefont
  {Scher}}\ and\ \bibinfo {author} {\bibfnamefont {E.~W.}\ \bibnamefont
  {Montroll}},\ }\bibfield  {title} {\bibinfo {title} {Anomalous transit-time
  dispersion in amorphous solids},\ }\href
  {https://doi.org/10.1103/PhysRevB.12.2455} {\bibfield  {journal} {\bibinfo
  {journal} {Phys. Rev. B}\ }\textbf {\bibinfo {volume} {12}},\ \bibinfo
  {pages} {2455} (\bibinfo {year} {1975})}\BibitemShut {NoStop}%
\bibitem [{\citenamefont {Bertin}\ and\ \citenamefont
  {Bouchaud}(2003)}]{bertin2003}%
  \BibitemOpen
  \bibfield  {author} {\bibinfo {author} {\bibfnamefont {E.~M.}\ \bibnamefont
  {Bertin}}\ and\ \bibinfo {author} {\bibfnamefont {J.-P.}\ \bibnamefont
  {Bouchaud}},\ }\bibfield  {title} {\bibinfo {title} {Subdiffusion and
  localization in the one-dimensional trap model},\ }\href
  {https://doi.org/10.1103/PhysRevE.67.026128} {\bibfield  {journal} {\bibinfo
  {journal} {Phys. Rev. E}\ }\textbf {\bibinfo {volume} {67}},\ \bibinfo
  {pages} {026128} (\bibinfo {year} {2003})}\BibitemShut {NoStop}%
\bibitem [{\citenamefont {Niblett}\ \emph {et~al.}(2017)\citenamefont
  {Niblett}, \citenamefont {Biedermann}, \citenamefont {Wales},\ and\
  \citenamefont {de~Souza}}]{niblett2017}%
  \BibitemOpen
  \bibfield  {author} {\bibinfo {author} {\bibfnamefont {S.~P.}\ \bibnamefont
  {Niblett}}, \bibinfo {author} {\bibfnamefont {M.}~\bibnamefont {Biedermann}},
  \bibinfo {author} {\bibfnamefont {D.~J.}\ \bibnamefont {Wales}},\ and\
  \bibinfo {author} {\bibfnamefont {V.~K.}\ \bibnamefont {de~Souza}},\
  }\bibfield  {title} {\bibinfo {title} {Pathways for diffusion in the
  potential energy landscape of the network glass former sio2},\ }\href
  {https://doi.org/10.1063/1.5005924} {\bibfield  {journal} {\bibinfo
  {journal} {J. Chem. Phys.}\ }\textbf {\bibinfo {volume} {147}},\ \bibinfo
  {pages} {152726} (\bibinfo {year} {2017})}\BibitemShut {NoStop}%
\bibitem [{\citenamefont {Lampo}\ \emph {et~al.}(2017)\citenamefont {Lampo},
  \citenamefont {Stylianidou}, \citenamefont {Backlund}, \citenamefont
  {Wiggins},\ and\ \citenamefont {Spakowitz}}]{lampo2017}%
  \BibitemOpen
  \bibfield  {author} {\bibinfo {author} {\bibfnamefont {T.~J.}\ \bibnamefont
  {Lampo}}, \bibinfo {author} {\bibfnamefont {S.}~\bibnamefont {Stylianidou}},
  \bibinfo {author} {\bibfnamefont {M.~P.}\ \bibnamefont {Backlund}}, \bibinfo
  {author} {\bibfnamefont {P.~A.}\ \bibnamefont {Wiggins}},\ and\ \bibinfo
  {author} {\bibfnamefont {A.~J.}\ \bibnamefont {Spakowitz}},\ }\bibfield
  {title} {\bibinfo {title} {Cytoplasmic rna-protein particles exhibit
  non-gaussian subdiffusive behavior},\ }\href
  {https://doi.org/10.1016/j.bpj.2016.11.3208} {\bibfield  {journal} {\bibinfo
  {journal} {Biophys. J.}\ }\textbf {\bibinfo {volume} {112}},\ \bibinfo
  {pages} {532} (\bibinfo {year} {2017})}\BibitemShut {NoStop}%
\bibitem [{\citenamefont {Dankel}(1991)}]{dankel1991}%
  \BibitemOpen
  \bibfield  {author} {\bibinfo {author} {\bibfnamefont {T.}~\bibnamefont
  {Dankel}},\ }\bibfield  {title} {\bibinfo {title} {On the distribution of the
  integrated square of the ornstein-uhlenbeck process},\ }\href@noop {}
  {\bibfield  {journal} {\bibinfo  {journal} {J. Appl. Math.}\ }\textbf
  {\bibinfo {volume} {51}},\ \bibinfo {pages} {568} (\bibinfo {year}
  {1991})}\BibitemShut {NoStop}%
\bibitem [{\citenamefont {Binder}(1981)}]{binder1981}%
  \BibitemOpen
  \bibfield  {author} {\bibinfo {author} {\bibfnamefont {K.}~\bibnamefont
  {Binder}},\ }\bibfield  {title} {\bibinfo {title} {Finite size scaling
  analysis of ising model block distribution functions},\ }\href
  {https://doi.org/10.1007/BF01293604} {\bibfield  {journal} {\bibinfo
  {journal} {Zeitschrift f{\"u}r Physik B Condensed Matter}\ }\textbf {\bibinfo
  {volume} {43}},\ \bibinfo {pages} {119} (\bibinfo {year} {1981})}\BibitemShut
  {NoStop}%
\bibitem [{\citenamefont {Rahman}(1964)}]{rahman1964}%
  \BibitemOpen
  \bibfield  {author} {\bibinfo {author} {\bibfnamefont {A.}~\bibnamefont
  {Rahman}},\ }\bibfield  {title} {\bibinfo {title} {Correlations in the motion
  of atoms in liquid argon},\ }\href {https://doi.org/10.1103/PhysRev.136.A405}
  {\bibfield  {journal} {\bibinfo  {journal} {Phys. Rev.}\ }\textbf {\bibinfo
  {volume} {136}},\ \bibinfo {pages} {A405} (\bibinfo {year}
  {1964})}\BibitemShut {NoStop}%
\bibitem [{\citenamefont {Kawasaki}\ and\ \citenamefont
  {Kim}(2017)}]{kawasaki2017}%
  \BibitemOpen
  \bibfield  {author} {\bibinfo {author} {\bibfnamefont {T.}~\bibnamefont
  {Kawasaki}}\ and\ \bibinfo {author} {\bibfnamefont {K.}~\bibnamefont {Kim}},\
  }\bibfield  {title} {\bibinfo {title} {Identifying time scales for
  violation/preservation of stokes-einstein relation in supercooled water},\
  }\href {https://doi.org/10.1126/sciadv.1700399} {\bibfield  {journal}
  {\bibinfo  {journal} {Sci. Adv.}\ }\textbf {\bibinfo {volume} {3}},\ \bibinfo
  {pages} {e1700399} (\bibinfo {year} {2017})}\BibitemShut {NoStop}%
\bibitem [{\citenamefont {Stillinger}(1988)}]{stillinger1988}%
  \BibitemOpen
  \bibfield  {author} {\bibinfo {author} {\bibfnamefont {F.~H.}\ \bibnamefont
  {Stillinger}},\ }\bibfield  {title} {\bibinfo {title} {Relaxation and flow
  mechanisms in ``fragile'' glass-forming liquids},\ }\href
  {https://doi.org/10.1063/1.455365} {\bibfield  {journal} {\bibinfo  {journal}
  {J. Chem. Phys.}\ }\textbf {\bibinfo {volume} {89}},\ \bibinfo {pages} {6461}
  (\bibinfo {year} {1988})}\BibitemShut {NoStop}%
\bibitem [{\citenamefont {Donati}\ \emph {et~al.}(1998)\citenamefont {Donati},
  \citenamefont {Douglas}, \citenamefont {Kob}, \citenamefont {Plimpton},
  \citenamefont {Poole},\ and\ \citenamefont {Glotzer}}]{donati1998}%
  \BibitemOpen
  \bibfield  {author} {\bibinfo {author} {\bibfnamefont {C.}~\bibnamefont
  {Donati}}, \bibinfo {author} {\bibfnamefont {J.~F.}\ \bibnamefont {Douglas}},
  \bibinfo {author} {\bibfnamefont {W.}~\bibnamefont {Kob}}, \bibinfo {author}
  {\bibfnamefont {S.~J.}\ \bibnamefont {Plimpton}}, \bibinfo {author}
  {\bibfnamefont {P.~H.}\ \bibnamefont {Poole}},\ and\ \bibinfo {author}
  {\bibfnamefont {S.~C.}\ \bibnamefont {Glotzer}},\ }\bibfield  {title}
  {\bibinfo {title} {Stringlike cooperative motion in a supercooled liquid},\
  }\href {https://doi.org/10.1103/PhysRevLett.80.2338} {\bibfield  {journal}
  {\bibinfo  {journal} {Phys. Rev. Lett.}\ }\textbf {\bibinfo {volume} {80}},\
  \bibinfo {pages} {2338} (\bibinfo {year} {1998})}\BibitemShut {NoStop}%
\bibitem [{\citenamefont {Kumar}\ \emph {et~al.}(2006)\citenamefont {Kumar},
  \citenamefont {Szamel},\ and\ \citenamefont {Douglas}}]{kumar2006}%
  \BibitemOpen
  \bibfield  {author} {\bibinfo {author} {\bibfnamefont {S.~K.}\ \bibnamefont
  {Kumar}}, \bibinfo {author} {\bibfnamefont {G.}~\bibnamefont {Szamel}},\ and\
  \bibinfo {author} {\bibfnamefont {J.~F.}\ \bibnamefont {Douglas}},\
  }\bibfield  {title} {\bibinfo {title} {Nature of the breakdown in the stokes-
  einstein relationship in a hard sphere fluid},\ }\href
  {https://doi.org/10.1063/1.2192769} {\bibfield  {journal} {\bibinfo
  {journal} {J. CHem. Phys.}\ }\textbf {\bibinfo {volume} {124}},\ \bibinfo
  {pages} {214501} (\bibinfo {year} {2006})}\BibitemShut {NoStop}%
\bibitem [{\citenamefont {Douglass}\ and\ \citenamefont
  {Dyre}(2022)}]{douglass2022}%
  \BibitemOpen
  \bibfield  {author} {\bibinfo {author} {\bibfnamefont {I.~M.}\ \bibnamefont
  {Douglass}}\ and\ \bibinfo {author} {\bibfnamefont {J.~C.}\ \bibnamefont
  {Dyre}},\ }\bibfield  {title} {\bibinfo {title} {Distance-as-time in physical
  aging},\ }\href@noop {} {\bibfield  {journal} {\bibinfo  {journal} {arXiv}\
  ,\ \bibinfo {pages} {2205.07658}} (\bibinfo {year} {2022})}\BibitemShut
  {NoStop}%
\bibitem [{\citenamefont {Niss}\ \emph {et~al.}(2020)\citenamefont {Niss},
  \citenamefont {Dyre},\ and\ \citenamefont {Hecksher}}]{niss2020}%
  \BibitemOpen
  \bibfield  {author} {\bibinfo {author} {\bibfnamefont {K.}~\bibnamefont
  {Niss}}, \bibinfo {author} {\bibfnamefont {J.~C.}\ \bibnamefont {Dyre}},\
  and\ \bibinfo {author} {\bibfnamefont {T.}~\bibnamefont {Hecksher}},\
  }\bibfield  {title} {\bibinfo {title} {Long-time structural relaxation of
  glass- forming liquids: Simple or stretched exponential?},\ }\href
  {https://doi.org/10.1063/1.5142189} {\bibfield  {journal} {\bibinfo
  {journal} {J. Chem. Phys.}\ }\textbf {\bibinfo {volume} {152}},\ \bibinfo
  {pages} {041103} (\bibinfo {year} {2020})}\BibitemShut {NoStop}%
\bibitem [{\citenamefont {Ciamarra}\ \emph {et~al.}(2023)\citenamefont
  {Ciamarra}, \citenamefont {Ji},\ and\ \citenamefont {Wyart}}]{massimo2023}%
  \BibitemOpen
  \bibfield  {author} {\bibinfo {author} {\bibfnamefont {M.~P.}\ \bibnamefont
  {Ciamarra}}, \bibinfo {author} {\bibfnamefont {W.}~\bibnamefont {Ji}},\ and\
  \bibinfo {author} {\bibfnamefont {M.}~\bibnamefont {Wyart}},\ }\bibfield
  {title} {\bibinfo {title} {The energy cost of local rearrangements, not
  cooperative effects, makes glasses solid},\ }\href
  {https://doi.org/10.48550/arXiv.2302.05150} {\bibfield  {journal} {\bibinfo
  {journal} {arXiv}\ ,\ \bibinfo {pages} {2302.05150}} (\bibinfo {year}
  {2023})}\BibitemShut {NoStop}%
\bibitem [{\citenamefont {Li}\ \emph {et~al.}(2020)\citenamefont {Li},
  \citenamefont {Lou}, \citenamefont {Kob},\ and\ \citenamefont
  {Granick}}]{li2020}%
  \BibitemOpen
  \bibfield  {author} {\bibinfo {author} {\bibfnamefont {B.}~\bibnamefont
  {Li}}, \bibinfo {author} {\bibfnamefont {K.}~\bibnamefont {Lou}}, \bibinfo
  {author} {\bibfnamefont {W.}~\bibnamefont {Kob}},\ and\ \bibinfo {author}
  {\bibfnamefont {S.}~\bibnamefont {Granick}},\ }\bibfield  {title} {\bibinfo
  {title} {Anatomy of cage formation in a two-dimensional glass-forming
  liquid},\ }\href {https://doi.org/10.1038/s41586-020-2869-5} {\bibfield
  {journal} {\bibinfo  {journal} {Nature}\ }\textbf {\bibinfo {volume} {587}},\
  \bibinfo {pages} {225} (\bibinfo {year} {2020})}\BibitemShut {NoStop}%
\bibitem [{\citenamefont {G{\"{o}}tze}(2008)}]{goetzebook}%
  \BibitemOpen
  \bibfield  {author} {\bibinfo {author} {\bibfnamefont {W.}~\bibnamefont
  {G{\"{o}}tze}},\ }\href@noop {} {\emph {\bibinfo {title} {Complex Dynamics of
  Glass-Forming Liquids}}}\ (\bibinfo  {publisher} {Oxford University Press},\
  \bibinfo {year} {2008})\BibitemShut {NoStop}%
\bibitem [{\citenamefont {Das}(2004)}]{spdas2004}%
  \BibitemOpen
  \bibfield  {author} {\bibinfo {author} {\bibfnamefont {S.~P.}\ \bibnamefont
  {Das}},\ }\bibfield  {title} {\bibinfo {title} {Mode-coupling theory and the
  glass transition in supercooled liquids},\ }\href
  {https://doi.org/10.1103/RevModPhys.76.785} {\bibfield  {journal} {\bibinfo
  {journal} {Rev. Mod. Phys.}\ }\textbf {\bibinfo {volume} {76}},\ \bibinfo
  {pages} {785} (\bibinfo {year} {2004})}\BibitemShut {NoStop}%
\bibitem [{\citenamefont {Biroli}\ \emph {et~al.}(2006)\citenamefont {Biroli},
  \citenamefont {Bouchaud}, \citenamefont {Miyazaki},\ and\ \citenamefont
  {Reichman}}]{IMCT}%
  \BibitemOpen
  \bibfield  {author} {\bibinfo {author} {\bibfnamefont {G.}~\bibnamefont
  {Biroli}}, \bibinfo {author} {\bibfnamefont {J.-P.}\ \bibnamefont
  {Bouchaud}}, \bibinfo {author} {\bibfnamefont {K.}~\bibnamefont {Miyazaki}},\
  and\ \bibinfo {author} {\bibfnamefont {D.~R.}\ \bibnamefont {Reichman}},\
  }\bibfield  {title} {\bibinfo {title} {Inhomogeneous mode-coupling theory and
  growing dynamic length in supercooled liquids},\ }\href
  {https://doi.org/10.1103/PhysRevLett.97.195701} {\bibfield  {journal}
  {\bibinfo  {journal} {Phys. Rev. Lett.}\ }\textbf {\bibinfo {volume} {97}},\
  \bibinfo {pages} {195701} (\bibinfo {year} {2006})}\BibitemShut {NoStop}%
\bibitem [{\citenamefont {Lubchenko}\ and\ \citenamefont
  {Wolynes}(2007)}]{lubchenko2007}%
  \BibitemOpen
  \bibfield  {author} {\bibinfo {author} {\bibfnamefont {V.}~\bibnamefont
  {Lubchenko}}\ and\ \bibinfo {author} {\bibfnamefont {P.~G.}\ \bibnamefont
  {Wolynes}},\ }\bibfield  {title} {\bibinfo {title} {Theory of structural
  glasses and supercooled liquids},\ }\href
  {https://doi.org/10.1146/annurev.physchem.58.032806.104653} {\bibfield
  {journal} {\bibinfo  {journal} {Ann. Rev. Phys. Chem.}\ }\textbf {\bibinfo
  {volume} {58}},\ \bibinfo {pages} {235} (\bibinfo {year} {2007})}\BibitemShut
  {NoStop}%
\bibitem [{\citenamefont {Adam}\ and\ \citenamefont
  {Gibbs}(1965)}]{adamgibbs1965}%
  \BibitemOpen
  \bibfield  {author} {\bibinfo {author} {\bibfnamefont {G.}~\bibnamefont
  {Adam}}\ and\ \bibinfo {author} {\bibfnamefont {J.~H.}\ \bibnamefont
  {Gibbs}},\ }\bibfield  {title} {\bibinfo {title} {On the temperature
  dependence of cooperative relaxation properties in glass‐forming liquids},\
  }\href {https://doi.org/10.1063/1.1696442} {\bibfield  {journal} {\bibinfo
  {journal} {J. Chem. Phys.}\ }\textbf {\bibinfo {volume} {43}},\ \bibinfo
  {pages} {139} (\bibinfo {year} {1965})}\BibitemShut {NoStop}%
\bibitem [{\citenamefont {Franz}\ \emph {et~al.}(2011)\citenamefont {Franz},
  \citenamefont {Parisi}, \citenamefont {Ricci-Tersenghi},\ and\ \citenamefont
  {Rizzo}}]{franz2011}%
  \BibitemOpen
  \bibfield  {author} {\bibinfo {author} {\bibfnamefont {S.}~\bibnamefont
  {Franz}}, \bibinfo {author} {\bibfnamefont {G.}~\bibnamefont {Parisi}},
  \bibinfo {author} {\bibfnamefont {F.}~\bibnamefont {Ricci-Tersenghi}},\ and\
  \bibinfo {author} {\bibfnamefont {T.}~\bibnamefont {Rizzo}},\ }\bibfield
  {title} {\bibinfo {title} {Field theory of fluctuations in glasses},\ }\href
  {https://doi.org/10.1140/epje/i2011-11102-0} {\bibfield  {journal} {\bibinfo
  {journal} {Euro. Phys. J. E}\ }\textbf {\bibinfo {volume} {34}},\ \bibinfo
  {pages} {102} (\bibinfo {year} {2011})}\BibitemShut {NoStop}%
\bibitem [{\citenamefont {Biroli}\ \emph {et~al.}(2022)\citenamefont {Biroli},
  \citenamefont {Charbonneau}, \citenamefont {Hu},\ and\ \citenamefont
  {Zamponi}}]{biroli2022}%
  \BibitemOpen
  \bibfield  {author} {\bibinfo {author} {\bibfnamefont {G.}~\bibnamefont
  {Biroli}}, \bibinfo {author} {\bibfnamefont {P.}~\bibnamefont {Charbonneau}},
  \bibinfo {author} {\bibfnamefont {G.~F.~Y.}\ \bibnamefont {Hu}},\ and\
  \bibinfo {author} {\bibfnamefont {F.}~\bibnamefont {Zamponi}},\ }\bibfield
  {title} {\bibinfo {title} {Local dynamical heterogeneity in simple glass
  formers},\ }\href {https://doi.org/10.1103/PhysRevLett.128.175501} {\bibfield
   {journal} {\bibinfo  {journal} {Phys. Rev. Lett.}\ }\textbf {\bibinfo
  {volume} {128}},\ \bibinfo {pages} {175501} (\bibinfo {year}
  {2022})}\BibitemShut {NoStop}%
\bibitem [{\citenamefont {Mari}\ and\ \citenamefont
  {Kurchan}(2011)}]{mari2011}%
  \BibitemOpen
  \bibfield  {author} {\bibinfo {author} {\bibfnamefont {R.}~\bibnamefont
  {Mari}}\ and\ \bibinfo {author} {\bibfnamefont {J.}~\bibnamefont {Kurchan}},\
  }\bibfield  {title} {\bibinfo {title} {Dynamical transition of glasses: From
  exact to approximate},\ }\href {https://doi.org/10.1063/1.3626802} {\bibfield
   {journal} {\bibinfo  {journal} {J. Chem. Phys.}\ }\textbf {\bibinfo {volume}
  {135}},\ \bibinfo {pages} {124504} (\bibinfo {year} {2011})}\BibitemShut
  {NoStop}%
\bibitem [{\citenamefont {Charbonneau}\ \emph {et~al.}(2014)\citenamefont
  {Charbonneau}, \citenamefont {Jin}, \citenamefont {Parisi},\ and\
  \citenamefont {Zamponi}}]{charbonneau2014}%
  \BibitemOpen
  \bibfield  {author} {\bibinfo {author} {\bibfnamefont {P.}~\bibnamefont
  {Charbonneau}}, \bibinfo {author} {\bibfnamefont {Y.}~\bibnamefont {Jin}},
  \bibinfo {author} {\bibfnamefont {G.}~\bibnamefont {Parisi}},\ and\ \bibinfo
  {author} {\bibfnamefont {F.}~\bibnamefont {Zamponi}},\ }\bibfield  {title}
  {\bibinfo {title} {Hopping and the stokes–einstein relation breakdown in
  simple glass formers},\ }\href {https://doi.org/10.1073/pnas.141718211}
  {\bibfield  {journal} {\bibinfo  {journal} {Proc. Natl. Acad. Sci.}\ }\textbf
  {\bibinfo {volume} {111}},\ \bibinfo {pages} {15025} (\bibinfo {year}
  {2014})}\BibitemShut {NoStop}%
\bibitem [{\citenamefont {Kob}\ and\ \citenamefont {Andersen}(1995)}]{kob1995}%
  \BibitemOpen
  \bibfield  {author} {\bibinfo {author} {\bibfnamefont {W.}~\bibnamefont
  {Kob}}\ and\ \bibinfo {author} {\bibfnamefont {H.~C.}\ \bibnamefont
  {Andersen}},\ }\bibfield  {title} {\bibinfo {title} {Testing mode-coupling
  theory for a supercooled binary lennard-jones mixture i: The van hove
  correlation function},\ }\href {https://doi.org/10.1103/PhysRevE.51.4626}
  {\bibfield  {journal} {\bibinfo  {journal} {Phys. Rev. E}\ }\textbf {\bibinfo
  {volume} {51}},\ \bibinfo {pages} {4626} (\bibinfo {year}
  {1995})}\BibitemShut {NoStop}%
\bibitem [{\citenamefont {Thompson}\ \emph {et~al.}(2022)\citenamefont
  {Thompson}, \citenamefont {Aktulga}, \citenamefont {Berger}, \citenamefont
  {Bolintineanu}, \citenamefont {Brown}, \citenamefont {Crozier}, \citenamefont
  {in~'t Veld}, \citenamefont {Kohlmeyer}, \citenamefont {Moore}, \citenamefont
  {Nguyen}, \citenamefont {Shan}, \citenamefont {Stevens}, \citenamefont
  {Tranchida}, \citenamefont {Trott},\ and\ \citenamefont {Plimpton}}]{LAMMPS}%
  \BibitemOpen
  \bibfield  {author} {\bibinfo {author} {\bibfnamefont {A.~P.}\ \bibnamefont
  {Thompson}}, \bibinfo {author} {\bibfnamefont {H.~M.}\ \bibnamefont
  {Aktulga}}, \bibinfo {author} {\bibfnamefont {R.}~\bibnamefont {Berger}},
  \bibinfo {author} {\bibfnamefont {D.~S.}\ \bibnamefont {Bolintineanu}},
  \bibinfo {author} {\bibfnamefont {W.~M.}\ \bibnamefont {Brown}}, \bibinfo
  {author} {\bibfnamefont {P.~S.}\ \bibnamefont {Crozier}}, \bibinfo {author}
  {\bibfnamefont {P.~J.}\ \bibnamefont {in~'t Veld}}, \bibinfo {author}
  {\bibfnamefont {A.}~\bibnamefont {Kohlmeyer}}, \bibinfo {author}
  {\bibfnamefont {S.~G.}\ \bibnamefont {Moore}}, \bibinfo {author}
  {\bibfnamefont {T.~D.}\ \bibnamefont {Nguyen}}, \bibinfo {author}
  {\bibfnamefont {R.}~\bibnamefont {Shan}}, \bibinfo {author} {\bibfnamefont
  {M.~J.}\ \bibnamefont {Stevens}}, \bibinfo {author} {\bibfnamefont
  {J.}~\bibnamefont {Tranchida}}, \bibinfo {author} {\bibfnamefont
  {C.}~\bibnamefont {Trott}},\ and\ \bibinfo {author} {\bibfnamefont {S.~J.}\
  \bibnamefont {Plimpton}},\ }\bibfield  {title} {\bibinfo {title} {{LAMMPS} -
  a flexible simulation tool for particle-based materials modeling at the
  atomic, meso, and continuum scales},\ }\href
  {https://doi.org/10.1016/j.cpc.2021.108171} {\bibfield  {journal} {\bibinfo
  {journal} {Comp. Phys. Comm.}\ }\textbf {\bibinfo {volume} {271}},\ \bibinfo
  {pages} {108171} (\bibinfo {year} {2022})}\BibitemShut {NoStop}%
\bibitem [{\citenamefont {Durian}(1995)}]{durian1995foam}%
  \BibitemOpen
  \bibfield  {author} {\bibinfo {author} {\bibfnamefont {D.~J.}\ \bibnamefont
  {Durian}},\ }\bibfield  {title} {\bibinfo {title} {Foam mechanics at the
  bubble scale},\ }\href@noop {} {\bibfield  {journal} {\bibinfo  {journal}
  {Physical review letters}\ }\textbf {\bibinfo {volume} {75}},\ \bibinfo
  {pages} {4780} (\bibinfo {year} {1995})}\BibitemShut {NoStop}%
\bibitem [{\citenamefont {Berthier}\ and\ \citenamefont
  {Witten}(2009)}]{berthier2009compressing}%
  \BibitemOpen
  \bibfield  {author} {\bibinfo {author} {\bibfnamefont {L.}~\bibnamefont
  {Berthier}}\ and\ \bibinfo {author} {\bibfnamefont {T.~A.}\ \bibnamefont
  {Witten}},\ }\bibfield  {title} {\bibinfo {title} {Compressing nearly hard
  sphere fluids increases glass fragility},\ }\href@noop {} {\bibfield
  {journal} {\bibinfo  {journal} {EPL (Europhysics Letters)}\ }\textbf
  {\bibinfo {volume} {86}},\ \bibinfo {pages} {10001} (\bibinfo {year}
  {2009})}\BibitemShut {NoStop}%
\bibitem [{\citenamefont {Charbonneau}\ \emph {et~al.}(2011)\citenamefont
  {Charbonneau}, \citenamefont {Ikeda}, \citenamefont {Parisi},\ and\
  \citenamefont {Zamponi}}]{charbonneau2011glass}%
  \BibitemOpen
  \bibfield  {author} {\bibinfo {author} {\bibfnamefont {P.}~\bibnamefont
  {Charbonneau}}, \bibinfo {author} {\bibfnamefont {A.}~\bibnamefont {Ikeda}},
  \bibinfo {author} {\bibfnamefont {G.}~\bibnamefont {Parisi}},\ and\ \bibinfo
  {author} {\bibfnamefont {F.}~\bibnamefont {Zamponi}},\ }\bibfield  {title}
  {\bibinfo {title} {Glass transition and random close packing above three
  dimensions},\ }\href@noop {} {\bibfield  {journal} {\bibinfo  {journal}
  {Physical review letters}\ }\textbf {\bibinfo {volume} {107}},\ \bibinfo
  {pages} {185702} (\bibinfo {year} {2011})}\BibitemShut {NoStop}%
\bibitem [{\citenamefont {Brown}\ and\ \citenamefont
  {Clarke}(1984)}]{brown1984comparison}%
  \BibitemOpen
  \bibfield  {author} {\bibinfo {author} {\bibfnamefont {D.}~\bibnamefont
  {Brown}}\ and\ \bibinfo {author} {\bibfnamefont {J.}~\bibnamefont {Clarke}},\
  }\bibfield  {title} {\bibinfo {title} {A comparison of constant energy,
  constant temperature and constant pressure ensembles in molecular dynamics
  simulations of atomic liquids},\ }\href@noop {} {\bibfield  {journal}
  {\bibinfo  {journal} {Molecular Physics}\ }\textbf {\bibinfo {volume} {51}},\
  \bibinfo {pages} {1243} (\bibinfo {year} {1984})}\BibitemShut {NoStop}%
\end{thebibliography}
%

\end{document}